\documentclass[letterpaper, 11pt]{article}
\pdfoutput=1 

\usepackage{jheppub}

\usepackage[T1]{fontenc}

\newcommand{\be}{\begin{equation}}
\newcommand{\ee}{\end{equation}}
\graphicspath{ {graphs/} }

\preprint{YITP-SB-2017-44}

\title{\boldmath 
How to Succeed at Holographic Correlators without Really Trying
}

\author[]{Leonardo Rastelli,}
\author[]{Xinan Zhou}

\affiliation[]{C. N. Yang Institute for Theoretical Physics, Stony Brook University, Stony Brook, 11794, NY, USA}

\abstract{We give a detailed account of the methods introduced in  \cite{Rastelli:2016nze} to calculate holographic four-point correlators
in IIB supergravity on $AdS_5 \times S^5$.   
Our approach relies entirely on general consistency conditions and maximal supersymmetry. We discuss two related methods, one in position space and the other in Mellin space.
The position space method is based on the observation that the holographic  four-point correlators of one-half  BPS single-trace operators can be written as   finite sums of contact Witten diagrams. We demonstrate in several examples that imposing the superconformal Ward identity is sufficient to fix the parameters of this ansatz uniquely,  avoiding  the need for a detailed knowledge of the supergravity effective action. The Mellin space approach is an ``on-shell method''
inspired by  the close analogy between holographic correlators and flat space scattering amplitudes. 
We conjecture a compact formula
for the four-point correlators of one-half BPS single-trace operators of arbitrary weights. Our general formula has the expected analytic structure, obeys the superconformal Ward identity, satisfies the appropriate asymptotic conditions and reproduces all the previously calculated cases.  We believe that these conditions determine it uniquely.
}

\begin{document}
\maketitle
\flushbottom

\section{Introduction}\label{intro}

Thanks to integrability, ${\cal N}=4$ super Yang-Mills  (SYM) theory should be completely tractable in the planar limit. However,  much work remains to turn this statement  of principle into a practical
computational recipe. A basic class of observables that still defy our technical abilities are the four-point correlation functions of one-half BPS local operators,
\begin{equation} \label{general4pt}
\langle   {\cal O}_{p_1} (x_1)  {\cal O}_{p_2} (x_2)   {\cal O}_{p_3} (x_3)  {\cal O}_{p_4} (x_4) \rangle \, , 
\end{equation}
with $\mathcal{O}_{p}(x) = {\rm Tr}\, X^{\{I_1}\ldots X^{I_{p}\} }(x)$, $I_k = 1, \dots 6$, 
  in the symmetric-traceless  representation of the $SO(6)$ R-symmetry. For general weights $p_i$ and arbitrary 't Hooft coupling $\lambda$
  these correlators
are extremely complicated functions of the conformal and R-symmetry cross ratios, 
encoding a large amount of non-protected spectral data and operator product coefficients.\footnote{On the other hand, two- and three point functions of one-half BPS operators obey non-renormalization theorems  \cite{Freedman:1998tz, Lee:1998bxa,Intriligator:1998ig,  Intriligator:1999ff,Eden:1999gh,Petkou:1999fv,Howe:1999hz,Heslop:2001gp,Baggio:2012rr}  and are easily evaluated in free field theory. A non-renormalization theorem also holds for extremal and next-to-extremal correlators \cite{DHoker:1999jke,Bianchi:1999ie,Eden:1999kw,Erdmenger:1999pz,Eden:2000gg},   
defined respectively by the conditions $p_1 = p_2 +  p_3 + p_4$ and  $p_1 = p_2 +  p_3 + p_4-2$.
} Finding a useful representation for these correlators at any value of the 't Hooft coupling $\lambda$ will be a crucial benchmark for the statement that planar ${\cal N}=4$ SYM has been exactly solved.\footnote{
Four-point functions are the current frontier of the ${\cal N}=4$ integrability program -- see,  {\it e.g.}, \cite{Fleury:2016ykk, Eden:2016xvg, Basso:2017khq} and references therein for  very interesting recent progress.}   

At strong coupling, planar ${\cal N}=4$ SYM has a dual description in terms of classical  IIB supergravity on $AdS_5 \times S^5$ \cite{Maldacena:1997re,Witten:1998qj,Gubser:1998bc}. Casual readers could be forgiven for supposing that a complete calculation of (\ref{general4pt}) in the supergravity limit must have been achieved
in the early days of  AdS/CFT. Far from it!  Kaluza-Klein supergravity is a devilishly complicated theory -- or so it {\it appears} in its effective action presentation -- and the standard methods of calculation run out of steam very quickly.
 Prior to our work only a few non-trivial  cases were known:
  \begin{enumerate}
  \item[(i)] The three simplest cases with four  identical weights, namely $p_i = 2$ \cite{Arutyunov:2000py},  $p_i = 3$ \cite{Arutyunov:2002fh}, and $p_i = 4$ \cite{Arutyunov:2003ae}.
  \item[(ii)]
 The next-to-next-to-extremal correlators with two equal weights, {\it i.e.} the cases $p_1 = n+k$, $p_2 = n -k$, $p_3 = p_4 = k+2$ \cite{Berdichevsky:2007xd,Uruchurtu:2008kp,Uruchurtu:2011wh}.\footnote{As we have remarked in the previous footnote, the extremal and next-to-extremal correlators do not depend on $\lambda$ and can thus be evaluated at $\lambda =0$ from Wick contractions in free field theory,
 yielding some simple rational functions of the cross ratios. It has been shown that the holographic calculation at $\lambda = \infty$ gives the same result \cite{DHoker:2000xhf,Arutyunov:1999fb,Arutyunov:2000ima}.}
 \end{enumerate}
The standard algorithm to evaluate holographic correlators is  straightforward but very cumbersome.
To the leading non-trivial order in the large $N$ expansion, one  is instructed to calculate a sum of tree-level Witten diagrams, with external legs given by bulk-to-boundary propagators and internal legs by bulk-to-bulk propagators. The vertices are read off from the effective action in $AdS_5$ obtained by Kaluza-Klein (KK) reduction
of IIB supergravity on $S^5$.  The evaluation  of the exchange Witten diagrams  is not immediate, but has been streamlined 
in a series of early papers \cite{DHoker:1998bqu,DHoker:1998ecp,Liu:1998th,DHoker:1999bve,DHoker:1999pj,DHoker:1999aa, Arutyunov:2002fh}. A key simplification \cite{DHoker:1999aa} that occurs for the $AdS_5 \times S^5$ background is that all the requisite exchange diagrams (see Figure \ref{exchange}) can be written as {\it finite} sums of contact  diagrams (Figure \ref{contact}),
the so-called $D$-functions. However, the supergravity effective action is extremely complicated \cite{Lee:1998bxa,Arutyunov:1999en,Arutyunov:1999fb}. The scalar quartic vertices were obtained by Arutyunov and Frolov \cite{Arutyunov:1999fb}
in a heroic undertaking and they fill 15 pages. Moreover, the number of exchange diagrams grows rapidly as the weights $p_i$ are increased,\footnote{Because of selection rules, the number of diagrams is vastly smaller for correlators near extremality, which explains why an explicit calculation is possible in those cases.} making it practically impossible to go beyond $p_i$ of the order of a few.
 What's worse, the final answer takes the  completely unintuitive form of a sum of $D$-functions. 
 It takes some work  to extract  from it even the leading OPE singularities.
  
  This sorry state of affairs is all the more embarrassing when contrasted with the beautiful progress in the field of flat space scattering amplitudes   (see, {\it e.g.}, \cite{Elvang:2015rqa, nima} for recent textbook presentations). Holographic correlators are the direct AdS analog of S-matrix amplitudes, to which in fact they reduce in a suitable limit,
  so we might hope to find for them analogous computational shortcuts and elegant geometric structures.
  A related motivation to 
   revisit this problem is our prejudice that  for the maximally supersymmetric $AdS_5 \times S^5$ background the holographic $n$-point functions of arbitrary KK modes must be completely fixed
   by general consistency conditions such as crossing symmetry and superconformal Ward identities. This is just a restatement of the on-shell uniqueness for the two-derivative action of IIB supergravity. 
    It should then be possible to directly  ``bootstrap'' the holographic correlators.   The natural language for this approach
    is the Mellin representation of conformal correlators, introduced by Mack  \cite{Mack:2009mi} for a general CFTs and advocated by Penedones  and others \cite{Penedones:2010ue, Paulos:2011ie, Fitzpatrick:2011ia, Costa:2012cb} as particularly
    natural in the   holographic context.    
 The analogy between  AdS correlators and flat space  scattering amplitudes becomes manifest in Mellin space:  holographic correlators  
    are   functions of Mandelstam-like invariants $s$, $t$, $u$, with poles and residues controlled by OPE factorization. (For the $AdS_5 \times S^5$ background, tree-level correlators are in fact {\it rational} functions -- 
    this is the Mellin counterpart of the fact that only a finite number  of $D$-functions are needed in position space.)
   However, most applications to date of the Mellin technology to holography ({\it e.g.}, \cite{Penedones:2010ue, Paulos:2011ie, Fitzpatrick:2011ia,Costa:2012cb, Aharony:2016dwx}) have focussed on the study
of individual Witten diagrams in toy models. This is not where the real simplification lies. The main message of our work is that one should focus on the total on-shell answer of the complete theory
and avoid the diagrammatic expansion
 altogether.

Our principal result is a compelling conjecture for the Mellin representation   
of the general one-half BPS four-point functions (\ref{general4pt}) in the supergravity limit.  We have found a very compact formula that obeys all the consistency conditions: Bose symmetry, expected analytic structure, correct asymptotic behavior at large $s$ and $t$, and superconformal Ward invariance. We have checked that our formula reproduces (in a more concise presentation) all the previously calculated examples. We believe
it is the unique solution of our set of algebraic conditions, but at present we can show uniqueness only for the simplest case ($p_i = 2$). 

We have also developed an independent position space method.
This method  mimics the conventional algorithm to calculate holographic correlators, writing the answer as a sum of exchange and contact Witten diagrams,
but it eschews  knowledge of the precise cubic and quartic couplings, which are left as undetermined parameters. 
 The exchange diagrams are expressed in terms of  ${D}$-functions, so that  all in all one is led
to an ansatz as a  finite sum of $D$-functions. Finally, the
undetermined couplings  are fixed by imposing the superconformal Ward identity. 
This method is completely rigorous, relying only on the structure of the supergravity calculation with no further assumptions. 
Despite being simpler than the conventional approach,
it also  becomes intractable as the weights $p_i$ increase.
We have obtained results for the cases with equal weights
 $p_i=2,3,4, 5$. The $p_i=5$ result is new and it agrees both with our Mellin formula
 and with a previous conjecture
by Dolan, Nirschl and Osborn \cite{Dolan:2006ec}.

The remainder of the paper is organized as follows. In Section \ref{sugrareview} we start with a quick review of the traditional method of calculation of four-point functions using supergravity. 
In Section {\ref{Mellinsection} we review and discuss the Mellin representation for CFT correlators and of Witten diagrams. We place a special emphasis on the simplifications expected 
in the large $N$ limit and when the operator dimensions take the special values that occur in our supergravity case.
 In Section \ref{mellinapproach}, after reviewing the constraints of superconformal invariance, we   formulate and solve an algebraic problem for the four-point Mellin amplitude
 of generic one-half BPS operators.
  We also discuss some technical subtleties about the relation between the Mellin and position space representations. The position space method is developed in Section \ref{positionspace}. We conclude in Section \ref{discussion} with a brief discussion.
Four Appendices collect some of the lengthier formulae and technical details.

\section{The traditional method}\label{sugrareview}
The standard recipe to
calculate holographic correlation functions follows from the most basic entry of the AdS/CFT dictionary \cite{Witten:1998qj,Maldacena:1997re,Gubser:1998bc}, which states that the generating functional of  boundary CFT correlators equals  the AdS path integral with boundary sources. Schematically,
\begin{equation}
\big<e^{ i \int_{\partial AdS} \bar{\varphi}_i \mathcal{O}_i }\big>_{\rm CFT}=Z[\bar{\varphi}_i]=\int_{AdS} \mathcal{D}\varphi_i \; e^{iS} \;\bigg|_{\varphi_i \big|_{z\to 0}=\bar{\varphi}_\Delta}\; .
\end{equation}
Here and throughout the paper we  are using the Poincar\'e coordinates
\begin{equation}
ds^2=R^2 \, \frac{dz^2+d\vec{x}^2}{z^2}\; .
\end{equation} 
The AdS radius $R$ will be   set to one by a choice of units, unless otherwise stated.

We  focus on the limit of the duality where the bulk theory becomes a weakly coupled gravity theory. As is familiar, for the canonical duality pair of $\mathcal{N}=4$ SYM and type IIB string theory on $AdS_5\times S^5$ this amounts to taking the number of colors $N$ large and further sending the 't Hooft coupling $\lambda = g_{YM}^2 N$ to infinity. In this limit, the bulk theory reduces to IIB supergravity with a small five-dimensional Newton constant $\kappa_5^2=4\pi^2 /N^2\ll 1$. The task of computing correlation functions in the strongly coupled planar gauge theory has thus become the task of computing suitably defined ``scattering amplitudes'' in the weakly coupled  supergravity on an $AdS_5$ background. The AdS supergravity amplitudes can be computed by a perturbative diagrammatic expansion,  in powers of the small Newton constant, where the so-called ``Witten diagrams'' play the role of position space Feynman diagrams. The Witten diagrams are ``LSZ reduced'', in the sense that their external legs (the bulk-to-boundary propagators) have been put ``on-shell'' with Dirichlet-like boundary conditions at the boundary $\partial AdS_{d+1}$.

In this paper we restrict ourselves to the evaluation of four-point correlation functions of the single-trace one-half BPS operators,
\be \label{superprimary}
\mathcal{O}^{(p)}_{I_1 \ldots I_p} \equiv  \textrm{Tr} X^{\{I_1}\ldots X^{I_p\}} \, , \quad p \geqslant 2 \, ,
\ee 
where $X^I$, $I = 1, \dots 6$ are the scalar fields  of the SYM theory,   in the $\bf 6$ representation of  $SO(6) \cong SU(4)$ R-symmetry.  
The symbol  $\{\ldots\}$ indicates the projection onto the symmetric traceless representation of $SO(6)$ --  in terms of $SU(4)$ Dynkin labels, this is the irrep denoted by $[0, p, 0]$.  In the notations of \cite{Dolan:2002zh},
the operators (\ref{superprimary})  are the superconformal primaries of the one-half BPS superconformal multiplets $\mathcal{B}^{(\frac{1}{2},\frac{1}{2})}_{[0,p,0]}$. They are annihilated 
 by half of the Poincar\'e supercharges and have protected dimensions $\Delta=p$. 
  By acting with the other half of the supercharges, one generates the full supermultiplet, which comprises a finite number of conformal primary operators  in various $SU(4)$
  representations and spin $\leqslant 2$ (see, {\it e.g.}, \cite{Dolan:2002zh} for a complete tabulation of the multiplet). 
Each conformal primary in the $\mathcal{B}^{(\frac{1}{2},\frac{1}{2})}_{[0,p,0]}$ multiplet is dual to a supergravity field in $AdS_5$,
  arising from the Kaluza-Klein reduction of IIB supergravity on $S^5$  \cite{Kim:1985ez}, with the integer $p$ corresponding to the KK level. For example,
 the  superprimary $\mathcal{O}^{(p)}$ is mapped to a bulk scalar field  $s_p$, which is a certain  linear combination of KK modes of the $10d$ metric and four-form with indices on the $S^5$.
 
 The traditional method evaluates the correlator of four operators (\ref{superprimary}) as the sum of all tree level diagrams with external legs $s_{p_1}$, $s_{p_2}$, $s_{p_3}$, $s_{p_4}$.
One needs the precise values of the cubic vertices responsible for  exchange diagrams (Figure \ref{exchange}), and of the quartic vertices responsible for the contact diagrams (Figure \ref{contact}). The relevant vertices  have been systematically worked out in the literature  \cite{Lee:1998bxa,Arutyunov:1999en,Lee:1999pj,Arutyunov:1999fb} and  take very complicated expressions.
Our methods, on the other hand, do not require the detailed form of these vertices, so we will only review some pertinent  qualitative features. 

Let us first focus on the cubic vertices. The only information that we  need  are selection rules, {\it i.e.},
which cubic vertices are non-vanishing.  An obvious constraint comes from the following product rule of $SU(4)$ representations, \begin{equation}
[0,p_1,0]\otimes [0,p_2,0]=\sum_{r=0}^{\min\{p_1,p_2\}}\sum_{s=0}^{\min\{p_1,p_2\}-r}[r,|p_2-p_1|+2s,r] \, ,
\end{equation}
which restricts the $SU(4)$ representations that can show up in an exchange diagram. We collect in Table \ref{exchangedmultiplets} (reproduced from \cite{Dolan:2002zh, Arutyunov:2003ae})
the list of bulk fields  $\{ \varphi_{\mu_1 \dots \mu_\ell  } \}$ that are  {\it a priori} allowed in an exchange diagram with external $s_{p_i}$ legs if one only imposes the R-symmetry selection rule.

\begin{table}[htp]
\begin{center}\begin{tabular}{|c|c|c|c|c|c|c|}\hline fields & $s_k$ & $A_{\mu,k}$ & $C_{\mu,k}$ & $\phi_k$ & $t_k$ & $\varphi_{\mu\nu,k}$ \\\hline $SU(4)$ irrep & $[0,k,0]$ & $[1,k-2,1]$ & $[1,k-4,1]$ & $[2,k-4,2]$ & $[0,k-4,0]$ & $[0,k-2,0]$ \\\hline $m^2$ & $k(k-4)$ & $k(k-2)$ & $k(k+2)$ & $k^2-4$ & $k(k+4)$ & $k^2-4$ \\\hline $\Delta$ & $k$ & $k+1$ & $k+3$ & $k+2$ & $k+4$ & $k+2$ \\\hline
$\Delta - \ell$ & $k$ & $k$ & $k+2$ & $k+2$ & $k+4$ & $k$ \\\hline \end{tabular} \caption{KK modes contributing to exchange diagrams with four external superprimary modes $s_k$.}
\end{center}
\label{exchangedmultiplets}
\end{table}
From the explicit expressions of the cubic vertices \cite{Arutyunov:1999en} one deduces two additional selection rules on the {\it twist} $\Delta- \ell$  of the field $\phi_{\mu_1 \dots \mu_\ell}$
in order for  the cubic vertex $s_{p_1} s_{p_2} \phi_{\mu_1 \dots \mu_\ell}$ to be non-vanishing,
\be
\Delta - \ell = p_1 + p_2 \quad ({\rm mod} \,2) \, , \qquad \Delta - \ell < p_1 + p_2  \,.
\ee
The selection rule on the parity of the twist can be understood as follows. In order for the cubic vertex $s_{p_1} s_{p_2} \phi_{\mu_1 \dots \mu_\ell}$ to be non-zero, it is necessary for 
the ``parent'' vertex $s_{p_1} s_{p_2} s_{p_3}$ be non-zero, where
 $s_{p_3}$ is the superprimary of which $\phi_{\mu_1 \dots \mu_\ell}$ is a descendant.  By $SU(4)$ selection rules,  $p_3$ must have the same parity as $p_1 + p_2$. One then checks that all descendants of  $s_{p_3}$ that are allowed to couple to $s_{p_1}$ and $s_{p_2}$ by $SU(4)$ selection rules have the same twist parity as $p_3$.  On the other hand, the selection rule  $\langle {\cal O}^{p_1}  {\cal O}^{p_2}  {\cal O}^{p_1+ p_2}\rangle$  is not fully explained by this kind of reasoning.  To understand it, we first need to recall that the cubic vertices obtained in \cite{Lee:1998bxa, Arutyunov:1999en} are cast in a ``canonical form''
 \be
 \int_{AdS_5}  c_{ikj}\, \varphi_i \varphi_j \varphi_k \, ,
 \ee
 by performing field redefinitions  that eliminate vertices with spacetime derivatives. This is harmless so long as the twists of the three fields satisfy a strict triangular inequality, but  subtle for the ``extremal case'' of one twist being equal  to the sum of the other two \cite{DHoker:1999jke}. For example, for the superprimaries, one finds that the cubic coupling $s_{p_1} s_{p_2} s_{p_1 + p_2}$ is absent, in apparent contradiction with the fact that the in ${\cal N}=4$ SYM three-point function $\langle {\cal O}^{p_1}  {\cal O}^{p_2}  {\cal O}^{p_1+ p_2}\rangle$ is certainly non-vanishing. One way to calculate  $\langle {\cal O}^{p_1}  {\cal O}^{p_2}  {\cal O}^{p_3 = p_1 + p_2}\rangle$ 
 is by analytic continuation in $p_3$ \cite{Lee:1998bxa, DHoker:1999jke}. One finds that while the coupling $c_{p_1 p_2 p_3} \sim (p_3 - p_2 - p_1)$, the requisite cubic contact Witten diagram diverges as $1/(p_3 - p_1 - p_2)$, so that their product yields the finite correct answer.\footnote{\label{footnotedt}
 If one wishes to work exactly at extremality $p_3 = p_1 + p_2$, one can understand  the finite three-point function as arising
 from  boundary terms that are thrown away by the field redefinition that brings the cubic vertex to the canonical non-derivative form \cite{DHoker:1999jke}. One can rephrase this phenomenon as follows \cite{Arutyunov:2000ima}:   the field redefinition on the supergravity side (which throws away boundary terms) amounts to a redefinition of the dual operators that adds  admixtures of multi-trace terms, ${\cal O}^p  \to {\cal O}^p + 1/N \sum_{k=2}^p\, c_{k}^p\,  {\cal O}^{p-k} {\cal O}^k + \dots$. The double-trace terms contribute  to the extremal three point functions
at leading large $N$ order, but 
 are subleading away from extremality. The operators dual to the redefined fields $s_p$ (which have only non-derivative cubic couplings) are  linear combinations of single and double-trace terms such all extremal three-point functions are zero,  in agreement with the vanishing of the extremal three-point vertices $s_{p_1} s_{p_2} s_{p_1 + p_2}
  $.
 } From this viewpoint, it is in fact {\it necessary}
 for the extremal coupling $c_{p_1 p_2 p_1+ p_2}$ to vanish,  or else one would find an infinite answer for the three-point function. This provides a rationale for the selection rule   
 $\Delta - \ell < p_1 + p_2$. When it is violated, the requisite three-point contact Witten diagram diverges, so the corresponding coupling must vanish.  We will see in Section \ref{truncationsection}, \ref{mellinwittendiagrams} that the selection rule has also a natural interpretation in Mellin space.

The requisite quartic vertices were obtained in \cite{Arutyunov:1999fb}.  The quartic terms in the effective action for the $s_k$ fields contain up to four spacetime derivatives, but we argued in \cite{Rastelli:2016nze} that compatibility with the flat space limit requires that holographic correlators can get contributions from vertices with at most {\it two} derivatives.
The argument is easiest to phrase in Mellin space and will be reviewed in Section \ref{flatspacelimit}. That is indeed the case in the handful of explicitly calculated examples \cite{Arutyunov:2000py,Arutyunov:2002fh,Arutyunov:2003ae,Berdichevsky:2007xd,Uruchurtu:2008kp,Uruchurtu:2011wh}.
Our claim has been recently proven  in full generality \cite{Arutyunov:2017dti}. These authors have shown that the four-derivative terms effectively cancel out 
 in all four-point correlators of one-half BPS operators, thanks to non-trivial group theoretic identities.

\begin{figure}[htbp]
\begin{center}
\includegraphics[scale=0.4]{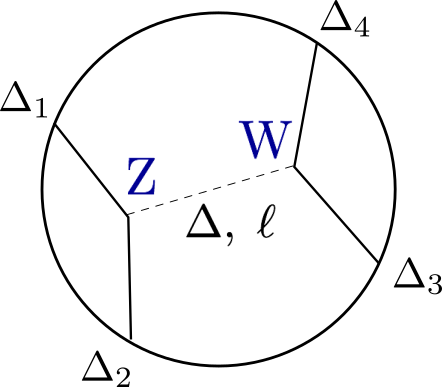}
\caption{An exchange Witten diagram.}
\label{exchange}
\end{center}
\end{figure}

\begin{figure}[htbp]
\begin{center}
\includegraphics[scale=0.25]{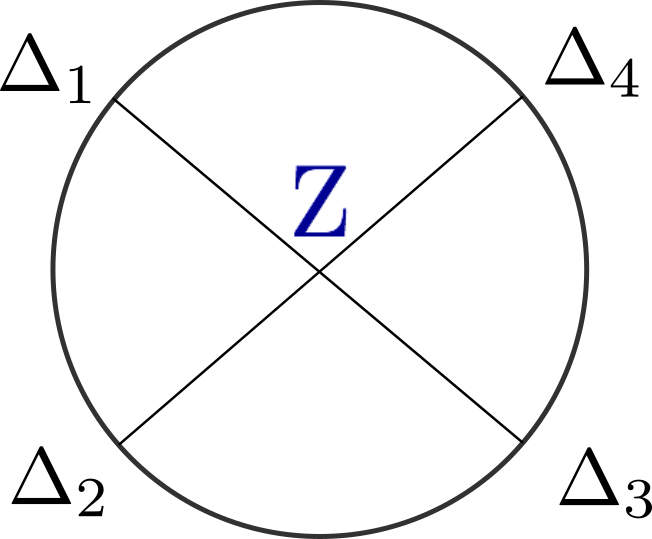}
\caption{A contact Witten diagram.}
\label{contact}
\end{center}
\end{figure}

The rules of evaluation of Witten diagrams are entirely analogous to the ones for position space Feynman diagrams: we assign a bulk-to-bulk propagator $G_{BB}(z,w)$ to each internal line connecting two bulk vertices at positions $z$ and $w$; and a bulk-to-boundary propagator $G_{B\partial }(z,\vec{x})$ to each external line connecting a bulk vertex at $z$ and a boundary point $\vec x$. These propagators are  Green's functions in AdS with appropriate boundary conditions. Finally, integrations over the bulk AdS
space are performed for each interacting vertex point. The simplest connected Witten diagram is a contact diagram of external scalars with no derivatives in the quartic vertex (Figure \ref{contact}). It is given by  the integral of the product of four scalar bulk-to-boundary propagators integrated over the common bulk point,
\begin{equation}
\mathcal{A}_{\rm contact}(\vec{x}_i)=\int_{AdS}dz\; G_{B\partial}(z,\vec{x}_1)\;G_{B\partial}(z,\vec{x}_2)\;G_{B\partial}(z,\vec{x}_3)\;G_{B\partial}(z,\vec{x}_4)\;.
\end{equation}
 Here, the scalar bulk-to-boundary propagator is \cite{Witten:1998qj}\footnote{Note that we are using the {\it unnormalized} propagator, to avoid cluttering of several formulae.
 In a complete calculation, care must be taken to add the well-known normalization factors \cite{Freedman:1998tz}.},
\begin{equation}
G_{B\partial}(z,\vec{x}_i)=\left(\frac{z_0}{z_0^2+(\vec{z}-\vec{x}_i)^2}\right)^{\Delta_i}
\end{equation}
 where $\Delta_i$ is the conformal dimension of the $i$th boundary CFT operator. The  integral can be evaluated in terms of derivatives of the dilogarithm function.
It is useful to give it a name, defining the so-called $D$-functions as the four-point scalar contact diagrams with external dimensions $\Delta_i$,
\begin{equation} \label{Dfunctions}
D_{\Delta_1\Delta_2\Delta_3\Delta_4}(x_1,x_2,x_3,x_4) \equiv \int_0^\infty \frac{dz_0}{z_0^{d+1}}\int d^dx \prod_{i=1}^4 \bigg(\frac{z_0}{z_0^2+(\vec z-\vec x_i)^2}\bigg)^{\Delta_i}\;.
\end{equation}
The other type of tree-level four-point diagrams are the exchange diagrams (Figure \ref{exchange}),
\begin{equation}
\mathcal{A}_{\rm exchange}(\vec x_i)=\int_{AdS}dz dw G_{B\partial}(z,\vec{x}_1)G_{B\partial}(z,\vec{x}_2)G_{BB}(z,w)G_{B\partial}(w,\vec{x}_3)G_{B\partial}(w,\vec{x}_4)
\end{equation}
Exchange diagrams are usually difficult to evaluate in closed form. In \cite{DHoker:1999aa} a technique was invented that allows, when  certain ``truncation conditions'' for the quantum numbers of the external and exchanged operators are met, to trade the propagator of an exchange diagram for  a {\it finite} sum of contact vertices. In such cases, one is able to evaluate an exchange Witten diagram as a finite sum of  $D$-functions. Fortunately, the spectrum and selection rules of  IIB supergravity on $AdS_5\times S^5$
are precisely such that all exchange diagrams obey the truncation conditions. We will exploit this fact in our position space method (Section \ref{positionspace}). The formulae for the requisite exchange diagrams have been collected in Appendix \ref{exch}.

\section{Mellin formalism}

\label{Mellinsection}

In this Section we review and discuss the Mellin amplitude formalism introduced by Mack  \cite{Mack:2009mi} and developed in \cite{Penedones:2010ue,Paulos:2011ie,Fitzpatrick:2011ia,Costa:2012cb,Fitzpatrick:2012cg,Costa:2014kfa,Goncalves:2014rfa}.\footnote{For other applications and recent developments, see \cite{Paulos:2012nu,Nandan:2013ip,Lowe:2016ucg,Rastelli:2016nze,Paulos:2016fap,Nizami:2016jgt,Gopakumar:2016cpb,Gopakumar:2016wkt,Dey:2016mcs,Aharony:2016dwx, Rastelli:2017ecj,Dey:2017fab, Yuan:2017vgp}). }
After introducing the basic formalism in Section \ref{mellinbasics}, we discuss the special features that occur large $N$ CFTs  in Section \ref{truncationsection} and 
review the application to tree-level four-point Witten diagrams in Section \ref{mellinwittendiagrams}. 
A remarkable simplification occurs  for  Witten diagrams with special values of the external and exchanged operator dimensions: the associated Mellin amplitude is a rational function
of the Mandelstam invariants $s$ and $t$. We explain that this is dictated by the consistency with the  structure of the operator product expansion at large $N$.
Finally, in Section \ref{flatspacelimit} we discuss the asymptotic behavior of the supergravity Mellin amplitude. Compatibility with the flat space limit 
gives an upper bound for the asymptotic growth of the supergravity Mellin amplitude at large $s$ and $t$.

\subsection{Mellin amplitudes for scalar correlators} \label{mellinbasics}

We consider a general  correlation function of $n$ scalar operators with conformal dimensions $\Delta_i$. Conformal symmetry restricts its form to be
\begin{equation}
G_{\Delta_1,\ldots,\Delta_n}(x_1,\ldots,x_n)=\prod_{i<j}(x^2_{ij})^{-\delta_{ij}^0}\mathcal{G}(\xi_r) \, ,
\end{equation}
where $\xi_r$ are the conformally invariant cross ratios constructed from $x_{ij}^2$,
\begin{equation}
\frac{(x_i-x_j)^2(x_k-x_l)^2}{(x_i-x_l)^2(x_k-x_j)^2}\;.
\end{equation}
Requiring that the correlator transforms with appropriate weights under conformal transformations, one finds the constraints
\begin{equation} \label{deltaconstraints}
\sum_{j\neq i}\delta_{ij}^0=\Delta_i\;.
\end{equation}
The number of independent cross ratios in a $d$-dimensional spacetime is given by
\begin{equation} \label{crossratiocounting}
\begin{split}
{}&n< d+1: \quad \frac{1}{2}n(n-3)\;,\\
{}&n\geqslant d+1: \quad nd-\frac{1}{2}(d+1)(d+2)\; ,
\end{split}
\end{equation}
as seen from a simple counting argument. We have a configuration space of $n$ points which is $nd$-dimensional, while the dimension of the conformal group $SO(d+1,1)$ is $\frac{1}{2}(d+1)(d+2)$. For sufficiently large $n$, the difference of the two gives the number of free parameters unfixed by the conformal symmetry, as in the second line of (\ref{crossratiocounting}). However this is incorrect for $n < d+1$ because we have overlooked a nontrivial stability group. To see this, we first use a conformal transformation to send two of the $n$ points to the origin and the infinity. If $n< d+1$, the remaining $n-2$ points will define a hyperplane and the stability group is the rotation group $SO(d+2-n)$ perpendicular to the hyperplane. After adding back the dimension of the stability group we get the first line of the counting. To phrase it differently, when the spacetime dimension $d$  is high enough, there are always $\frac{1}{2}n(n-3)$ conformal cross ratios, independent of the spacetime dimension. But when $n\geq d+1$ there exist nontrivial algebraic relations among the $\frac{1}{2}n(n-3)$ conformal cross ratios.

The constraints (\ref{deltaconstraints}) admit $\frac{1}{2}n(n-3)$ solutions, in correspondence with the $\frac{1}{2}n(n-3)$ cross ratios (ignoring the algebraic relations that  exist for small $n$). Mack \cite{Mack:2009mi} suggested instead of taking $\delta^0_{ij}$ to be fixed, we should view them as variables $\delta_{ij}$ satisfying the same constraints, 
\begin{equation}\label{deltaconstr}
\delta_{ij} = \delta_{ji}  \, ,\quad \sum_{j}\delta_{ij}=\Delta_i \, , 
\end{equation}
and write 
the correlator as an integral transform with respect to these variables. More precisely, one defines the following (inverse) Mellin transform for the {\it connected}\footnote{The disconnected part is a sum of powers of $x_{ij}^2$ and
its Mellin transform is singular.} part of the correlator,
\begin{equation}\label{mack}
G_{\Delta_1,\ldots,\Delta_n}^{\rm conn}(x_1,\ldots,x_n)=\int [d\delta_{ij}] M(\delta_{ij})\prod_{i<j}(x^2_{ij})^{-\delta_{ij}}
\end{equation}
The integration is performed with respect to the $\frac{1}{2}n(n-3)$ independent variables along the imaginary axis. We will be more specific about the integration in a moment. The correlator $\mathcal{G}(\xi_r)_{\rm conn}$ is captured by the function $M(\delta_{ij})$, which following Mack we shall call the {\it reduced} Mellin amplitude.

The constraints (\ref{deltaconstr}) can be solved by introducing some fictitious  ``momentum'' variables $k_i$ living in a $D$-dimensional spacetime,
\begin{equation}
\delta_{ij}=k_i\cdot k_j\;.
\end{equation}
These variables obey ``momentum conservation''
\begin{equation}
\sum_{i=1}^n k_i=0
\end{equation}
and the ``on-shell'' condition
\begin{equation}
k_i^2=-\Delta_i\;.
\end{equation}
The number of  independent  Lorentz invariants $\delta_{ij}$ (``Mandelstam variables'')   in a $D$-dimensional spacetime is given by 
\begin{equation} \label{mandelstamcounting}
\begin{split}
{}&n<D:\quad \frac{1}{2}n(n-3)\;,\\
{}&n\geqslant D:\quad n(D-1)-\frac{1}{2}D(D+1)\;.
\end{split}
\end{equation}
The counting goes as follows. The configuration space of $n$ on-shell momenta in $D$ dimensions is $n(D-1)$-dimensional, while the Poincar\'e group has dimension $\frac{1}{2}D(D+1)$. Assuming that the stability group is trivial, there will be $n(D-1)-\frac{1}{2}D(D+1)$ free parameters,  giving the second line of (\ref{mandelstamcounting}). However for $n< D$ there is a nontrivial stability group $SO(D-n+1)$. This can be seen by using momentum conservation to make the $n$ momenta lie in a $n-1$ dimensional hyperplane -- the rotations orthogonal to the hyperplane generate the stability group $SO(D-n+1)$. Adding  back the dimension of the stability group  we obtain the first line of (\ref{mandelstamcounting}). Again we see when $D$ is high enough, the number of independent Mandelstam variables is a $D$-independent number $\frac{1}{2}n(n-3)$. When $n\geq D$, the $\frac{1}{2}n(n-3)$ Mandelstam variables are subject to further relations. This is the counterpart of the statement we made about the conformal cross ratios. We conclude that the counting of independent Mandelstam variables in $D$ dimensions coincides precisely with the counting of independent conformal cross ratios in $d$ dimensions if we set $D=d+1$. 

The virtue of the integral representation (\ref{mack}) is to encode the consequences of the operator product expansion into   simple analytic properties for $M(\delta_{ij})$. Indeed, consider the OPE
\begin{equation}
{\cal O}_i (x_i) {\cal O}_j (x_j) = \sum_k  c_{ij}^{\; k} \, \left(  (x_{ij}^2)^{-\frac{\Delta_i + \Delta_j - \Delta_k}{2} } {\cal O}_k (x_k)    \,  + {\rm  descendants} \right) \, ,
\end{equation}
where for simplicity ${\cal O}_k$ is taken to be a scalar operator.  To reproduce the leading behavior as $x_{ij}^2 \to 0$, $M$ must have a pole at $\delta_{ij} =  \frac{\Delta_i + \Delta_j  - \Delta_k}{2}$, as can be seen by closing the $\delta_{ij}$ integration
contour to the left of the complex plane. More generally,
the location of the leading pole is controlled by the twist $\tau$ of the exchanged operator ($\tau \equiv \Delta- \ell$, the conformal dimension minus the spin). Conformal descendants contribute an infinite sequence of satellite poles,
so that all in all for any primary operator ${\cal O}_k$ of twist $\tau_k$  that contributes to the ${\cal O}_i {\cal O}_j$ OPE
the reduced Mellin amplitude $M(\delta_{ij})$ has poles at
\begin{equation}  \label{Mellinpoles}
\delta_{ij} =  \frac{\Delta_i + \Delta_j  - \tau_k - 2n}{2} \, , \quad n = 0, 1, 2 \dots \, .
\end{equation}
Mack further defined \textit{Mellin amplitude} $\mathcal{M}(\delta_{ij})$ by stripping off a product of Gamma functions,
\begin{equation}\label{Mackmellin}
\mathcal{M}(\delta_{ij}) \equiv \frac{M(\delta_{ij})}{\prod_{i<j}\Gamma[\delta_{ij}]}\, .
\end{equation} 
This is a convenient definition because ${\cal M}$ has simpler factorization properties. In particular, for the four-point function, the s-channel OPE $(x_{12} \to 0)$ implies
that the Mellin amplitude ${\cal M} (s, t)$ has poles in $s$ with residues that are {\it polynomials} of $t$. These {\it Mack polynomials}
depend on the spin of the exchanged operator, in analogy with the familiar partial wave expansion of  a flat-space S-matrix. (The analogy is not perfect, because each operator contributes an infinite of satellite poles, and because
Mack polynomials are significantly more involved than the Gegenbauer polynomials that appear in the usual flat-space partial wave expansion.) 
We will see in Section \ref{truncationsection} that Mack's definition of ${\cal M}$ is particularly natural for large $N$ theories. 

Finally let us comment on the integration  contours in (\ref{mack}). The prescription given in \cite{Mack:2009mi} is  that the real part of the arguments in the stripped off Gamma functions be
all positive along the integration contours. To be more precise, one is instructed to integrate $\frac{1}{2}n(n-3)$ independent variables $s_k$ along the imaginary axis,
where $s_k$ are related
  to $\delta_{ij}$ via
\begin{equation} \label{deltac}
\delta_{ij}=\delta_{ij}^0+\sum_{k=1}^{\frac{1}{2}n(n-3)} c_{ij,k}s_k\;.
\end{equation}
Here $\delta_{ij}^0$ is a special solution of the constraints (\ref{deltaconstr}) with $\Re(\delta_{ij}^0)>0$. The coefficients $c_{ij,k}$ are any solution of 
 \begin{equation}
\begin{split}
c_{ii,k}={}&0\;,\\
\sum_{j=1}^n c_{ij,k}={}&0\;,
\end{split}
\end{equation}
which is just the homogenous version (\ref{deltaconstr}). There are $\frac{1}{2}n(n-3)$ independent coefficients $c_{ij,k}$ for each  $k$. We can choose to integrate over
$c_{ij,k}$ with $2\leq i<j\leq n$
except for $c_{23,k}$, so that the chosen  $c_{ij,k}$ forms a $\frac{n(n-3)}{2}\times \frac{n(n-3)}{2}$ square matrix (the row index are the independent elements of the pair $(ij)$ and the column index is $k$).  We normalize this matrix to satisfy
\begin{equation} \label{detc}
|\det c_{ij,k}|=1\;.
\end{equation}

For four-point amplitudes, which are the focus of this paper, it is convenient to introduce
 ``Mandelstam'' variables $s$, $t$, $u$, and write
\begin{equation} \label{stu}
\begin{split}
\delta_{12}={}&-\frac{s}{2}+\frac{\Delta_1+\Delta_2}{2}\;,\;\;\;\;\;\;\;\;\delta_{34}=-\frac{s}{2}+\frac{\Delta_3+\Delta_4}{2}\;,\\
\delta_{23}={}&-\frac{t}{2}+\frac{\Delta_2+\Delta_3}{2}\;,\;\;\;\;\;\;\;\;\delta_{14}=-\frac{t}{2}+\frac{\Delta_1+\Delta_4}{2}\;,\\
\delta_{13}={}&-\frac{u}{2}+\frac{\Delta_1+\Delta_3}{2}\;,\;\;\;\;\;\;\;\;
\delta_{24}=-\frac{u}{2}+\frac{\Delta_2+\Delta_4}{2}\;.\\
\end{split}
\end{equation}
With this parametrization, the constraints obeyed by $\delta_{ij}$ translate into the single constraint
\be
s+t+u=\Delta_1+\Delta_2+\Delta_3+\Delta_4 \, . 
 \ee
We can take $s$ and $t$ as the independent integration variables, and rewrite the integration measure  as
\begin{equation}
\int [d\delta_{ij}]=\frac{1}{4}\int_{s_0-i\infty}^{s_0 + i\infty} ds \int_{t_0-i\infty}^{t_0+i\infty} dt\;  .
\end{equation}
 In fact this simple  contour prescription will need some modification.  In the context of the AdS supergravity calculations, we will find it  necessary to break the connected correlator into several terms and associate different contours to each term, instead of using a universal contour.   The are usually poles inside the region specified by $\Re(\delta_{ij}^0)>0$, and the answer given by the correct modified prescription differs from the naive one by the residues that are crossed in  deforming the contours.

\subsection{Large $N$} \label{truncationsection}

The Mellin formalism is ideally suited for large $N$ CFTs. While in a general CFT the analytic structure of  Mellin amplitudes is rather intricate, it becomes much simpler at large $N$.
To appreciate this point, we recall the remarkable theorem about the spectrum of CFTs in dimension $d >2$ proven in \cite{Fitzpatrick:2012yx,Komargodski:2012ek}. For any two primary operators ${\cal O}_1$ and ${\cal O}_2$ of twists $\tau_1$ and $\tau_2$, and for each non-negative integer $k$, 
the CFT must contain an infinite family of  so-called ``double-twist'' operators with increasing spin $\ell$ and twist approaching $\tau_1 + \tau_2 + 2k$ as $\ell \to \infty$ \cite{Komargodski:2012ek,Fitzpatrick:2012yx}.
This  implies that  the  Mellin amplitude   has infinite sequences of poles   accumulating at these asymptotic values of the twist, so  it is {\it not} a meromorphic function.\footnote{In two dimensions, there are no double-twist families,
but one encounters a different pathology: the existence of infinitely many operators of the same twist, because Virasoro generators have twist zero.}

As emphasized by Penedones \cite{Penedones:2010ue}, 
a key simplification occurs in large $N$ CFTs, where the double-twist operators  are recognized as  the  usual double-trace operators. Thanks to large $N$ factorization, spin $\ell$ conformal primaries of the schematic form 
$: {\cal O}_1 \Box^n \partial^\ell {\cal O}_2 :$, where ${\cal O}_1$ and ${\cal O}_2$ are single-trace operators,
 have twist  $\tau_1 + \tau_2 + 2n + O(1/N^2)$\footnote{For definiteness, we are using the large $N$ counting appropriate to a theory with matrix degrees of freedom, {\it e.g.}, a $U(N)$
gauge theory. In other kinds of large $N$ CFTs the leading correction would have a different power -- for example, $O(1/N^3)$ in the $A_N$ six-dimensional (2, 0) theory, and $O(1/N)$ in two-dimensional
symmetric product orbifolds.}  for any $\ell$.   Recall also that the Mellin amplitude 
 is defined in terms of the {\it connected} part of the $k$-point correlator, which is of order $O(1/N^{k-2})$ for unit-normalized single-trace operators. 
The contribution of intermediate double-trace operators arises precisely at  $O(1/N^2)$, 
 so that to this order we can use their uncorrected dimensions. Remarkably, the poles corresponding to the exchanged  double-trace operators are precisely captured by the product of Gamma functions $\prod_{i<j} \Gamma(\delta_{ij})$ that Mack stripped
 off to define the Mellin amplitude ${\cal M}$.  All in all, we conclude that the $O(1/N^{k-2})$ Mellin amplitude ${\cal M}$   {\it is} a meromorphic function, whose poles are controlled by just the exchanged {\it single-trace}  operators.

 Let us analyze in some detail the case of the four-point function. For four scalar operators ${\cal O}_i$ of dimensions $\Delta_i$, conformal covariance implies
\begin{equation}
\langle\mathcal{O}_1 (x_1) \mathcal{O}_2 (x_2) \mathcal{O}_3 (x_3) \mathcal{O}_4 (x_4) \rangle=\frac{1}{(x_{12}^2)^{\frac{\Delta_1+\Delta_2}{2}}(x_{34}^2)^{\frac{\Delta_3+\Delta_4}{2}}}\left(\frac{x_{24}^2}{x_{14}^2}\right)^{\frac{\Delta_1-\Delta_2}{2}}\left(\frac{x_{14}^2}{x_{13}^2}\right)^{\frac{\Delta_3-\Delta_4}{2}}\mathcal{G}(U,V)\; ,
\end{equation}
where $U$ and $V$ are the usual conformal cross-ratios\footnote{We use capital letters because  the symbol $u$ is already taken to denote the Mandelstam invariant, (\ref{stu}).}
\be
U = \frac{x_{12}^2 x_{34}^2}{x_{13}^2 x_{24}^2} \, ,\quad V = \frac{x_{14}^2 x_{23}^2}{x_{13}^2 x_{24}^2} \,.
\ee
Taking the operators ${\cal O}_i$ to be unit-normalized single-trace operators, and separating out the disconnected and connected terms,\footnote{The disconnected term  $\mathcal{G}_{\rm disc}$ will of course vanish unless the four operators are pairwise identical.}
\be
{\cal G} =  \mathcal{G}_{\rm disc} +\mathcal{G}_{\rm conn}\, ,
\ee
we have the following familiar large $N$ counting:
\be
 \mathcal{G}_{\rm disc} = O(1) \, ,  \quad \mathcal{G}_{\rm conn} =  \frac{1}{N^2} {\cal G}^{(1)} + \frac{1}{N^4}  {\cal G}^{(2)} + \dots
\ee
The Mellin amplitude ${\cal M}$ is defined by the integral transform
\begin{equation}
\begin{split} \label{GMellin}
\mathcal{G}_{\rm conn}(U,V)
={}&\int_{-i\infty}^{i\infty} \frac{ds}{2}\frac{dt}{2} U^{\frac{s}{2}}V^{\frac{t}{2}-\frac{\Delta_2+\Delta_3}{2}} \mathcal{M}(s,t)\Gamma[\frac{\Delta_1+\Delta_2-s}{2}]\Gamma[\frac{\Delta_3+\Delta_4-s}{2}]\\
{}&\times \Gamma[\frac{\Delta_1+\Delta_4-t}{2}]\Gamma[\frac{\Delta_2+\Delta_3-t}{2}]\Gamma[\frac{\Delta_1+\Delta_3-u}{2}]\Gamma[\frac{\Delta_2+\Delta_4-u}{2}]\;,
\end{split}
\end{equation}
with $s+t+u=\Delta_1+\Delta_2+\Delta_3+\Delta_4$.  

Let us first assume that the dimensions $\Delta_i$ are generic.  In the s-channel OPE, we expect contributions to ${\cal G}_{\rm conn}$
from the tower of double-trace operators of the  form\footnote{In fact for fixed $n$ and $\ell$, there are in general multiple conformal primaries
of this schematic form, which differ in the way the derivatives are distributed.}
 $:\mathcal{O}_1\square^n \partial^\ell\mathcal{O}_2:$, 
with twists $\tau = \Delta_1 + \Delta_2 + 2n + O(1/N^2)$, and from the tower $:\mathcal{O}_{3}\square^n \partial^\ell \mathcal{O}_{4}:$, which have twists $\tau = \Delta_1 + \Delta_2 + 2n + O(1/N^2)$. 
The OPE coefficients scale as 
\begin{eqnarray}
\langle {\cal O}_1 \; {\cal O}_2   \; :\mathcal{O}_1\square^n \partial^\ell\mathcal{O}_2: \rangle = O(1) \, &,& \quad \langle {\cal O}_3  \; {\cal O}_4  \; :\mathcal{O}_1 \square^n \partial^\ell\mathcal{O}_2 : \rangle  = O(1/N^2) \, , \\
\langle {\cal O}_3 \; {\cal O}_4   \; :\mathcal{O}_3 \square^n \partial^\ell\mathcal{O}_4: \rangle = O(1) \, &,& \quad \langle {\cal O}_1 \; {\cal O}_2  \;:\mathcal{O}_{3}\square^n \partial^\ell\mathcal{O}_{4}: \rangle  = O(1/N^2) \, ,\nonumber
\end{eqnarray}
so that to leading $\mathcal{O}(1/N^2)$ order, we can neglect the $1/N^2$ corrections to  the conformal dimensions of the double-trace operators. All in all, we expect that these towers of double-trace operators 
contribute  poles in $s$ at
\begin{equation}
\begin{split} \label{sequence}
s={}&\Delta_1+\Delta_2+2m_{12}\;,\quad\quad\quad m_{12}\in \mathbb{Z}_{\geqslant 0}        \;,\\
s={}&\Delta_3+\Delta_4+2m_{34}\;,\quad\quad\quad  m_{34}\in \mathbb{Z}_{\geqslant 0}.  \;. \end{split}
\end{equation}
These are precisely the locations of the poles of the first two Gamma functions in (\ref{GMellin}). In complete analogy, the poles in $t$ and $u$ in the other Gamma functions account for the contributions
of the double-trace operators exchanged in the $t$ and $u$ channels.

If  $\Delta_1+\Delta_2-(\Delta_3+\Delta_4)=0\;{\rm mod}\; 2$,  the two sequences of poles in (\ref{sequence}) (partially) overlap, giving rise to a sequence of {\it double} poles
at 
\be
s = {\rm max}\{ \Delta_1 + \Delta_2, \Delta_3 + \Delta_4\} + 2n \, , \qquad n \in   \mathbb{Z}_{\geqslant 0} \,.
\ee
 A double pole at $s = s_0$ gives a contribution to ${\cal G}_{\rm conn}(U, V)$ of the from $U^{s_0/2}\log 
U$. This has a natural interpretation in terms of the $O(1/N^2)$  anomalous dimensions of the exchanged double-trace operators. Indeed, a little thinking shows that  in this
case both OPE coefficients in the s-channel conformal block expansion are of order one (in contrast with the generic case (\ref{sequence})), so that the $O(1/N^2)$ correction to the dilation operator gives
a leading contribution to the connected four-point function. 

 Let's see this more explicitly. Let's take for definiteness  $\Delta_1 + \Delta_2  \leqslant \Delta_3 + \Delta_4$, so that 
$\Delta_3 + \Delta_4 = \Delta_1 + \Delta_2 + 2 k$ for some non-negative integer $k$.   Then the double-trace operators of the schematic form 
\be \label{dtbasis}
:\mathcal{O}_1\square^{n+k} \partial^\ell\mathcal{O}_2: \quad {\rm and} \quad :\mathcal{O}_3\square^{n} \partial^\ell\mathcal{O}_4:
\ee
have the same conformal dimension to leading large $N$ order, as well as the same Lorentz quantum numbers. 
 They are then expected to mix under the action of the $O(1/N^2)$ dilation operator. It is important to realize that  the mixing matrix that relates the
 basis (\ref{dtbasis}) to the double-trace dilation eigenstates ${\cal O}^{\rm DT}_{\alpha}$   is of order one.  The OPE coefficients $\langle {\cal O}_1  {\cal O}_2 {\cal O}^{\rm DT}_{\alpha} \rangle = c_{12 \alpha}$ and $\langle {\cal O}_3  {\cal O}_4 {\cal O}^{\rm DT}_{\alpha} \rangle= c_{34 \alpha}$ are then both $O(1)$, as claimed. The twist $\tau_\alpha = \Delta_\alpha - \ell$ has a large $N$ expansion of the form $\tau_\alpha = \Delta_3 + \Delta_4 + 2n + \gamma^{(1)}_\alpha/N^2 + O(1/N^4)$. All is all, we find a contribution to ${\cal G}_{\rm conn}$ of the form
 \be
\frac{c_{12\alpha} c_{34 \alpha} \, \gamma^{(1)}_\alpha }{N^2} \; U^{\frac{\Delta_3 + \Delta_4}{2} + n} \log U\, .
 \ee
In Mellin space, this corresponds to a double-pole at $s = \Delta_3 + \Delta_4 + 2n$, just as needed.  In summary, the explicit Gamma functions that appear in Mack's definition 
provide precisely the analytic structure expected in
a large $N$ CFT, if we take the $O(1/N^2)$ Mellin amplitude ${\cal M}$ to have poles associated with just
 the exchanged {\it single-trace} operators. The upshot is that to leading  $O(1/N^2)$ order, fixing  the single-trace contributions to the OPE is sufficient determine the double-trace contributions as well.\footnote{This is particularly apparent in Mellin space but can also be argued by more abstract CFT reasoning \cite{Caron-Huot:2017vep, Li:2017lmh, Alday:2017gde, Kulaxizi:2017ixa}. }

\smallskip
By following a similar reasoning, we will now argue that compatibility with the large $N$ OPE imposes some further constraints on the analytic structure of ${\cal M}$. We have seen that to leading $O(1/N^2)$ order the Mellin amplitude ${\cal M}(s, t, u)$ is a meromorphic function with only simple poles
 associated to the exchanged single-trace operators. In the generic case, a single-trace operator  ${\cal O}^{\rm ST}$ of twist $\tau$ contributing to the s-channel OPE is responsible for an infinite sequence of simple poles at $s = \tau + 2n$,  $n \in \mathbb{Z}_{\geqslant 0}$ (and similarly for the other channels). But this rule needs to be modified if this sequence of ``single-trace poles'' overlaps with the ``double-trace poles'' from the explicit Gamma functions in (\ref{GMellin}).
 This happens
 if  $\tau = \Delta_1 + \Delta_2$ mod 2, or if $\tau = \Delta_3 + \Delta_4$ mod 2. (We assume for now that  $\Delta_1 + \Delta_2 \neq \Delta_3 + \Delta_4$ mod 2, so that only one of  the two options is realized.) 
   In the first  case, the  infinite sequence of poles in ${\cal M}$ must truncate to the set $\{\tau, \tau+2,\ldots, \tau+ \Delta_1 + \Delta_2-2 \}$,
 and in the second case to the set $\{\tau, \tau+2,\ldots, \tau+ \Delta_3 + \Delta_4-2 \}$\footnote{Note that the first set empty if $\Delta_1 + \Delta_2  < \tau$ (again we are assuming $\Delta_1 + \Delta_2 = \tau$ mod 2) and the second is empty if $\Delta_3 + \Delta_4  < \tau$  (with $\Delta_3 + \Delta_4 = \tau$ mod 2). In these cases, ${\cal O}^{\rm ST}$ 
 does not contribute {\it any} poles to ${\cal M}$.}. This truncation must happen  because double poles in $s$, translating to $\sim \log U$ terms in  ${\cal G}_{\rm conn}$, are incompatible with the large $N$ counting. 
Indeed, the OPE coefficients  already 
provide an $O(1/N^2)$ suppression, so that we should use the $O(1)$ dilation operator, and no logarithmic terms can arise in ${\cal G}_{\rm conn}$ to leading $O(1/N^2)$ order.\footnote{In the even more fine-tuned case  $\tau = \Delta_1 + \Delta_2 = \Delta_3 + \Delta_4$
mod 2, clearly the poles in $s$ in the $O(1/N^2)$ Mellin amplitude ${\cal M}$ must truncate to the set $\{\tau, \tau+2,\ldots, \tau+ {\rm min}\{ \Delta_1 + \Delta_2 , \Delta_3 + \Delta_4 \} -2 \}$. The double poles at
$\{   {\rm min}\{ \Delta_1 + \Delta_2 , \Delta_3 + \Delta_4 \}, {\rm min}\{ \Delta_1 + \Delta_2 , \Delta_3 + \Delta_4 \}+2 \dots,      {\rm max}\{ \Delta_1 + \Delta_2 , \Delta_3 + \Delta_4 \} -2  \}$ can be ruled out by the same reasoning, while the {\it triple} poles at $s =   {\rm max}\{ \Delta_1 + \Delta_2 , \Delta_3 + \Delta_4 \} + 2n$ would give rise to $\sim (\log U)^2$  terms, which absolutely cannot appear to $O(1/N^2)$. }

\subsection{Mellin amplitudes for Witten diagrams} \label{mellinwittendiagrams}

The effectiveness of Mellin formalism is best illustrated by its application to the calculation  of Witten diagrams.  Conceptually,
Mellin space makes transparent the analogy of holographic correlators and  $S$-matrix amplitudes. Practically, 
Mellin space expressions for Witten diagrams
are much simpler than their position space counterparts. For starters, the Mellin amplitude of a four-point contact diagram, which is the building blocks of AdS four-point correlators as we reviewed in Section \ref{sugrareview}, is just a constant, 
\begin{equation}
D_{\Delta_1\Delta_2\Delta_3\Delta_4}=\int [d\delta_{ij}] \left(\frac{\pi^{d/2}\Gamma[\frac{\sum\Delta_i}{2}-d/2]}{\prod \Gamma[\Delta_i]}\right)\times \prod_{i<j}\Gamma[\delta_{ij}](x^2_{ij})^{-\delta_{ij}}\;.
\end{equation}
As was shown in \cite{Penedones:2010ue}, this generalizes to $n$-point contact  diagram with a non-derivative vertex: their   Mellin amplitude is again a constant. Contact diagrams with derivative vertices are also easily evaluated.
It will be important in the following that the Mellin amplitude for a contact diagram arising from a vertex with $2n$ derivatives is an order $n$ polynomial in the Mandelstam variables $\delta_{ij}$.

Exchange diagrams are also much simpler in Mellin space. The s-channel exchange Witten diagram  with an exchanged field of conformal dimension $\Delta$ and spin $J$  has a Mellin amplitude with the following simple analytic structure \cite{Costa:2012cb},
\begin{equation}\label{mellinexchnage}
\mathcal{M}(s,t)= \sum_{m=0}^{\infty}\frac{Q_{J,m}(t)}{s-\tau-2m}+P_{J-1}(s,t) \, ,
\end{equation}
where $\tau=\Delta-J$ is the twist. Here $Q_{J,m}(t)$ are polynomials in $t$ of degree $J$ and $P_{J-1}(s,t)$ polynomials in $s$ and $t$ of degree $J-1$. These polynomials depend on the dimensions $\Delta_{1,2,3,4}$, $\Delta$, as well as the spin $J$. The detailed expressions for these polynomials are quite complicated but will not be needed
for our analysis. The $m=0$ pole at $s=\tau$ is called the leading pole, corresponding to the primary operator that is dual to the exchanged field, while the $m > 0$ poles are called satellite poles, and they are associated with  conformal descendants.

It has been observed (see, {\it e.g.}, \cite{Penedones:2010ue}) that the infinite series of poles in (\ref{mellinexchnage}) truncates to a finite sum if
 $\tau = \Delta_1+\Delta_2$ mod 2 or  if $\tau = \Delta_3+\Delta_4$ mod 2. One finds that the upper limit of the sum $m_{\rm max}$ is given by $\tau - \Delta_1 - \Delta_2 = 2 (m_{\rm max}+1)$ in the first case
 and by $\tau - \Delta_3 - \Delta_4 = 2 (m_{\rm max}+1)$ in the second case.
 This is the Mellin space  version of the phenomenon described in Section \ref{sugrareview}: an exchange Witten diagram with these special values of quantum numbers can be written as a {\it finite} sum of contact Witten diagrams. 
As we have explained in the previous subsection,  this remarkable simplification  is dictated by compatibility 
with the large $N$ OPE in the dual CFT.

\subsection{Asymptotics and the flat space limit} \label{flatspacelimit}

In the next section we will determine the supergravity four-point Mellin amplitude using general consistency principles. A crucial constraint will be provided by the asymptotic behavior of ${\cal M}(s, t)$ 
when $s$ and $t$ are simultaneously scaled to infinity. 
On general grounds, one can argue  \cite{Penedones:2010ue} that in this limit the Mellin amplitude should reduce to the flat-space bulk S-matrix (in $\mathbb{R}^{d, 1}$).

A precise prescription for relating the massless\footnote{For massive external particles, see the discussion in \cite{Paulos:2016fap}.} flat-space scattering amplitude $\mathcal{T}(K_i)$    to the asymptotic form of the holographic Mellin amplitude   was given in \cite{Penedones:2010ue} and justified in \cite{Fitzpatrick:2011hu},
\begin{equation} \label{joao}
{\cal M} (\delta_{ij}) \approx \frac{R^{n(1-d)/2+d+1}}{\Gamma(\frac{1}{2}\sum_i\Delta_i-\frac{d}{2})}\int_0^\infty d\beta \beta^{\frac{1}{2}\sum_i\Delta_i-\frac{d}{2}-1}e^{-\beta}\mathcal{T}\left(S_{ij}=\frac{2\beta}{R^2}s_{ij}\right)
\end{equation}
where $S_{ij} = -(K_i + K_j)^2$ are the Mandelstam invariants of the flat-space scattering process. We have a precise opinion for asymptotic behavior of the flat-space four-point amplitude ${\cal T}(S, T)$ --  it can grow at most linearly
for large $S$ and $T$. Indeed, a spin $\ell$ exchange
diagrams grows with power $\ell -1$, and the highest spin state is of course the graviton with $\ell = 2$. Similarly, contact interactions with $2n$ derivatives give a power $n$ growth, and IIB supergravity (in ten-dimensional flat space) contains contact interactions with at most two derivatives. From (\ref{joao}) we then deduce 
\begin{equation} \label{largelambda}
{\cal M}(\beta s, \beta t) \sim O(\beta) \quad {\rm for} \; \beta \to \infty \,.
\end{equation}
It is of course crucial to this argument that we are calculating within the standard two-derivative supergravity theory. Stringy $\alpha'$-corrections would introduce higher derivative terms and invalidate this conclusion.\footnote{In a perturbative $\alpha'$-expansion,
we expect increasing polynomial growth, but for finite $\alpha'$ the behavior should be very soft, as in string theory.}

Curiously, the asymptotic behavior (\ref{largelambda}) is not immediately obvious  if one computes holographic correlators in $AdS_5 \times S^5$ by the standard diagrammatic approach. Exchange Witten diagrams have the expected behavior,
with growth at most linear from spin two exchanges, see   (\ref{mellinexchnage}).\footnote{The $AdS_5$ effective theory contains an infinite tower of spin two massive states that arise from the Kaluza-Klein reduction of the ten-dimensional graviton, and of course
{\it no} states of spin higher than two.}
 However, the $AdS_5$ effective action \cite{Arutyunov:1999fb}  obtained by Kaluza-Klein reduction of IIB supergravity on $S^5$ contains quartic vertices
with four derivatives (or fewer). The four-derivative vertices are in danger of producing an $O(\beta^2)$ growth, which would ruin the expected flat space  asymptotics. On this basis, we made the assumption in  \cite{Rastelli:2016nze} that the
total contribution of the four-derivative vertices to a holographic correlator must also grow at most linearly for large $\beta$. Indeed, this was experimentally the case in all the explicit supergravity calculations performed at the time. 
Fortunately, the conjectured cancellation of the $O(\beta^2)$ terms has been recently proved in full generality  \cite{Arutyunov:2017dti}.

\section{The general one-half BPS four-point amplitude in Mellin space}\label{mellinapproach}

As we have just reviewed, holographic correlators are most naturally evaluated in Mellin space. Mellin amplitudes have an intuitive interpretation as  scattering processes in AdS space, and their analytic structure is simple and well understood.
We have also discussed the additional simplification that occurs for one-half BPS correlators in $AdS_5 \times S^5$ supergravity.    The  Kaluza-Klein spectrum satisfies the ``truncation conditions'' that allow exchange Witten diagrams to be expressed 
as finite sum of contact diagrams. This translates into the statement that the Mellin amplitude for these correlators is a {\it rational} function, with poles at predictable locations controlled by the single-particle spectrum. We have not yet imposed the constraints
of {\it super}conformal invariance. They turn out to be so stringent that when combined with the analytic structure of the Mellin amplitude they appear to completely fix the answer! In this Section we derive a set of algebraic and analytic conditions on the Mellin amplitude
for one-half BPS correlators with arbitrary weights. We have found a simple solution of these constraints, which we believe to be unique.

We start  in Section \ref{scfwisection} by reviewing the superconformal Ward identity in position space. A useful technical step is the introduction of auxiliary variables $\sigma$ and $\tau$ to keep track of the R-symmetry quantum numbers.
We translate the Ward identity in Mellin space in Section \ref{manifestscfwi}. The Mellin amplitude ${\cal M} (s, t ; \sigma, \tau)$ is written in terms 
of a difference operator acting on an auxiliary  object $\widetilde{\mathcal{M}} (s, t ; \sigma, \tau)$.
A purely algebraic problem is then formulated in Section \ref{bootstrapproblem} by imposing a set of consistency conditions on  ${\cal M} (s, t ; \sigma, \tau)$. We find a simple elegant solution to this problem in  Section \ref{bootstrapsolution}. While we lack a general proof, we believe that this is the unique solution, and we do show uniqueness in Section \ref{peq2proof} in the simplest case where all $p_i=2$. Finally, in Section \ref{subtlefree} we discuss some subtleties with the contour prescription in the inverse Mellin transform. We show in particular how the
``free'' piece of the correlator  can arise as a regularization effect.

\subsection{Superconformal Ward identity: position space}\label{scfwisection}

The global symmetry group of   $\mathcal{N}=4$ SYM is $PSU(2,2|4)$, which contains as subgroups the four-dimensional conformal group $SO(4,2)\cong SU(2,2)$ and the R-symmetry group $SO(6)\cong SU(4)$.  As we have already mentioned, the one-half BPS operators $\mathcal{O}^{(p_i)}_{I_1\ldots I_{p_i}}$ transform in the symmetric traceless representation of $SO(6)$. Their R-symmetry structure can therefore be conveniently kept track of by contracting the $SO(6)$ indices with a null vector $t_i$,
 \begin{equation}
\mathcal{O}^{(p_i)}(x_i,t_i)\equiv t_{i}^{I_1}\ldots t_{i}^{I_{p_i}}\mathcal{O}^{(p_i)}_{I_1\ldots I_{p_i}}(x_i),\;\;\;\;\;\;\;\;\;\;\; t_i\cdot t_i=0\; .
\end{equation}
The four-point function
\begin{equation}
G(x_i,t_i)\equiv \langle\mathcal{O}^{(p_1)}(x_1,t_1)\mathcal{O}^{(p_2)}(x_2,t_2)\mathcal{O}^{(p_3)}(x_3,t_3)\mathcal{O}^{(p_4)}(x_4,t_4)\rangle
\end{equation}
is thus a function of the spacetime coordinates $x_i$ as well as the ``internal'' coordinates $t_i$. The R-symmetry covariance and null property requires that the $t_i$ variables can only show up as sum of monomials $\prod_{i<j} (t_{ij})^{\gamma_{ij}}$ with integer powers $\gamma_{ij} \geqslant 0$, where we have defined $t_{ij} \equiv t_i\cdot t_j$. Moreover the exponents $\gamma_{ij}$ are constrained by $\sum_{i\neq j}\gamma_{ij}=p_j$, as seen by requiring the correct homogeneity under independent scaling of each null vector $t_i\to \zeta_i t_i$. We can solve this set of constraints by using the following parameterization,
\begin{equation} \label{para}
\begin{split}
\gamma_{12}={}&-\frac{a}{2}+\frac{p_1+p_2}{2}\;,\;\;\;\;\;\;\;\;\;\;\;\;\gamma_{34}=-\frac{a}{2}+\frac{p_3+p_4}{2}\;,\\
\gamma_{23}={}&-\frac{b}{2}+\frac{p_2+p_3}{2}\;,\;\;\;\;\;\;\;\;\;\;\;\;\gamma_{14}=-\frac{b}{2}+\frac{p_1+p_4}{2}\;,\\
\gamma_{13}={}&-\frac{c}{2}+\frac{p_1+p_3}{2}\;,\;\;\;\;\;\;\;\;\;\;\;\;\gamma_{24}=-\frac{c}{2}+\frac{p_2+p_4}{2}\;,\\
\end{split}
\end{equation}
with the additional condition $a+b+c=p_1+p_2+p_3+p_4$.

\begin{figure}[htbp]
\begin{center}
\includegraphics[scale=0.5]{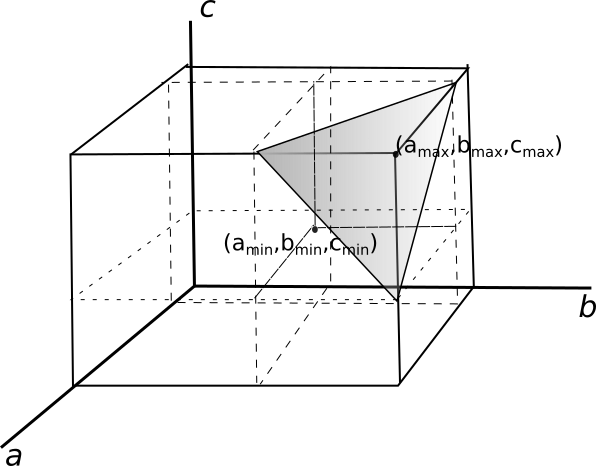} 
\caption{Solution to the $\gamma_{ij}$ constraints.}
\label{cube}
\end{center}
\end{figure}
Without  loss of generality we can assume $p_1\geqslant p_2\geqslant p_3\geqslant p_4$. Then we should distinguish two possibilities, 
\be
p_1+p_4\leqslant p_2+p_3  \quad ({\rm case \; I}) \quad {\rm and} \quad
 p_1+p_4> p_2+p_3 \quad ({\rm case \; II}) \,.
\ee
In either case the inequality constraints $\gamma_{ij} \geqslant 0$  
define a cube inside the parameter space  $(a,b,c)$, as shown in Figure \ref{cube}. The solution is further restricted by the  condition $a+b+c=p_1+p_2+p_3+p_4$,
which carves out the
 equilateral triangle inside the cube shown shaded in the figure.  It is useful to find the coordinates of  vertices of the cube closest and furthest from the origin,
which we denote as $(a_{\mathrm{min}},b_{\mathrm{min}},c_{\mathrm{min}})$ and $(a_{\mathrm{max}},b_{\mathrm{max}},c_{\mathrm{max}})$.  Then in case I,
\begin{equation}
\begin{split}
a_{\mathrm{max}}=p_3+p_4\;,\;\;\; {}& a_{\mathrm{min}}=p_3-p_4\;,\\
b_{\mathrm{max}}=p_1+p_4\;,\;\;\; {}& a_{\mathrm{min}}=p_1-p_4\;,\\
c_{\mathrm{max}}=p_2+p_4\;,\;\;\; {}& a_{\mathrm{min}}=p_2-p_4\;,
\end{split}
\end{equation}
and in case II,
\begin{equation}
\begin{split}
a_{\mathrm{max}}=p_3+p_4\;,\;\;\; {}& a_{\mathrm{min}}=p_1-p_2\;,\\
b_{\mathrm{max}}=p_2+p_3\;,\;\;\; {}& a_{\mathrm{min}}=p_1-p_4\;,\\
c_{\mathrm{max}}=p_2+p_4\;,\;\;\; {}& a_{\mathrm{min}}=p_1-p_3\;.
\end{split}
\end{equation}
Denoting by $2L$  the length of each side of the cube, we find in the two cases
\begin{equation}
\begin{split}
L={}& p_4 \;\;\;\;\;\;\;\; \;\;\;\;\;\;\;\; \;\;\;\;\;\;\;\; \;\;\;\;\;\;\;\; \mathrm{(case\;\; I)}\;,\\
L={}& \frac{p_2+p_3+p_4-p_1}{2} \;\;\;\;\;\;\;\; \mathrm{(case\;\; II)}\;.
\end{split}
\end{equation} 
From the parametrization (\ref{para}) we see that  $\gamma_{ij} \geq \gamma_{ij}^0$, where $\gamma_{ij}^0$ are obtained by
substituting the maximal values $(a_{\mathrm{max}},b_{\mathrm{max}},c_{\mathrm{max}})$,
\begin{equation}
\begin{split}
\gamma^0_{12}={}& \frac{p_1+p_2-p_3-p_4}{2}\;,\\
\gamma^0_{13}={}& \frac{p_1+p_3-p_2-p_4}{2}\;,\\
\gamma^0_{34}={}&\gamma^0_{24}=0\;,\\
\gamma^0_{14}={}&0\;\;\;\;\mathrm{(case\;\; I)},\;\;\;\;\;\;  \frac{p_1+p_4-p_2-p_3}{2}\;\;\mathrm{(case\;\; II)}\;,\\
\gamma^0_{23}={}& \frac{p_2+p_3-p_1-p_4}{2}\;\;\;\;\mathrm{(case\;\; I)},\;\;\;\;\;\; 0\;\;\mathrm{(case\;\; II)}\;.
\end{split}
\end{equation}
Factoring out the product $\prod_{i<j}\left(\frac{t_{ij}}{x_{ij}^2}\right)^{\gamma^0_{ij}}$ from the correlator, we are left with an object with the same scaling properties of a
four-point function with  equal weights $L$. This motivates the definition
\begin{equation}\label{GandcurlyG}
G(x_i,t_i)=\prod_{i<j}\left(\frac{t_{ij}}{x_{ij}^2}\right)^{\gamma^0_{ij}}\left(\frac{t_{12}t_{34}}{x^2_{12}x^2_{34}}\right)^L \mathcal{G}(U,V;\sigma,\tau)\; ,
\end{equation}
where besides the usual conformal cross ratios
\begin{equation}
      U = \frac{(x_{12})^2(x_{34})^2}{(x_{13})^2(x_{24})^2}\;,\;\;\;\;\;\;\;\; V =   \frac{(x_{14})^2(x_{23})^2}{(x_{13})^2(x_{24})^2}
\end{equation}
we have introduced analogous R-symmetry cross ratios 
\begin{equation}
       \sigma = \frac{(t_{13}) (t_{24})}{(t_{12}) (t_{34})}\;,\;\;\;\;\;\;\;\;\;\;\;\;\;\;\;\tau=  \frac{(t_{14}) (t_{23})}{(t_{12}) (t_{34})}\;.
\end{equation}
 It is easy to see   that $\mathcal{G}(U,V;\sigma,\tau)$ is a {\it polynomial} of degree $L$  in $\sigma$ and $\tau$.
So far we have only imposed covariance under the bosonic subgroups of the  supergroup $PSU(2,2|4)$. The fermionic generators impose further constraints on the four-point function. It is useful to introduce the following change of variables
\begin{equation}
\begin{split}
{}&U=z \bar{z}\;,\\
{}&V=(1-z)(1-\bar{z})\;,\\
{}&\sigma=\alpha \bar{\alpha}\;,\\
{}&\tau=(1-\alpha)(1-\bar{\alpha})\;.
\end{split} \label{changeofvariable}
\end{equation}
In terms of these variables, the superconformal Ward identity reads  \cite{Eden:2000bk, Nirschl:2004pa}
\begin{equation}\label{scfwi}
\partial_{\bar{z}}[\mathcal{G}(z\bar{z},(1-z)(1-\bar{z});\alpha\bar{\alpha},(1-\alpha)(1-\bar{\alpha}))\big|_{\bar{\alpha}\to1/\bar{z}}]=0\;.
\end{equation} 
Its solution can be written as \cite{Eden:2000bk, Nirschl:2004pa}\footnote{There is an implicit  regularity assumption for $\mathcal{H}(U,V;\sigma,\tau)$ as
$ \bar \alpha \to 1/\bar z$, otherwise the following equation would be an empty statement.}
\begin{equation}\label{splitform} 
\mathcal{G}(U,V;\sigma,\tau)=\mathcal{G}_{\rm free}(U,V;\sigma,\tau)+
R\,\mathcal{H}(U,V;\sigma,\tau) \, ,
\end{equation}
where $\mathcal{G}_{\rm free}$ is the 
answer in free SYM theory and
\begin{eqnarray}\label{rfactor}
 R & =&\tau \, 1+(1-\sigma-\tau)\, V+(-\tau-\sigma\tau+\tau^2)\, U+(\sigma^2-\sigma-\sigma\tau)\, UV+ \sigma V^2+\sigma\tau \, U^2 \nonumber  \\
&=&(1-z\alpha)(1-\bar{z}\alpha)(1-z\bar{\alpha})(1-\bar{z}\bar{\alpha})\,.
\end{eqnarray}
All dynamical information is contained in the
 {\it a priori} unknown function $\mathcal{H}(U,V;\sigma,\tau)$.   Note that
$\mathcal{H}(U,V;\sigma,\tau)$ is a polynomial in $\sigma$, $\tau$ of degree $L-2$.

\subsection{Superconformal Ward identity: Mellin space}\label{manifestscfwi}

We now turn to analyze the constraints of superconformal symmetry in Mellin space. We rewrite (\ref{splitform}) for the connected correlator,
\begin{equation} \label{wardconnected}
\mathcal{G}_{\rm{conn}}(U,V;\sigma,\tau)=\mathcal{G}_{\rm free, conn}(U,V;\sigma,\tau)+
R(U, V ; \sigma, \tau) \,\mathcal{H}(U,V;\sigma,\tau) \; ,
\end{equation}
and take the Mellin transform of both sides of this equation. The transform\footnote{This definition should be taken with a grain of salt. In general,
the integral transform of the full connected correlator is divergent. In the supergravity limit, there is a natural decomposition of $\mathcal{G}_{\rm{conn}}$ into a sum of $\bar D$ functions, each of which has a well-defined Mellin transform in a certain
region of the $s$ and $t$ complex domains. However, it is often the case that there is no common region such that the transforms of the $\bar D$ functions are  all convergent. On the other hand, the {\it inverse} Mellin transform (\ref{inverseG}) is well-defined, but
care must be taken in specifying the integration contours.  We will come back to this subtlety in Section \ref{subtlefree}.}  of the left-hand side gives the reduced Mellin amplitude $M$,
\begin{equation}\label{mellinleft}
M(s,t;\sigma,\tau)=\int_0^\infty dU U^{-\frac{s}{2}+\frac{p_3+p_4}{2}-L-1}\int_0^\infty dV V^{-\frac{t}{2}+\frac{\min\{p_1+p_4,p_2+p_3\}}{2}-1}\mathcal{G}_{\rm conn}(U,V;\sigma,\tau)\; ,
\end{equation}
from which we define  the Mellin amplitude ${\cal M}$,
\begin{equation}
\mathcal{M}(s,t;\sigma,\tau) \equiv \frac{M(s,t;\sigma,\tau)}{\Gamma_{p_1p_2p_3p_4}} \, ,
\end{equation}
where as always
\begin{equation}
\begin{split}
\Gamma_{p_1p_2p_3p_4} \equiv {}&\Gamma[-\frac{s}{2}+\frac{p_1+p_2}{2}]\Gamma[-\frac{s}{2}+\frac{p_3+p_4}{2}]\Gamma[-\frac{t}{2}+\frac{p_2+p_3}{2}]\\
\times{}&\Gamma[-\frac{t}{2}+\frac{p_1+p_4}{2}]\Gamma[-\frac{u}{2}+\frac{p_1+p_3}{2}]\Gamma[-\frac{u}{2}+\frac{p_2+p_4}{2}]\; , \\
& u \equiv p_1+p_2+p_3+p_4-s-t\,.
\end{split}
\end{equation}
On the right-hand side of (\ref{wardconnected}), the first term is the free part of the correlator. It consists of a sum of terms of the form $\sigma^a \tau^b U^m V^n$, where $m$, $n$ are integers and $a$, $b$  non-negative integers.
The Mellin transform of any such term is ill-defined. As we shall explain in Section \ref{subtlefree}, there is a consistent sense in which it can be defined to be {\it zero}. The function 
$\mathcal{G}_{\rm free, conn}(U,V;\sigma,\tau)$ 
will be recovered as a regularization effect in transforming back from Mellin space to position space.\footnote{Our treatment for the free part of the correlator also turns out to be consistent in the context of holographic higher spin theory, as is discussed in v3  of \cite{Ponomarev:2017qab}.} 

We then turn to the second term on the 
 on the right-hand side of  (\ref{wardconnected}). We define an auxiliary amplitude $\widetilde {\cal M}$ from the Mellin transform of the dynamical function 
 $\mathcal{H}$,  
 \begin{equation}
\widetilde{\mathcal{M}}(s,t;\sigma,\tau)=\frac{\int_0^\infty dU U^{-\frac{s}{2}+\frac{p_3+p_4}{2}-L-1}\int_0^\infty dV V^{-\frac{t}{2}+\frac{\min\{p_1+p_4,p_2+p_3\}}{2}-1}\mathcal{H}(U,V;\sigma,\tau)}{\tilde{\Gamma}_{p_1p_2p_3p_4}} \, ,
\end{equation}
with
\begin{equation}
\begin{split}
\tilde{\Gamma}_{p_1p_2p_3p_4} \equiv {}&\Gamma[-\frac{s}{2}+\frac{p_1+p_2}{2}]\Gamma[-\frac{s}{2}+\frac{p_3+p_4}{2}]\Gamma[-\frac{t}{2}+\frac{p_2+p_3}{2}]\\
\times{}&\Gamma[-\frac{t}{2}+\frac{p_1+p_4}{2}]\Gamma[-\frac{\tilde{u}}{2}+\frac{p_1+p_3}{2}]\Gamma[-\frac{\tilde{u}}{2}+\frac{p_2+p_4}{2}] \, .\\
\end{split}
\end{equation}
Note that we have introduced a ``shifted''  Mandelstam variable $\tilde u$,
\begin{equation}
\tilde{u} \equiv u-4=p_1+p_2+p_3+p_4-4-s-t \, .
\end{equation}
This shift is motived by the desire to keep the crossing symmetry properties of ${\cal H}$ as simple as possible, as we shall explain shortly.
Let us also record the expressions of the inverse transforms,
\begin{eqnarray}\label{inverseG}
&& \mathcal{G}_{\rm conn}(U,V;\sigma,\tau) = \int \frac{ds}{2} \frac{dt}{2}\; U^{\frac{s}{2}-\frac{p_3+p_4}{2}+L}V^{\frac{t}{2}-\frac{\min\{p_1+p_4,p_2+p_3\}}{2}}\mathcal{M}(s,t;\sigma,\tau) \Gamma_{p_1p_2p_3p_4}\quad \\
\label{inverseH}
&& \mathcal{H}(U,V;\sigma,\tau) = \int \frac{ds}{2} \frac{dt}{2} U^{\frac{s}{2}-\frac{p_3+p_4}{2}+L}V^{\frac{t}{2}-\frac{\min\{p_1+p_4,p_2+p_3\}}{2}}\widetilde{\mathcal{M}}(s,t;\sigma,\tau) \tilde{\Gamma}_{p_1p_2p_3p_4} \, ,
\end{eqnarray}
where the precise definition of the integration contours will require a careful discussion in Section \ref{subtlefree} below.

We are now ready to write down the Mellin translation of (\ref{wardconnected}). It takes  the simple form
\begin{equation}\label{MRMt}
\mathcal{M} (s,t; \sigma, \tau) =\widehat{R}\circ\widetilde{\mathcal{M}} (s, t, ; \sigma, \tau) \, .
\end{equation}
The multiplicative factor $R$ has turned into a {\it difference operator} $\widehat R$, 
\begin{equation}\label{hatR}
\widehat R= \tau \, 1+(1-\sigma-\tau)\, \widehat V+(-\tau-\sigma\tau+\tau^2)\, \widehat U+(\sigma^2-\sigma-\sigma\tau)\,\widehat{UV}+ \sigma \widehat{V^2}+\sigma\tau \, \widehat{U^2}\; ,
\end{equation}
where the hatted monomials in $U$ and $V$ are defined to act as follows,
\begin{equation}\label{roperator}
\begin{split}
\widehat{U^mV^n}\circ \widetilde{\mathcal{M}}(s,t;\sigma,\tau)\equiv{}&  \widetilde{\mathcal{M}}(s-2m,t-2n);\sigma,\tau)\\
\times{}&\left(\frac{p_1+p_2-s}{2}\right)_m\left(\frac{p_3+p_4-s}{2}\right)_m\left(\frac{p_2+p_3-t}{2}\right)_n\\
\times{}&\left(\frac{p_1+p_4-t}{2}\right)_n\left(\frac{p_1+p_3-u}{2}\right)_{2-m-n}\left(\frac{p_2+p_4-u}{2}\right)_{2-m-n}\; ,
\end{split}
\end{equation}
with $(a)_n \equiv \Gamma[a+n]/\Gamma[a]$  the usual Pochhammer symbol.

\subsubsection{Crossing symmetry and $\tilde u$}

The Mellin amplitude $\mathcal{M}$  satisfies  Bose symmetry, namely, it is invariant under  permutation of  the Mandelstam variables $s$, $t$, $u$ if the external quantum numbers are also permuted accordingly.
The auxiliary amplitude $\widetilde{\mathcal{M}}$ has been defined to enjoy the same symmetry  under permutation of the {\it shifted} Mandelstam variables $s$, $t$, $\tilde{u}$. The point is that the factor $R$ multiplying $\mathcal{H}$ is not crossing-invariant, and the shift in $u$  precisely compensates for this asymmetry. Let us see this in some detail.

To make expressions more compact, we introduce some shorthand notations for the following combinations of coordinates,
\begin{equation}
\begin{split}
A={}&x_{12}^2x_{34}^2\;,\;\;\;\;B=x_{13}^2x_{24}^2\;,\;\;\;\;C=x_{14}^2x_{23}^2\;,\\
a={}&t_{12}t_{34}\;,\;\;\;\;\;\;b=t_{13}t_{24}\;,\;\;\;\;\;\;c=t_{14}t_{23}\;.
\end{split}
\end{equation}
In the equal-weights case (on which we focus for simplicity), the four-point function $G(x_i,t_i)$ is related to $\mathcal{G}(U,V;\sigma,\tau)$ by 
\begin{equation}
G(x_i,t_i)=\left(\frac{a}{A}\right)^L\mathcal{G}(U,V;\sigma,\tau)\;.
\end{equation} 
Substituting into this expression the inverse Mellin transformation (\ref{inverseG}), one finds
\begin{equation}
\begin{split}
G(x_i,t_i)=\int_{i \infty}^{i\infty} ds dt {}& \sum_{I+J+K=L}A^{\frac{s}{2}-L}B^{\frac{u}{2}-L}C^{\frac{t}{2}-L}a^Kb^Ic^J \mathcal{M}_{IJK}(s,t)\\
{}&\times\Gamma^2[-\frac{s}{2}+L]\Gamma^2[-\frac{t}{2}+L]\Gamma^2[-\frac{u}{2}+L]\, ,
\end{split}
\end{equation} 
where we defined $ \sum_{I+J+K=L}a^Kb^Ic^J \mathcal{M}_{IJK}(s,t)\equiv a^L\mathcal{M}(s,t;\sigma,\tau)$. In terms of these new variables, crossing amounts to permuting simultaneously $(A, B, C)$ and $(a, b, c)$: 
\begin{equation}
\begin{split}
{\rm 1\leftrightarrow 4:}\;\quad\quad{}& \left\{\begin{array}{c}\sigma\leftrightarrow 1/\sigma,\; \tau\leftrightarrow\sigma/\tau,\; \\U\leftrightarrow 1/U,\; V\leftrightarrow V/U\end{array}\right\} \quad{\rm or}\quad \left\{\begin{array}{c}A\leftrightarrow B \\a\leftrightarrow b\end{array}\right\}\;,\\
{\rm 1\leftrightarrow 3:}\;\quad\quad{}& \left\{\begin{array}{c}\sigma\leftrightarrow \sigma/\tau,\; \tau\leftrightarrow 1/\tau,\; \\U\leftrightarrow V,\; V\leftrightarrow U\end{array}\right\} \quad{\rm or}\quad \left\{\begin{array}{c}A\leftrightarrow C \\a\leftrightarrow c\end{array}\right\}\;.\\
\end{split}
\end{equation}
Invariance of the four-point function under crossing implies that the Mellin amplitude $\mathcal{M}(s,t;\sigma,\tau)$  must obey 
\begin{equation}
\begin{split}
\sigma^{L} {\mathcal{M}}( u,t;1/\sigma,\tau/\sigma)={}& {\mathcal{M}}(s,t;\sigma,\tau)\;,\\
\tau^{L} {\mathcal{M}}(t,s;\sigma/\tau,1/\tau)={}& {\mathcal{M}}(s,t;\sigma,\tau)\,.
\end{split}
\end{equation}
On the other hand, a similar representation exists for $R\mathcal{H}$. The factor $R$ can be expressed  as
\begin{equation}\label{Rabc}
\begin{split}
R={}&\frac{1}{a^2 B^2}\big(a^2 B C+b^2A  C+c^2A B-ab A C -a b B C+a b C^2\\
{}&\;\;\;\;\;\;\;\;\;-ac A B+ac B^2 -ac B C+b cA^2 -bcA  B - b cA C \big)\\
\equiv {}& \frac{\mathfrak{R}}{a^2 B^2} \, ,
\end{split}
\end{equation}
with a crossing-invariant numerator $\mathfrak{R}$ but a non-invariant denominator. When we go to the Mellin representation of $\left(\frac{a}{A}\right)^LR\mathcal{H}$ by substituting in (\ref{inverseH}), we find that the power of $B$ receives an additional $-2$ from the denominator of $R$ in (\ref{Rabc}), explaining the shift from $u$ to $\tilde{u}$,
\begin{equation}
\begin{split}
\left(\frac{a}{A}\right)^LR\mathcal{H}=
\int_{i \infty}^{i\infty} ds dt {}&\sum_{i+j+k=L-2}A^{\frac{s}{2}-L}B^{\frac{\tilde{u}}{2}-L}C^{\frac{t}{2}-L}a^k\;b^i\;c^j\; \mathfrak{R}\; \widetilde{\mathcal{M}}_{ijk}(s,t)\\
{}&\times\Gamma^2[-\frac{s}{2}+L]\Gamma^2[-\frac{t}{2}+L]\Gamma^2[-\frac{\tilde{u}}{2}+L]\;.
\end{split}
\end{equation}
Here we have similarly defined
\begin{equation}
 \sum_{i+j+k=L-2}a^k\;b^i\;c^j\; \mathfrak{R}\; \widetilde{\mathcal{M}}_{ijk}(s,t)=a^{L-2}\widetilde{\mathcal{M}}(s,t;\sigma,\tau)\;.
\end{equation}
Invariance of this expression under crossing implies the following  transformation rules for $\widetilde{\mathcal{M}}$,
\begin{equation}
\begin{split}
\sigma^{L-2} {\widetilde{\mathcal{M}}}( \tilde{u},t;1/\sigma,\tau/\sigma)={}& {\widetilde{\mathcal{M}}}(s,t,;\sigma,\tau)\;,\\
\tau^{L-2} {\widetilde{\mathcal{M}}}(t,s;\sigma/\tau,1/\tau)={}& {\widetilde{\mathcal{M}}}(s,t;\sigma,\tau)\,.
\end{split}
\end{equation}
We see that in the auxiliary amplitude $\widetilde{\mathcal{M}}$, 
the role of $u$ is played by $\tilde{u}$. This generalizes to the unequal-weight cases.

\subsection{An algebraic problem}\label{bootstrapproblem}

Let us now take stock and summarize  the properties of  $\mathcal{M}$ that we have demonstrated so far:

\begin{enumerate}
\item \textit{Superconformal symmetry.} The Mellin amplitude  $\mathcal{M}$ can be expressed in terms of an auxiliary amplitude  $\widetilde{\mathcal{M}}$,
\begin{equation} \label{Wardagain}
\mathcal{M}(s, t ; \sigma, \tau) =\widehat{R}\circ\widetilde{\mathcal{M}}  (s, t ; \sigma, \tau) \, ,
\end{equation}
with the help of the difference operator $\widehat{R}$ defined in (\ref{hatR}).
\item \textit{{Bose symmetry.}} $\mathcal{M}$ is
 invariant under permutation of the Mandelstam variables, if the quantum numbers of the external operators  are permuted accordingly.
For  example, when the conformal dimensions of the four half-BPS operators are set to equal $p_i = L$,  Bose symmetry  gives the usual crossing relations
\begin{equation}
\begin{split}
\sigma^{L} {\mathcal{M}}( u,t;1/\sigma,\tau/\sigma)={}& {\mathcal{M}}(s,t,;\sigma,\tau)\;,\\
\tau^{L} {\mathcal{M}}(t,s;\sigma/\tau,1/\tau)={}& {\mathcal{M}}(s,t;\sigma,\tau)\, .
\end{split}
\end{equation}
\item\textit{{Asymptotics.}} The asymptotic behavior of the Mellin amplitude ${\cal M}$ is bounded by the flat space scattering amplitude. At large values of the Mandelstam variables, ${\cal M}$ should grow linearly
\begin{equation}
\label{scaling}
{\mathcal{M}}(\beta s,\beta t;\sigma,\tau) \sim O(\beta) \, \quad {\rm for}\; \beta \to \infty \,.
\end{equation}
\item\textit{Analytic structure.}  $\mathcal{M}$   has only simple poles and there are a finite number of such simple poles in variables $s$, $t$, $u$, located at
\begin{eqnarray}
\nonumber s_0 &= & s_M-2 a \, ,  \quad s_0 \geq  2\;,\\
\nonumber t_0 &= & t_M -2 b \, ,  \, \,\quad t_0 \geq 2\;,\\
u_0 &= &  u_M - 2 c \, ,  \quad u_0 \geq 2
\end{eqnarray}
where
\begin{eqnarray} \label{mpolerange}
\nonumber{}&s_M=\min\{p_1+p_2,p_3+p_4\}-2\;,\\  
\nonumber{}&t_M=\min\{p_1+p_4,p_2+p_3\}-2\;,\\
{}&{u}_M=\min\{p_1+p_3,p_2+p_4\}-2\;,
\end{eqnarray}
and $a$, $b$, $c$ are non-negative integers. The position of these poles are determined by the twists of the exchanged single-trace operators in the three channels -- see  Table 1 and related discussion in Section 2.
  Moreover, at each simple pole,  the residue of the amplitude $\mathcal{M}$ must be a polynomial in the other Mandelstam variable.
\end{enumerate}
These conditions define a very constraining ``bootstrap''  problem. To start unpacking their content,
let us recall that the dependence on the R-symmetry variables $\sigma$ and $\tau$ is polynomial, of degree $L$ and $L-2$ for $\mathcal{M}$ and $\widetilde{\mathcal{M}}$, respectively,
\begin{equation}
\begin{split}
\mathcal{M}(s,t;\sigma,\tau)={}&\sum_{I+J+K=L}\sigma^I\tau^J\mathcal{M}_{IJK}(s,t)\;,\\
\widetilde{\mathcal{M}}(s,t;\sigma,\tau)={}&\sum_{i+j+k=L-2}\sigma^i\tau^j\widetilde{\mathcal{M}}_{ijk}(s,t)\;.
\end{split}
\end{equation}
Bose symmetry amounts to the invariance of $\mathcal{M}_{IJK}(s,t)$ under permutation of $(I, J, K)$ accompanied by simultaneous permutation of $(s, t, u)$, with  $u \equiv \sum_{i=1}^4 p_i - s - t$. Analogously, $\widetilde{\mathcal{M}}_{ijk}(s,t)$
is invariant under simultaneous permutation of $(i, j, k)$ and  $(s, t, \tilde u)$, with $\tilde u \equiv  \sum_{i=1}^4 p_i - s - t -4$. A little combinatoric argument shows that the number  ${\cal N}_L$ of independent $\mathcal{M}_{IJK}$ functions
is given by
\begin{equation}
{\cal N}_L = \frac{(L+5)(L+1)}{12} + \frac{17}{72} + \frac{(-1)^L}{8} + \frac{2}{9} \cos \left(\frac{2 \pi L}{3} \right)\,.  
\end{equation}
The superconformal Ward identity (\ref{Wardagain}) expresses the  ${\cal N}_L$ functions  $\mathcal{M}_{IJK}(s,t)$  in terms of the
   ${\cal N}_{L-2}$ functions $\widetilde{\mathcal{M}}_{ijk}(s,t)$.  Clearly since ${\cal N}_{L} > {\cal N}_{L-2}$ 
    the difference operator $\widehat R$ cannot  be invertible,
{\it i.e.}, (\ref{Wardagain}) represents a non-trivial constraint on ${\cal M}$. By assumption 4,  $\mathcal{M}_{IJK}(s,t)$  are rational functions of $s$ and $t$. We will now show  that  compatibility with  (\ref{Wardagain}) requires that $\widetilde{\mathcal{M}}_{ijk}(s,t)$ must also be rational functions. (The  argument that follows is elementary but slightly elaborate and can be safely skipped on first reading.)

\begin{figure}[htbp]
\begin{center}
\includegraphics[scale=0.20]{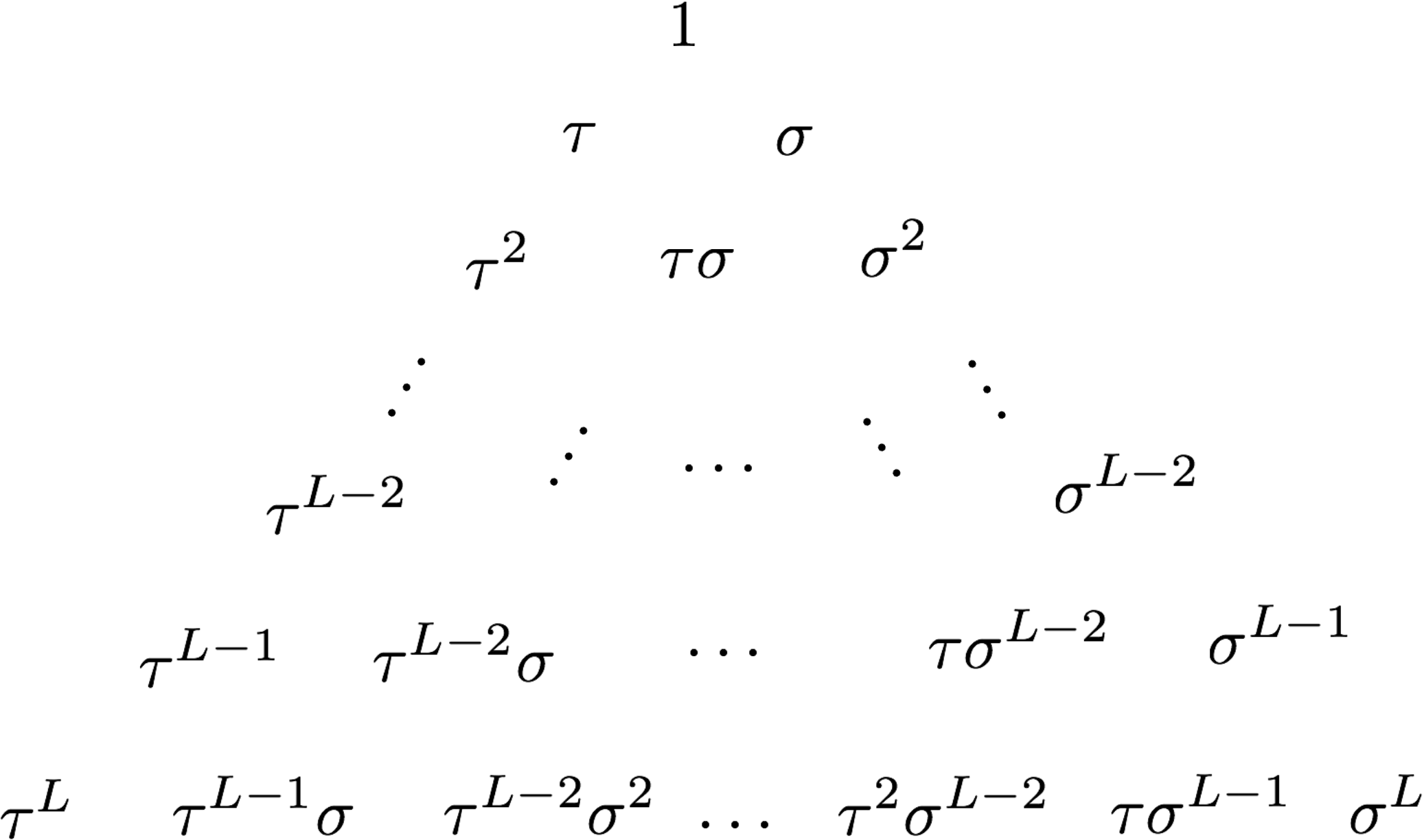}
\caption{R-symmetry monomials in $\mathcal{M}$.}
\label{R1}
\end{center}
\end{figure}

\begin{figure}[htbp]
\begin{center}
\includegraphics[scale=0.20]{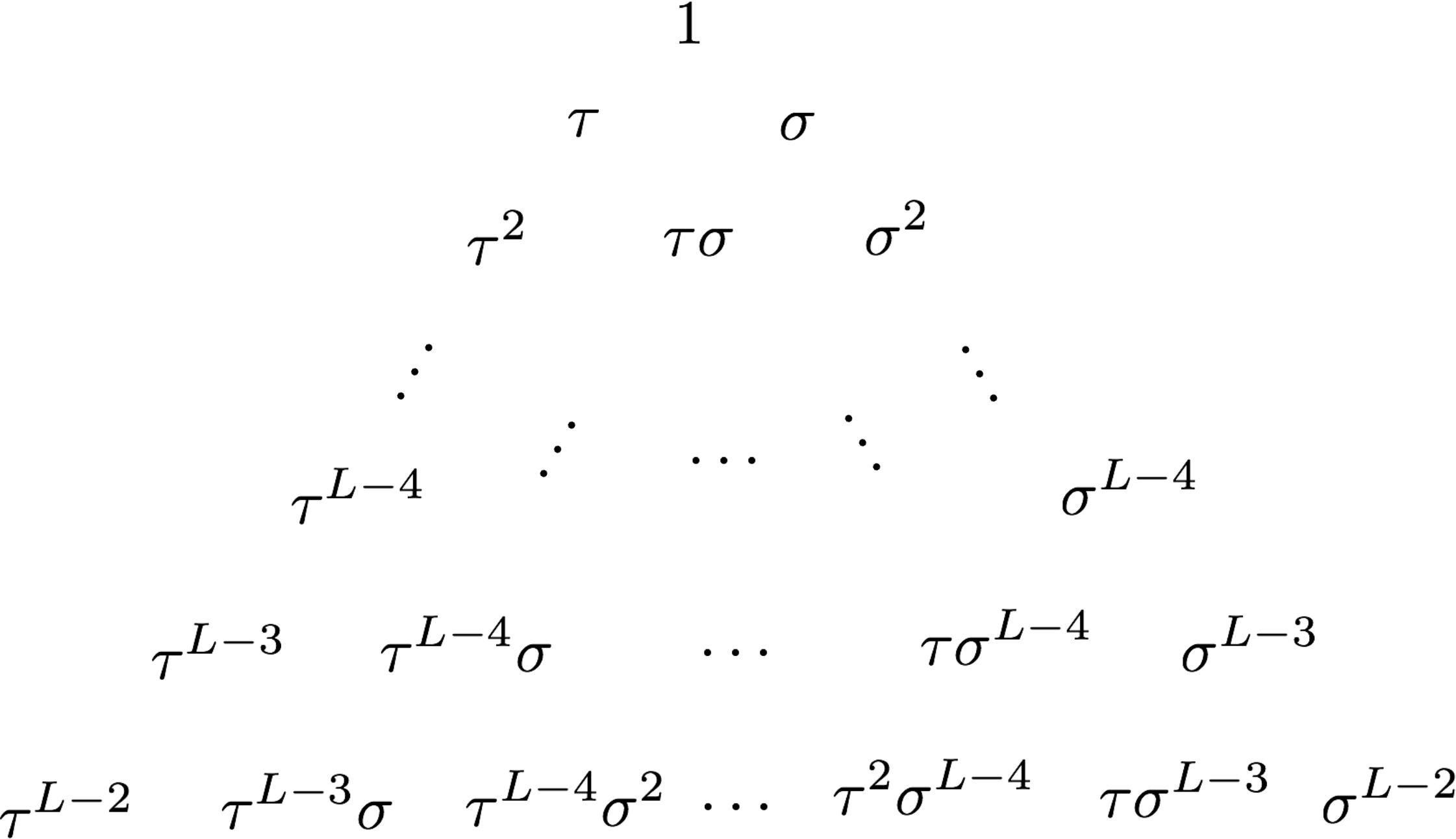}
\caption{R-symmetry monomials in $\widetilde{\mathcal{M}}$.}
\label{R2}
\end{center}
\end{figure}

The two sets of R-symmetry monomials $\{\sigma^I\tau^J\}$ and $\{\sigma^i\tau^j\}$ can be conveniently arranged into two equilateral triangles, illustrated  respectively by Figure \ref{R1} and Figure \ref{R2}. The Bose symmetry that relates different R-symmetry monomials corresponds to the $S_3$  
the symmetry of the equilateral triangle. Let us start by considering the monomial $1$ in $\mathcal{M}$, which is associated to the coefficient $\mathcal{M}_{0,0,L}(s,t)$. This monomial can only be reproduced by the monomial $1$ in $\widetilde{\mathcal{M}}$, {\it i.e.}, the term $\widetilde{\mathcal{M}}_{0,0,L-2}(s,t)$, via the action of the operator $\widehat{V}$ in $\widehat{R}$,
\begin{equation}
\begin{split}
\mathcal{M}_{0,0,L}(s,t)={}&\widehat{V}\circ\widetilde{\mathcal{M}}_{0,0,L-2}(s,t)\\
={}&\widetilde{\mathcal{M}}_{0,0,L-2}(s,t-2)\left(\frac{p_2+p_3-t}{2}\right)\left(\frac{p_1+p_4-t}{2}\right)\left(\frac{p_1+p_3-u}{2}\right)\left(\frac{p_2+p_4-u}{2}\right)\;.
\end{split}
\end{equation}
We can then $\widetilde{\mathcal{M}}_{0,0,L-2}(s,t)$ in terms  of $\mathcal{M}_{0,0,L}(s,t)$
\begin{equation}
\widetilde{\mathcal{M}}_{0,0,L-2}(s,t)=\frac{\mathcal{M}_{0,0,L}(s,t)}{\left(\frac{p_2+p_3-t}{2}\right)\left(\frac{p_1+p_4-t}{2}\right)\left(\frac{p_1+p_3-u}{2}\right)\left(\frac{p_2+p_4-u}{2}\right)}\bigg|_{t\to t+2}\; ,
\end{equation}
which makes it clear that  $\widetilde{\mathcal{M}}_{0,0,L-2}(s,t)$ is rational given that  $\mathcal{M}_{0,0,L}(s,t)$ is assumed to be rational. Similarly, one can easily see that $\sigma^L\mathcal{M}_{L,0,0}(s,t)$ can only be reproduced from $\sigma^{L-2}\widetilde{\mathcal{M}}_{L-2,0,0}(s,t)$ via the action of $\sigma^2\widehat{UV}$ and $\tau^L\mathcal{M}_{0,L,0}(s,t)$ can only come from $\tau^{L-2}\widetilde{\mathcal{M}}_{0,L-2,0}(s,t)$ with the action $\tau^2\widehat{U}$. These two sets of $\mathcal{M}_{IJK}$ and $\widetilde{\mathcal{M}}_{ijk}$  correspond to the six corners of the two triangles and are in the same orbit under the action of the Bose symmetry.  Using the explicit form of the operators $\widehat{UV}$ and $\widehat{U}$ it is apparent that both $\widetilde{\mathcal{M}}_{L-2,0,0}(s,t)$ and $\widetilde{\mathcal{M}}_{0,L-2,0}(s,t)$ can be analogously solved and have finitely many poles in the Mandelstam variables.  Now let us move on to consider $\sigma\mathcal{M}_{1,0,L-1}(s,t)$ which receives contribution from $\widetilde{\mathcal{M}}_{0,0,L-2}(s,t)$ with the action of $-\sigma\widehat{V}-\sigma\widehat{UV}+\sigma\widehat{V^2}$ as well as from $\sigma\widetilde{\mathcal{M}}_{1,0,L-3}(s,t)$ with the action of $\widehat{V}$
\begin{equation}
\mathcal{M}_{1,0,L-1}(s,t)=(-\widehat{V}-\widehat{UV}+\widehat{V^2})\circ \widetilde{\mathcal{M}}_{0,0,L-2}(s,t)+\widehat{V}\circ \widetilde{\mathcal{M}}_{1,0,L-3}(s,t)\;.
\end{equation}
Since we have deduced the finiteness of the number of poles in $\widetilde{\mathcal{M}}_{0,0,L-2}(s,t)$, it is obvious from the above equation that $\widetilde{\mathcal{M}}_{1,0,L-3}(s,t)$ also has a finite number of poles. By the same logic, one can easily convince oneself  that the number of poles in $\widetilde{\mathcal{M}}_{0,1,L-3}(s,t)$, $\widetilde{\mathcal{M}}_{L-3,1,0}(s,t)$, $\widetilde{\mathcal{M}}_{L-3,0,1}(s,t)$, $\widetilde{\mathcal{M}}_{0,L-3,1}(s,t)$, $\widetilde{\mathcal{M}}_{1,L-3,0}(s,t)$ is also finite. The strategy  is now clear. We start from the corners of the triangle and move along the edges. Each time we encounter a new element of $\widetilde{\mathcal{M}}_{i,j,k}(s,t)$ multiplied by a single difference operator of the type $\widehat{U^mV^n}$ and by recursion we can prove this new term has finitely many poles. After finishing the outer layer of the R-symmetry triangle, we move onto the adjacent layer, again starting from the three corners and then moving along the edges. It is not hard to see that at each step the same situation occurs and we only need to deal with one new element at a time. For example, $\sigma\tau \mathcal{M}_{1,1,L-2}(s,t)$, which is on the top corner of the second layer, is generated by $\widetilde{\mathcal{M}}_{0,0,L-2}(s,t)$ with the action of $-\sigma\tau \widehat{U}-\sigma\tau\widehat{UV}+\sigma\tau\widehat{U^2}$,  $\sigma\widetilde{\mathcal{M}}_{1,0,L-3}(s,t)$ with $-\tau\widehat{V}-\tau\widehat{U}+\tau\widehat{1}$,  $\tau\widetilde{\mathcal{M}}_{0,1,L-3}(s,t)$ with $-\sigma\widehat{V}-\sigma\widehat{UV}+\sigma\widehat{V^2}$ and $\sigma\tau\widetilde{\mathcal{M}}_{1,1,L-3}(s,t)$ with $\widehat{V}$. Among these four elements of the auxiliary amplitude $\widetilde{\mathcal{M}}_{0,0,L-2}(s,t)$, $\sigma\widetilde{\mathcal{M}}_{1,0,L-3}(s,t)$, $\tau\widetilde{\mathcal{M}}_{0,1,L-3}(s,t)$ belong to the outer layer which are determined to be rational in the previous round. Only the element $\sigma\tau\widetilde{\mathcal{M}}_{1,1,L-3}(s,t)$ belongs to the inner layer and is acted on by the simple difference operator $\widehat{V}$. This concludes by recursion that $\widetilde{\mathcal{M}}_{1,1,L-3}(s,t)$ is also rational. In finitely many steps, we can exhaust all the elements of $\widetilde{\mathcal{M}}_{ijk}$. This concludes the proof of rationality of $\widetilde{\mathcal{M}}$.  It might at first sight appear that this procedure amounts to an algorithm to invert the difference operator $\widehat R$, but  of course this is not the case. For general ${\cal M}_{IJK}$, one would find contradictory results for some element
$\widetilde {\cal M}_{ijk}$ applying the recursion procedure by following different paths in the triangle.

\subsection{Solution}\label{bootstrapsolution}

Experimentation with low-weight examples  led us to  the following ansatz for $\widetilde{\mathcal{M}}$, 
\begin{equation}\label{mtildesolution}
\widetilde{\mathcal{M}}(s,t,\tilde{u};\sigma,\tau)=\sum_{{\tiny\begin{split}{}&i+j+k=L-2,\\{}&0\leq i,j,k\leq L-2\end{split}}} \frac{a_{ijk}\sigma^i\tau^j}{(s-s_M+2k)(t-t_M+2j)(\tilde{u}-\tilde{u}_M+2i)}
\end{equation}
where
\begin{equation}
\begin{split}
s_M={}&\min\{p_1+p_2, p_3+p_4\}-2\;,\\
t_M={}&\min\{p_1+p_4,p_2+p_3\}-2\;,\\
\tilde{u}_M={}&\min\{p_1+p_3, p_2+p_4\}-2\;.
\end{split}
\end{equation}
The reader can check that this ansatz leads to an $\mathcal{M}$ that satisfies the asymptotic requirement, obeys  Bose symmetry and has simple poles at the required location. The further requirements that the poles have polynomials residues fixes the coefficients $a_{ijk}$  uniquely 
up to  normalization,
\begin{equation} \label{aijk}
a_{ijk}=\frac{C_{p_1 p_2 p_3 p_4}\; {L-2 \choose i,j,k}}{(1+\frac{|p_1-p_2+p_3-p_4|}{2})_i(1+\frac{|p_1+p_4-p_2-p_3|}{2})_j(1+\frac{|p_1+p_2-p_3-p_4|}{2})_k}\; ,
\end{equation}
where ${L-2 \choose i,j,k}$ is the trinomial coefficient. The overall normalization 
\begin{equation}
C_{p_1 p_2 p_3 p_4} = \frac{ f(p_1, p_2, p_3, p_4)}{N^2}
\end{equation}
 cannot  be fixed from our homogenous consistency conditions. In principle, it can be determined by transforming back to the position-space expression (\ref{wardconnected}).
  As we shall show below, the term $\mathcal{G}_{\rm free, conn}$ arises as a regularization effect in the inverse Mellin transformation. The constant  $f(p_1, p_2, p_3, p_4)$ is fixed by requiring that the regularization procedure gives the correctly normalized  free-field  correlator.
  In practice, this is very cumbersome, and it is easier to take instead $\mathcal{G}_{\rm free, conn}$  as an input from free-field theory. The overall normalization of ${\cal M}$ is then fixed by imposing the cancellation of spurious singularity associated to single-trace long operators \cite{Dolan:2006ec},
  which are separately present in $\mathcal{G}_{\rm free, conn}$ and in $ R \, {\cal H}$ but must cancel in the sum. This method has been used in \cite{Aprile:2017xsp} to determine $f(p, p, q, q)$, the normalization  in all cases with pairwise equal weights. The normalization for arbitrary weights $f(p_1, p_2, p_3, p_4)$ has been recently determined in \cite{Aprile:2018efk} by further taking a light-like limit.

\subsubsection{Uniqueness for $p_i=2$}\label{peq2proof}
Uniqueness of the ansatz (\ref{mtildesolution}) is in general difficult to prove. However in simple examples it is possible to solve the algebraic problem directly, thereby proving that the answer is unique. In this subsection we demonstrate it for the simplest case,
the equal-weights case with  $p=2$. This case is particularly simple because $\widetilde{\mathcal{M}}$ has no $\sigma$, $\tau$ dependence. 

Recall that the Mellin amplitude $\mathcal{M}$ has simple poles in $s$, $t$ and $u$ whose positions are restricted by the condition (\ref{mpolerange}). Specifically in the case of $p_i=2$, it means that the Mellin amplitude can only have simple poles at $s=2$, $t=2$ and $u=2$. On the other hand, $\widetilde{\mathcal{M}}$ must also have poles because the Pochhammer symbols in the difference operators (\ref{roperator}) do not introduce additional poles. 
To fix the position of these poles in $\widetilde{\mathcal{M}}$, let us look at the R-symmetry monomial $\sigma^I\tau^J$ in $\mathcal{M}(s,t;\sigma,\tau)$ with $I=J=0$. 
The $\sigma^I\tau^J$ term in $\mathcal{M}(s,t;\sigma,\tau)$ with $I=J=0$ can then only be produced from  $\widetilde{\mathcal{M}}(s,t)$ with the action of the term $\widehat{V}$ in (\ref{hatR})
\begin{equation}\label{vmtilde}
\widehat{V}\circ\widetilde{\mathcal{M}}(s,t)=\widetilde{M}(s,t-2)\left[\left(\frac{4-t}{2}\right)_1\left(\frac{4-u}{2}\right)_1\right]^2\;.
\end{equation}
For $s$ to have simple pole at $s=2$ in $\mathcal{M}$, it is easy to see that the only possible $s$-pole in $\widetilde{\mathcal{M}}(s,t)$ is a simple pole at $s=2$. For $t$, a simple pole at $t=0$ in  $\widetilde{\mathcal{M}}(s,t)$ is allowed, which after the shift on the right side of (\ref{vmtilde}) gives a simple pole at $t=2$ in $\mathcal{M}$. But there is also an additional pole in $t$ allowed due to the presence of the Pochhammer symbol. Since the Pochhammer symbol gives a degree-two zero at $t=4$  we can have  a pole at $t=2$ in $\widetilde{\mathcal{M}}(s,t)$  with pole degree up to two. These two possibilities exhaust all the allowed $t$-poles in $\widetilde{\mathcal{M}}(s,t)$ that are compatible with the pole structure of $\mathcal{M}$. Now the story for $\tilde{u}$-poles is exactly the same as $t$. To see this, we note that under the shift $t\to t-2$,  
\begin{equation}
\tilde{u}\to \tilde{u}+2=(u-4)+2=u-2\;.
\end{equation}
By the same argument  $\tilde{u}$ can have in $\widetilde{\mathcal{M}}(s,t)$ a simple pole at $\tilde{u}=0$ and at most a double pole at $\tilde{u}=2$. 

Now we use the constraints from Bose symmetry (actually crossing symmetry in this case) and the asymptotic condition to further narrow down the possibilities. Bose symmetry requires
\begin{equation}
\widetilde{\mathcal{M}}(s,t)=\widetilde{\mathcal{M}}(s,\tilde{u})=\widetilde{\mathcal{M}}(t,s)\;.
\end{equation}
Since $\widetilde{\mathcal{M}}(s,t)$ cannot have a pole at $s=0$, the poles at $t=0$, $\tilde{u}=0$ are prohibited. On the other hand the asymptotic condition further requires  ${\mathcal{M}}(s,t)$ to have growth rate one at large $s$, $t$, $u$. Consequently by simple power counting $\widetilde{\mathcal{M}}(s,t)$ should have growth rate $-3$. This leaves us with the unique crossing symmetric ansatz 
\begin{equation}
\widetilde{\mathcal{M}}(s,t)\propto\frac{1}{(s-2)(t-2)(\tilde{u}-2)}
\end{equation} 
which is just our solution (\ref{mtildesolution}). 

\subsection{Contour subtleties and the free correlator} \label{subtlefree}

In this section we address some subtleties related to $s$ and $t$ integration contours  in the Mellin representation. These subtleties are related to the decomposition of the position space correlator into a ``free'' and a dynamical term.
In transforming to Mellin space, we have ignored the  term   $\mathcal{G}_{\rm free, conn}$. We are going to see how this term can be recovered by taking the inverse Mellin transform  with proper integration contours.

The four-point function calculated from supergravity with the traditional method is a sum of four-point contact diagrams, known as $\bar D$-functions. (Their precise definition is given in  (\ref{dbar})). 
Through the repeated use of identities obeyed the  $\bar{D}$ functions, the supergravity answer can be massaged into a form that agrees with the solution to the superconformal Ward identity -- with a singled-out ``free'' piece. Manipulations of this sort can be found in, {\it e.g.}, \cite{Dolan:2001tt,Arutyunov:2002fh,Arutyunov:2003ae,Uruchurtu:2011wh}. Most of the requisite identities have an elementary proof either in position space or in Mellin space, but the 
  crucial identity which is key to the separation of the free term, namely
\begin{equation}\label{ddd}
\left(
\bar{D}_{\Delta_1+1\Delta_2\Delta_3+1\Delta_4}+U \bar{D}_{\Delta_1+1\Delta_2+1\Delta_3\Delta_4}+V\bar{D}_{\Delta_1\Delta_2+1\Delta_3+1\Delta_4}
\right)
\bigg|_{\Delta_4=\Delta_1+\Delta_2+\Delta_3}=\prod_{i=1}^3\Gamma(\Delta_i),
\end{equation}
 requires additional care. The Mellin transform of the rhs is clearly ill-defined. We will now show that the Mellin transform of the lhs is also ill-defined, because while each of the three terms has a perfectly good transform for  a finite domain of $s$ and $t$ (known as the ``fundamental domain''),
 the three domains have no common overlap. A suitable regularization procedure is required to make sense of this identity. 
 Let us see this in detail.

Recall that the Mellin transform of an individual 
 $\bar{D}$-function is just a product of Gamma functions. Its fundamental  domain can be characterized by the condition that 
 all the arguments of Gamma functions are positive \cite{Symanzik:1972wj}. For the three $\bar{D}$-functions appearing on the lhs of (\ref{ddd}), 
 we have
 \begin{eqnarray}
{}&\bar{D}_{\Delta_1+1\Delta_2\Delta_3+1\Delta_4}= \frac{1}{4}\int_{\mathcal{C}_1}ds dt U^{s/2}V^{t/2}\Gamma[-\frac{s}{2}]\Gamma[-\frac{t}{2}]\Gamma[\frac{s+t+\Delta_1+\Delta_2+\Delta_3-\Delta_4+2}{2}] \nonumber \\
{}&\;\;\;\;\;\;\;\;\;\;\;\;\;\;\;\times\Gamma[-\frac{s}{2}+\frac{\Delta_4+\Delta_3-\Delta_1-\Delta_2}{2}]\Gamma[-\frac{t}{2}+\frac{\Delta_4+\Delta_1-\Delta_2-\Delta_3}{2}]\Gamma[\frac{s+t}{2}+\Delta_2] \, ,\nonumber \\
{}&\bar{D}_{\Delta_1+1\Delta_2+1\Delta_3\Delta_4}= \frac{1}{4}\int_{\mathcal{C}_2}ds dt U^{s/2}V^{t/2}\Gamma[-\frac{s}{2}]\Gamma[-\frac{t}{2}]\Gamma[\frac{s+t+\Delta_1+\Delta_2+\Delta_3-\Delta_4+2}{2}] \nonumber \\
{}&\;\;\;\;\;\;\;\;\;\;\;\;\;\times\Gamma[-\frac{s}{2}+\frac{\Delta_4+\Delta_3-\Delta_1-\Delta_2}{2}-1]\Gamma[-\frac{t}{2}+\frac{\Delta_4+\Delta_1-\Delta_2-\Delta_3}{2}]\Gamma[\frac{s+t}{2}+\Delta_2+1] \, ,\nonumber \\
{}&\bar{D}_{\Delta_1\Delta_2+1\Delta_3+1\Delta_4}= \frac{1}{4}\int_{\mathcal{C}_3}ds dt U^{s/2}V^{t/2}\Gamma[-\frac{s}{2}]\Gamma[-\frac{t}{2}]\Gamma[\frac{s+t+\Delta_1+\Delta_2+\Delta_3-\Delta_4+2}{2}]. \nonumber \\
{}&\;\;\;\;\;\;\;\;\;\;\;\;\;\times\Gamma[-\frac{s}{2}+\frac{\Delta_4+\Delta_3-\Delta_1-\Delta_2}{2}]\Gamma[-\frac{t}{2}+\frac{\Delta_4+\Delta_1-\Delta_2-\Delta_3}{2}-1]\Gamma[\frac{s+t}{2}+\Delta_2+1] \, .
\end{eqnarray}
Here
\begin{equation}
\int_{\mathcal{C}_i}ds dt =\int_{s_{0i}-i\infty}^{s_{0i}+i\infty}ds\int_{t_{0i}-i\infty}^{t_{0i}+i\infty}dt\;,
\end{equation}
so the contours are specified by selecting a point inside the fundamental domains, $(s_{0i},t_{0i})\in \mathcal{D}_i$. 
With $\Delta_4=\Delta_1+\Delta_2+\Delta_3$, one finds that the fundamental domains are given by
\begin{equation}
\mathcal{D}_1=\mathcal{D}_2=\mathcal{D}_3=\{(s_0,t_0)|\Re(s)<0,\Re(t)<0,\Re(s)+\Re(t)>-2\}\;.
\end{equation} 
 Multiplication by $U$ and $V$ in the second and the third terms, respectively, shifts\footnote{To absorb $U^mV^n$ outside the integral into $U^{s/2}V^{t/2}$ inside the integral and then shift $s$ and $t$ to bring it back to the form  $U^{s/2}V^{t/2}$. Doing so amounts to shift $\mathcal{D}$ to $\mathcal{D}'$ by a vector $(2m, 2n)$.} the domains $\mathcal{D}_2$ and $\mathcal{D}_3$ into new domains $\mathcal{D}'_2$ and $\mathcal{D}'_3$,
\begin{equation}
\begin{split}
\mathcal{D}'_2={}&\{(s_0,t_0)|\Re(s)<2,\Re(t)<0,\Re(s)+\Re(t)>0\}\;,\\
\mathcal{D}'_3={}&\{(s_0,t_0)|\Re(s)<0,\Re(t)<2,\Re(s)+\Re(t)>0\}\;.
\end{split}
\end{equation} 
This is problematic because
\begin{equation}
\mathcal{D}_1\bigcap\mathcal{D}'_2\bigcap\mathcal{D}'_3=\emptyset\;.
\end{equation}
 Clearly it makes no sense to add up the integrands if the contour integrals share no common domain. On the other hand, if one is being cavalier and sums up the integrands anyway, one finds that the total integrand vanishes. 
 This is ``almost'' the correct result, since the rhs of the identity (\ref{ddd}) is simply a constant, whose Mellin transform is ill-defined and was indeed set to zero in our analysis in the previous section. We can however do better 
 and reproduce the exact identity if we adopt the following ``regularization'' prescription: we shift $s+t\to s+t+\epsilon$, with $\epsilon$  a small positive real number. After this shift,  the three domains develop a small common domain of size $\epsilon$,
\begin{figure}[htbp]
\begin{center}
\includegraphics[scale=0.4]{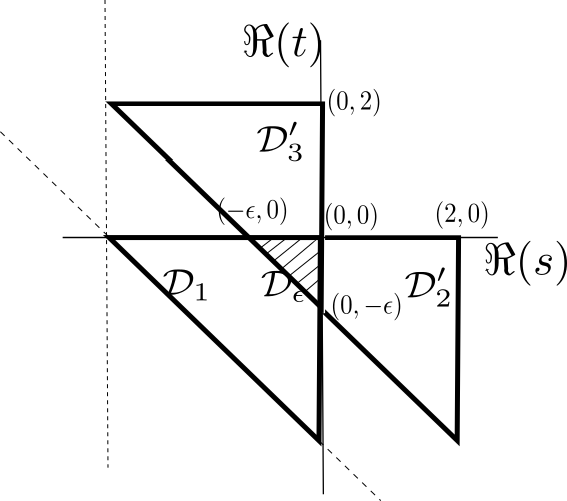}
\caption{The regularized domains. The common domain of size $\epsilon$ is depicted as the shaded region.}
\label{identity}
\end{center}
\end{figure}
\begin{equation}
\mathcal{D}_1\bigcap\mathcal{D}'_2\bigcap\mathcal{D}'_3\equiv\mathcal{D}_\epsilon= \{(s_0,t_0)|\Re(s)<0,\Re(t)<0,\Re(s)+\Re(t)>-\epsilon\}\;.
\end{equation}
We can therefore place the common integral contour inside $\mathcal{D}_\epsilon$ and combine the integrands,
\begin{equation}
\begin{split}
{\rm LHS}={}&\frac{1}{4}\int_{\mathcal{D}_\epsilon}dsdt U^{s/2}V^{t/2}(\frac{s+t+\epsilon}{2}-\frac{s}{2}-\frac{t}{2})\Gamma[-\frac{s}{2}]\Gamma[-\frac{t}{2}]\Gamma[\frac{s+t+\epsilon}{2}]\\
\;\;\;\;\;\;\;\;\;\;\;\;\;\;\;\;\;\;\;\;\;\;\;\;\;{}&\times\Gamma[-\frac{s}{2}+\Delta_3]\Gamma[-\frac{t}{2}+\Delta_1]\Gamma[\frac{s+t}{2}+\Delta_2]\\
={}&\frac{1}{4}\int_{\mathcal{D}_\epsilon}dsdt U^{s/2}V^{t/2}\frac{\epsilon}{2}\Gamma[-\frac{s}{2}]\Gamma[-\frac{t}{2}]\Gamma[\frac{s+t+\epsilon}{2}]\\
{}&\times\Gamma[-\frac{s}{2}+\Delta_3]\Gamma[-\frac{t}{2}+\Delta_1]\Gamma[\frac{s+t}{2}+\Delta_2]\;.
\end{split}
\end{equation}
As $\epsilon\to 0$, we can just substitute $s=t=0$ into the non-singular part of the integrand. The  resulting integral is easily evaluated,
\begin{equation}
{\rm LHS}=\frac{1}{2}\Gamma[\Delta_1]\Gamma[\Delta_2]\Gamma[\Delta_3]\int_{\mathcal{D}_\epsilon}\frac{ds}{2}\frac{dt}{2} \epsilon \Gamma[-\frac{s}{2}]\Gamma[-\frac{t}{2}]\Gamma[\frac{s+t+\epsilon}{2}]=\Gamma[\Delta_1]\Gamma[\Delta_2]\Gamma[\Delta_3]={\rm RHS}\,.
\end{equation}
This amounts to a ``proof'' of the identity (\ref{ddd}) directly in Mellin space. This exercise contains a useful general lesson. As we have already remarked, the identity  (\ref{ddd})  is responsible for generating the term $\mathcal{G}_{\rm free, conn}$ by collapsing sums of $\bar D$ functions in the supergravity answer. We have shown that it is consistent to treat the Mellin transform of $\mathcal{G}_{\rm free, conn}$ as ``zero'', provided that we are careful about the $s$, $t$ integration contours in the inverse Mellin transform. In general,
when one is adding up integrands, one should make sure the integrals share the same contour, which may require a regularization procedure of the kind we have just used. A naively ``zero'' Mellin amplitude 
can then  give nonzero contributions to the integral if the contour is pinched to an infinitesimal domain where the integrand has a pole. In Appendix \ref{peq2data} we illustrate in the simplest case of equal weights $p_i=2$ how the free field correlator is correctly 
reproduced by this mechanism. 

We conclude by alerting the reader about another small subtlety. The  free term $\mathcal{G}_{\rm{free, conn}}$ depends on the precise identification of the operators dual to the supergravity modes $s_p$. As explained
in footnote \ref{footnotedt}, if one adopts the scheme where the fields $s_p$  contain no derivative cubic couplings, the dual operators are necessarily admixtures of single- and multi-trace operators.
While the multi-trace pieces are in general subleading, they can affect the free-field four-point function if the four weights are sufficiently ``unbalanced''. This phenomenon was encountered in \cite{Uruchurtu:2008kp,Uruchurtu:2011wh},
where the four-point functions with weights $(2, 2, p, p)$ were evaluated from supergravity. A discrepancy was found for $p \geq 4$ between the function $\mathcal{G}_{\rm{free, conn}}$ obtained by writing the supergravity result in the split form (\ref{wardconnected})
and the free-field result obtained in free field theory from Wick contractions, assuming that the operators are pure single-traces. The resolution   is that  supergravity  is really computing the four-point function of  more complicated operators with multi-trace admixtures. Note that the contribution to the four-point functions from the multi-trace terms takes the form of a product of two- and three-point functions of one-half BPS operators, and is thus protected \cite{Arutyunov:2000ima}. The ambiguity in the precise identification of the 
dual operators can then only affect $\mathcal{G}_{\rm{free, conn}}$ and not the dynamical part.

\section{The position space method}\label{positionspace}

We now switch gears and describe a logically independent  position space method. This methods mimics the traditional recipe for computing four-point functions in supergravity, but eschews detailed knowledge supergravity effective action and complicated
combinatorics.  

The idea is to write the write full amplitude as a sum of exchange diagrams and contact diagrams, but parametrizing the vertices with undetermined coefficients. The spectrum of  IIB supergravity on $AdS_5\times S^5$ is such that all the exchange diagrams can be written as a finite sum of contact diagrams, {\it i.e.}, $D$-functions, making the whole amplitude a sum of $D$-functions. We then use the property of $D$-functions to decompose the amplitude into a basis of independent functions. The
full amplitude is encoded into four rational coefficient functions. Imposing the superconformal Ward identity we find a large number of relations among the undetermined coefficients. Uniqueness of the maximally supersymmetric Lagrangian guarantees that all the coefficients in the ansatz can be fixed up to overall rescaling. Finally the overall constant can be determined by comparing with the free field result after restricting the R-symmetry cross ratios to a special slice \cite{Nirschl:2004pa} (this is related 
the chiral algebra twist \cite{Beem:2013sza}). We emphasize that there is no guesswork anywhere. The position space method is {\it guaranteed} to give the same results as  a direct supergravity calculation, but it is technically much simpler.

We will discuss the method  only for the equal-weight  case $p_i=p$, but its generalization to the unequal-weight case is straightforward. In addition to reproducing the known examples $p=2, 3, 4$, we computed the new case of $p=5$ and found it to be an agreement both with the conjecture of \cite{Dolan:2006ec} and with our Mellin amplitude conjecture (\ref{mtildesolution}, \ref{aijk}).  We have included these results in the form of tables of coefficients in  Appendix \ref{p345}. Explicit form of the results as sums of ${D}$-functions is also available as a Mathematica notebook included in the ArXiv version of this paper.

We start by reviewing some facts about exchange and contact Witten diagrams, some of which have already been mentioned in the previous sections. We then explain in detail how to decompose the position space ansatz into a basis and how to implement the superconformal Ward identity. We end the section with a demonstration of the method in the simplest $p_i=2$ case.  Explicit formulae and technical details are given in the Appendices.  

\subsection{Exchange diagrams}

The Kaluza-Klein fields that can appear in the internal propagator of an exchanged diagrams
 have been listed in Table \ref{exchangedmultiplets}. The allowed fields are restricted by the R-symmetry selection rule,
\begin{equation}
[0,p,0]\otimes [0,p,0]=\oplus_{0\leq m\leq n\leq p}[n-m,2m,n-m] \, ,
\end{equation}
and  by a twist cut-off,
\begin{equation}
\tau<2p\;.
\end{equation}
This origin of the twist cut-off has been discussed in Section \ref{sugrareview}, and an alternative explanation from the Mellin amplitude perspective was given in Section \ref{mellinwittendiagrams}.

The requisite exchange diagrams have been computed in the early days of the AdS/CFT correspondence.  They can all  be represented as  finite sums of $\bar D$-functions. We have summarized the relevant formulae in Appendix \ref{exch}.
As each exchanged field belongs to a certain R-symmetry representation $[n-m,2m,n-m]$, we should multiply the exchange Witten diagrams by the corresponding R-symmetry polynomial $Y_{nm}$. These polynomials were derived in \cite{Nirschl:2004pa},
\begin{equation}\label{Ynm}
Y_{nm}(\alpha,\bar{\alpha})= \frac{P_n(\alpha)P_m(\bar{\alpha})-P_m(\alpha)P_n(\bar{\alpha})}{\alpha-\bar{\alpha}}\;, 
\end{equation}
with $P_n(\alpha)$ are the usual Legendre polynomials. The R-symmetry polynomials  are eigenfunctions of the R-symmetry Casimir operator and are thus the ``compact'' analogue of the conformal partial waves.

\subsection{Contact diagrams}
In addition to the exchange diagrams, the four-point functions receive contribution from contact diagrams. The contact vertices in the effective Lagrangian have been explicitly worked out in \cite{Arutyunov:1999fb}, with the number of derivatives going up to four. However as we argued, the requirement of a good flat space limit forbids genuine   four-derivative contributions. This was recently confirmed in \cite{Arutyunov:2017dti} by explicit computation. Therefore only zero-derivative vertices and two-derivative vertices will effectively contribute to the four-point function. We also observe a further simplification that for the equal-weight case: the zero-derivative contributions can  be absorbed into the two-derivative ones when the external dimension satisfies $p\neq 4$. The proof of this statement is presented in Appendix \ref{contactvertices}.

\subsection{Reducing the amplitude to four rational coefficient functions}

As always, it will be convenient to write the amplitude as function of the conformal and R-symmetry cross-ratios, pulling out an overall kinematic factor.
The $\bar{D}$-functions are defined in terms of contact Witten diagrams (known as $D$-functions, see (\ref{mtildesolution})) by the extraction of such a kinematic  factor,
\begin{equation}\label{dbar}
\frac{ \prod_{i=1}^4\Gamma(\Delta_i)}{\Gamma(\Sigma-\frac{1}{2}d)}\frac{2}{\pi^{\frac{d}{2}}}D_{\Delta_1\Delta_2\Delta_3\Delta_4}(x_1,x_2,x_3,x_4)=\frac{r_{14}^{\Sigma-\Delta_1-\Delta_4}r_{34}^{\Sigma-\Delta_3-\Delta_4}}{r_{13}^{\Sigma-\Delta_4}r_{24}^{\Delta_2}}\bar{D}_{\Delta_1\Delta_2\Delta_3\Delta_4} (U,V)\, ,
\end{equation}
where $2\Sigma \equiv \sum_{i=1}^4 \Delta_i$.

The set of $\bar{D}$-functions  is overcomplete, as they are related by several identities, but our method requires a basis of independent functions.
To remove this redundancy we will represent the $\bar{D}$-functions in a way that makes the identities manifest. We use the fact that all $\bar{D}_{\Delta_1\Delta_2\Delta_3\Delta_4}$ can be obtained from $\Phi(U,V)\equiv \bar{D}_{1111}(U,V)$ with the action of differential operators of $U$ and $V$. The following six differential operators  allow to move around in the weight space $(\Delta_1,\Delta_2,\Delta_3,\Delta_4)$ of $\bar{D}$-functions (see, {\it e.g.,} \cite{Arutyunov:2002fh}),
\begin{equation}
\begin{split}
\bar{D}_{\Delta_1+1,\Delta_2+1,\Delta_3,\Delta_4}={}&\mathcal{D}_{12}\bar{D}_{\Delta_1,\Delta_2,\Delta_3,\Delta_4}:=-\partial_U \bar{D}_{\Delta_1,\Delta_2,\Delta_3,\Delta_4}\;,\\
\bar{D}_{\Delta_1,\Delta_2,\Delta_3+1,\Delta_4+1}={}&\mathcal{D}_{34}\bar{D}_{\Delta_1,\Delta_2,\Delta_3,\Delta_4}:=(\Delta_3+\Delta_4-\Sigma-U\partial_U )\bar{D}_{\Delta_1,\Delta_2,\Delta_3,\Delta_4}\;,\\
\bar{D}_{\Delta_1,\Delta_2+1,\Delta_3+1,\Delta_4}={}&\mathcal{D}_{23}\bar{D}_{\Delta_1,\Delta_2,\Delta_3,\Delta_4}:=-\partial_V \bar{D}_{\Delta_1,\Delta_2,\Delta_3,\Delta_4}\;,\\
\bar{D}_{\Delta_1+1,\Delta_2,\Delta_3,\Delta_4+1}={}&\mathcal{D}_{14}\bar{D}_{\Delta_1,\Delta_2,\Delta_3,\Delta_4}:=(\Delta_1+\Delta_4-\Sigma-V\partial_V )\bar{D}_{\Delta_1,\Delta_2,\Delta_3,\Delta_4}\;,\\
\bar{D}_{\Delta_1,\Delta_2+1,\Delta_3,\Delta_4+1}={}&\mathcal{D}_{24}\bar{D}_{\Delta_1,\Delta_2,\Delta_3,\Delta_4}:=(\Delta_2+U\partial_U+V\partial_V )\bar{D}_{\Delta_1,\Delta_2,\Delta_3,\Delta_4}\;,\\
\bar{D}_{\Delta_1+1,\Delta_2,\Delta_3+1,\Delta_4}={}&\mathcal{D}_{13}\bar{D}_{\Delta_1,\Delta_2,\Delta_3,\Delta_4}:=(\Sigma-\Delta_4+U\partial_U+V\partial_V )\bar{D}_{\Delta_1,\Delta_2,\Delta_3,\Delta_4}\;.
\end{split}
\end{equation}
The ``seed'' $\bar{D}$-function $\Phi(U,V)$ is the famous scalar one-loop box integral in four dimensions and can be expressed in closed form in terms logarithms and dilogarithms. After the change of variable into $U=z \bar{z}$ and $V=(1-z)(1-\bar{z})$, the integral can be written as
\begin{equation}
\Phi(z,\bar{z})=\frac{1}{z-\bar{z}}\bigg(\log(z\bar{z})\log(\frac{1-z}{1-\bar{z}})+2{\rm Li}(z)-2\rm{Li}(\bar{z})\bigg)\;.
\end{equation}
The function $\Phi$ obeys the following differential relations \cite{Eden:2000bk},
\begin{equation}
\begin{split}
\partial_z\Phi={}&-\frac{1}{z-\bar{z}}\Phi-\frac{1}{z(z-\bar{z})}\log(-1+z)(-1+\bar{z})+\frac{1}{(-1+z)(z-\bar{z})}\log(z\bar{z})\;,\\
\partial_{\bar{z}}\Phi={}&\frac{1}{z-\bar{z}}\Phi+\frac{1}{\bar{z}(z-\bar{z})}\log(-1+z)(-1+\bar{z})-\frac{1}{(-1+\bar{z})(z-\bar{z})}\log(z\bar{z})\;.
\end{split}
\end{equation}
The recursive use of these two identities makes it clear that each $\bar{D}$-function can be uniquely written as 
\begin{equation}
\bar{D}_{\Delta_1\Delta_2\Delta_3\Delta_4}=R_\Phi \Phi(U,V)+R_{V} \log V+R_{U} \log U+R_0
\end{equation}
where $R_{\Phi,  U, V,0}$ are rational functions of $z$ and $\bar{z}$. As a result, the supergravity amplitude also admits such a unique decomposition as
\begin{equation}
\mathcal{A}_{\rm sugra}(z,\bar{z};\alpha,\bar{\alpha})=R^{\mathrm{sugra}}_\Phi \Phi(z,\bar{z})+R^{\mathrm{sugra}}_{V}\log V+R^{\mathrm{sugra}}_{U}\log U+R^{\mathrm{sugra}}_0\, .
\end{equation}
The   coefficient functions $R^{\mathrm{sugra}}_{\Phi, U, V, 0}$ are now polynomials of the R-symmetry variables $\alpha$ and $\bar{\alpha}$, where each $R$-symmetry monomial  $\alpha^m \bar \alpha^n$ is multiplied by a rational function of $U$ and $V$.
The coefficients functions $R^{\mathrm{sugra}}_{\Phi, U, V, 0}$ depend linearly on the undetermined coefficients that we have used to parameterize the vertices.

Our ansatz $\mathcal{A}_{\rm sugra}$ must satisfy superconformal Ward identity. The solution can be simply written as
\begin{equation}
\mathcal{A}_{\rm sugra}(z,\bar{z};\alpha,1/\bar{z})=G_{\mathrm{free}}(z,\bar{z};\alpha,1/\bar{z})
\end{equation}
with $G_{\mathrm{free}}(z,\bar{z};\alpha,1/\bar{z})$ being  a rational function that depends only on $z$ and $\alpha$ and can be obtained from free field theory \cite{Nirschl:2004pa}. Under our decomposition, the superconformal Ward identity becomes a set of conditions on the rational coefficient functions,
\begin{equation}
\begin{split}
R_\Phi(z,\bar{z};\alpha,1/\bar{z}){}&=0\;,\\
R_{V}(z,\bar{z};\alpha,1/\bar{z}){}&=0\;,\\
R_{U}(z,\bar{z};\alpha,1/\bar{z}){}&=0\;.
\end{split}
\end{equation}
These conditions imply a large set of linear equations for the undetermined parameters. Uniqueness of  two-derivative IIB supergravity strongly suggests that these conditions must admit a unique solution, up to overall rescaling. This is indeed what we have found
in all examples.
 Finally, the overall normalization is determined by comparing the coefficient function of $R_0$ with the free-field result,
\begin{equation}
R_0(z,\bar{z};\alpha,1/\bar{z})=\mathcal{G}_{\rm free, conn}(z,\bar{z};\alpha,1/\bar{z})\,.  
\end{equation}

\subsection{An example: $p_i =2$}
We now  illustrate the position space method in the simplest case  $p_i = 2$. The four-point amplitude with four identical external scalar has an $S_3$ crossing symmetry. Since the total amplitude is a sum over all Witten diagrams, we can just compute one channel and use crossing symmetry to relate to the other two channels.  In the s-channel, we know from Table  \ref{exchangedmultiplets}  and the twist cut-off $\tau<4$ that there are only three fields which can be exchanged: there is an exchange of scalar with dimension two and in the representation $[0,2,0]$,
\begin{equation}
\mathcal{A}_{\rm scalar}=\frac{1}{8} \pi ^2 \lambda_s U (3 \sigma +3 \tau -1) \bar{D}_{1122}\, ,
\end{equation}
a vector of dimension three in the representation $[1,0,1]$
\begin{equation}
\mathcal{A}_{\rm vector}=\frac{3}{8} \pi ^2 \lambda_v U (\sigma -\tau ) \left(\bar{D}_{1223}-\bar{D}_{2123}+\bar{D}_{2132}-V \bar{D}_{1232}\right) \, ,
\end{equation}
and a massless symmetric graviton in the singlet representation,
\begin{equation}
\mathcal{A}_{\rm graviton}=\frac{1}{3} (-2) \pi ^2 \lambda_g U \left(2 \bar{D}_{1122}-3 \left(\bar{D}_{2123}+\bar{D}_{2132}-\bar{D}_{3133}\right)\right)\;.
\end{equation}
In the above expressions we have used the formulae for exchange Witten diagrams from Appendix \ref{exch} and multiplied with the explicit expression of R-symmetry polynomials $Y_{00}$, $Y_{11}$, $Y_{10}$ given by (\ref{Ynm}). The constants $\lambda_s$, $\lambda_v$ and $\lambda_g$ are undetermined parameters.

For the contact diagram, following the discussion of Appendix \ref{contactvertices}, we only need to consider two-derivative vertices. The most general contribution is as follows (only in the s-channel, as we will sum over the channels in the next step),
\begin{equation}
\mathcal{A}_{\rm contact}=-\bigg(\sum_{0\leq a+b\leq 2} c_{ab} \sigma^a \tau^b \bigg) 2\pi^2 U^2 (-2 \bar{D}_{2222}+\bar{D}_{2233}+U \bar{D}_{3322})
\end{equation}
where $c_{ab}=c_{ba}$ because the s-channel is symmetric under the exchange of $1$ and $2$.

We can obtain the amplitudes in $t$- and $u$-channels by crossing. The total amplitude is the sum of the contributions from the three channels. Denoting the s-channel contribution as $\mathcal{A}_s$,
\begin{equation}
\mathcal{A}_s=\mathcal{A}_{\rm scalar}+\mathcal{A}_{\rm vector}+\mathcal{A}_{\rm graviton}+\mathcal{A}_{\rm contact}\;,
\end{equation}
the ansatz for the crossing-symmetric total amplitude is 
\begin{equation}
\begin{split}
{}&\;\;\;\;\;\;\mathcal{A}_{\rm sugra}(U,V;\sigma,\tau)=\mathcal{A}_{s}+\mathcal{A}_{t}+\mathcal{A}_{u}\\
{}&\;\;\;\;\;\;=\mathcal{A}_{s}(U,V;\sigma,\tau)+\big(\frac{U\tau}{V}\big)^2\mathcal{A}_{s}(V,U;\sigma/\tau,1/\tau)+\big(U\sigma\big)^2\mathcal{A}_{s}(1/U,V/U;1/\sigma,\tau/\sigma)\;.
\end{split}
\end{equation}
Being a sum of $\bar{D}$-functions, $\mathcal{A}_{\rm sugra}$ can be systematically decomposed into $\Phi$, $\ln U$, $\ln V$ and the rational part. For example, the coefficient function of $\Phi$ is of the form
\begin{equation}
R_\phi(z,\bar{z},\alpha,\bar{\alpha})=\frac{T(z,\bar{z},\alpha,\bar{\alpha})}{(z-\bar{z})^6}
\end{equation} 
where the numerator $T(z,\bar{z},\alpha,\bar{\alpha})$ a polynomial of degree 2 in $\alpha$, $\bar{\alpha}$ and of degree 12 
in $z$ and $\bar{z}$. The superconformal Ward identity then requires $T(z,\bar{z};\alpha,1/\bar{z})=0$ and reduces to a set of homogenous linear equations. Their solution is
\begin{equation}
\begin{split}
{}&\lambda_s= \xi,\;\;\;\;\;\; \lambda_v= -\frac{1}{2}\xi,\;\;\;\;\;\;\lambda_g=\frac{3}{16} \xi\;,\\
{}& c_{00}=\frac{3}{32}\xi,\;\;\;c_{01}=-\frac{3}{8}\xi,\;\;\;c_{02}= \frac{3}{32}\xi,\;\;\;c_{11}= -\frac{3}{16}\xi\; ,
\end{split} \label{peq2}
\end{equation}
where $\xi$ is an arbitrary overall constant. 
We then compute  ``twisted''  correlator \begin{equation}
\mathcal{A}_{\rm sugra}(\alpha,1/\bar{z},z,\bar{z})=-\frac{3\pi^2\zeta \left(\alpha ^2 z^2-2 \alpha  z^2+2 \alpha  z-z\right)}{8N^2 (z-1)}\;,
\end{equation}
and compare it to the free field result 
\begin{equation}
\mathcal{G}_{\rm free, conn}(\alpha,1/\bar{z},z,\bar{z})=-\frac{4 \left(\alpha ^2 z^2-2 \alpha  z^2+2 \alpha  z-z\right)}{N^2 (z-1)}\;.
\end{equation}
The functional agreement of the two expressions provides a consistency check, and fixes the value of the last undetermined constant,
\begin{equation}
\xi=\frac{32}{3N^2\pi^2}\; .
\end{equation}
The final answer agrees with the result in the literature \cite{Arutyunov:2000py}.

\section{Conclusion}\label{discussion}

The striking simplicity of the general Mellin formula (\ref{mtildesolution}, \ref{aijk})  is a real surprise.
 Like the Parke-Taylor formula for tree-level MHV gluon scattering amplitudes, it encodes in a succinct expression the sum of an intimidating number of diagrams. 
 The authors of \cite{Alday:2017xua,Aprile:2017bgs,Aprile:2017xsp} have used the information contained in our Mellin formula
  to disentangle the degeneracies and compute the $O(1/N^2)$  anomalous dimensions of double-trace operators of the form ${\cal O}_p \Box^n \partial^\ell {\cal O}_q$.
The solution of this mixing problem turns out to be remarkably simple, giving further evidence for some hidden elegant structure.

 An interesting question is whether our results could be recovered by a more constructive approach, perhaps in the form of a Mellin version of the BCFW recursion relations\footnote{A BCFW-inspired formalism for holographic correlators has been developed in momentum space \cite{Raju:2010by,Raju:2012zr}.}. Such an approach would  lend itself more easily to the generalization to higher $n$-point correlators. 
A preliminary step in this direction  is setting up the Mellin formalism for operators with spin (see \cite{Paulos:2011ie,Goncalves:2014rfa} for the state of the art of this problem).

Our work admits several natural extensions. At  tree level, a direct generalization of the methods developed here has led to structurally similar results for holographic correlators in $AdS_7\times S^4$, which will be described in an upcoming paper \cite{longads7}. The extension  to {$AdS_3 \times S^3 \times {\cal M}_4$} also appears within reach \cite{ads3}.\footnote{See also  \cite{Galliani:2017jlg} for recent progress on $AdS_3$ holographic four-point functions with two ``light'' and two ``heavy'' external operators.}
In all these backgrounds, the KK spectrum obeys
 the truncation conditions, and Mellin amplitudes for tree level correlators are rational functions. This is not the case for a generic holographic background. The most important example  that violates the truncation conditions is the maximally supersymmetric case {$AdS_4 \times S^7$}.
 New techniques will have to be developed to handle such cases \cite{Zhou:2017zaw}. At the loop level, impressive  progress has been made recently by several authors \cite{Aharony:2016dwx, Alday:2017xua,Aprile:2017bgs,Aprile:2017xsp} and it will be interesting to push this program using the insights of our methods.

In conclusion, holographic correlators in ${\cal N}=4$ SYM theory appear to be much simpler and elegant than previously understood. We believe that this warrants 
their renewed exploration, following the spirit of the modern approach to perturbative gauge theory amplitudes.

\acknowledgments
Our work is  supported in part by NSF Grant PHY-1620628.
We are grateful to Gleb Arutyunov, Sergey Frolov, Carlo Meneghelli, Jo\~ao Penenones and Volker Schomerus  for useful conversations.

\appendix

\section{Formulae for exchange Witten diagrams}\label{exch}
We are interested in here the case where the exchange diagrams truncate to a finite number of $D$-functions, as a result of the conspiracy of the spectrum and the space-time dimension. A simple general
method for calculating such exchange diagrams in $AdS_{d+1}$ was found \cite{DHoker:1999aa}.  We collect in this Appendix the relevant formulae needed in the computation of four-point function of identical scalars. The external operators have conformal dimension $\Delta$ and the exchanged operator  conformal dimension $\delta$.

\subsection*{Scalar exchanges}
\begin{equation}
S(x_1,x_2,x_3,x_4)=\sum_{k=\delta/2}^{\Delta-1}a_k |x_{12}|^{-2\Delta+2k}D_{k,k,\Delta,\Delta}\, ,
\end{equation}
where
\begin{equation}
a_{k-1}=\frac{(k-\frac{\delta}{2})(k-\frac{d}{2}+\frac{\delta}{2})}{(k-1)^2}a_k \end{equation}
and
\begin{equation}
a_{\Delta-1}=\frac{1}{4(\Delta-1)^2}\,.
\end{equation}
\subsection*{Vector exchanges}
\begin{equation}
  \begin{split}
V(x_1,x_2,x_3,x_4)= \sum_{k=k_{\rm min}}^{k_{\rm max}}{}& |x_{12}|^{-2\Delta+2k} a_k \Delta\bigg( {x_{24}}^2 D_{k, k+1, \Delta,\Delta+1}+{x_{13}}^2D_{k+1,k,\Delta+1,\Delta}\\
{}& - {x_{23}}^2 D_{k,k+1\Delta+1, \Delta}-{x_{14}}^2D_{k+1,k,\Delta,\Delta+1}\bigg)\, ,
\end{split}
\end{equation}
where
\begin{equation}
   \begin{split}
      k_{\rm min} = {}& \frac{d-2}{4}+\frac{1}{4}\sqrt{(d-2)^2+4(\delta-1)(\delta-d+1)}\;, \\
      k_{\rm max}=   {}& \Delta-1\;, \\
      a_{k-1}= {}& \frac{2k(2k+2-d)-(\delta-1)(\delta-d+1)}{4(k-1)k}a_k\;,\\
      a_{\Delta-1}={}&\frac{1}{2(\Delta-1)}\;.
   \end{split}
\end{equation}
\subsection*{Graviton exchanges}
 \begin{equation}
   \begin{split}
      G(x_1,x_2,x_3,x_4) =   \sum _{k=k_{\rm min}}^{k_{\rm max}}  {}&{x_{12}}^{-2 \Delta+2 k } a_k\bigg((\Delta ^2+\frac{1}{d-1}  \Delta(\Delta -d) ) D_{k,k,\Delta ,\Delta}\\
       {}& -2 \Delta ^2 \big({x_{13}}^2 D_{k+1,k,\Delta +1,\Delta}+{x_{14}}^2 D_{k+1,k,\Delta ,\Delta +1}\big)\\
      {}&+4 \Delta ^2 {x_{13}}^2 {x_{14}}^2 D_{k+2,k,\Delta +1,\Delta +1}\bigg)\, ,
         \end{split}
\end{equation}
where
 \begin{equation}
   \begin{split}
      k_{\rm min} = {}& \frac{d}{2}-1\;, \\
      k_{\rm max}=   {}& \Delta-1\;, \\
      a_{k-1}= {}& \frac{k+1-\frac{d}{2}}{k-1}a_k\;,\\
      a_{\Delta-1}={}&-\frac{\Delta}{2(\Delta-1)}\;.
   \end{split}
\end{equation}

\subsection*{Massive symmetric tensor exchanges}

The Witten diagrams for massive symmetric tensor exchange were worked out in \cite{Arutyunov:2002fh} for the general case\footnote{For easier comparison with the equations of \cite{Arutyunov:2002fh}, in this subsection we change our conventions such that {\bf $d$ is the bulk dimension.  In this subsection and in this subsection only, we are working in $AdS_{d}$ rather than in $AdS_{d+1}$.}.}  of $AdS_d$, and applied to the $AdS_5$ case. We fixed a small error  in \cite{Arutyunov:2002fh}, which only affects the results for $d \neq 5$
and thus leaves the conclusions of  \cite{Arutyunov:2002fh} unaltered. For future reference, we reproduce the general calculation here. Due to the complexity of the explicit form of the general solution, we will not present here the answer as a sum of $D$-functions. Instead we will break down the evaluation into a few parts and give the prescription of how to assemble them into a sum of $D$-functions.

The four-point amplitude $T(x_1,x_2,x_3,x_4)$ due to the exchange of a massive symmetric tensor of dimension $\delta$ is
\begin{equation}
T(x_1,x_2,x_3,x_4)=\int_{AdS}dw A_{\mu\nu}(w,x_1,x_2)T^{\mu\nu}(x_3,x_4,w)
\end{equation}
where
\begin{equation}
T_{\mu\nu}=\partial_\mu K_\Delta(x_3)\partial_\nu K_\Delta(x_4)-\frac{g_{\mu\nu}}{2}\big(\partial_\rho K_\Delta(x_3)\partial_\rho K_\Delta(x_4)+\frac{1}{2}(2\Delta(\Delta-d+1)-f)K_\Delta(x_3)K_\Delta(x_4)\big).
\end{equation}
Here $f=\delta(\delta-d+1)$ is the $m^2$ of the exchanged massive tensor and 
\begin{equation}
K_n(x_i)=\bigg(\frac{w_0}{(w-x)^2}\bigg)^n\;
\end{equation}
is the scalar bulk-to-boundary propagator. By conformal inversion and translation $A_{\mu\nu}$ can be rewritten as
\begin{equation}
A_{\mu\nu}(w,0,x)=\frac{1}{x^{2\Delta}w^4}J_{\mu\lambda}J_{\nu\rho}I_{\lambda\rho}(w'-x')\;,
\end{equation}
with $w'_\mu=\frac{w_\mu}{w^2}$, $x'_\mu=\frac{x_\mu}{x^2}$. The ansatz is 
\begin{equation}
I_{\mu\nu}(w)=g_{\mu\nu}h(t)+P_\mu P_\nu \phi(t)+\triangledown_\mu\triangledown_\nu X(t)+\triangledown_{(\mu}P_{\nu)}Y(t)\,.
\end{equation}
For any scalar function $b(t)$,
\begin{equation}
\begin{split}
\triangledown_\mu\triangledown_\nu b(t)={}&\frac{2w_\mu w_\nu}{w^4}t b'(t)+6(P_\mu-\frac{w_\mu}{w^2})(P_\nu-\frac{w_\nu}{w^2})t b'(t)\\
{}&+4(P_\mu-\frac{w_\mu}{w^2})(P_\nu-\frac{w_\nu}{w^2})t^2 b''(t)-2g_{\mu\nu}t b'(t)\;,
\end{split}
\end{equation}
\begin{equation}
\triangledown_{(\mu}P_{\nu)}b(t)=2(P_\mu P_\nu-g_{\mu\nu})b(t)+2(2P_\mu P_\nu -\frac{P_\mu w_\nu+P_\nu w_\mu}{w^2}) t b'(t)\,.
\end{equation}
 Here, as  standard in the literature, we have denoted
\begin{equation}
P_\mu=\frac{\delta_{0\mu}}{w_0}\;,  \quad
t=\frac{(w_0)^2}{w^2}\;.
\end{equation}
The functions $h(t)$, $\phi(t)$, $X(t)$, $Y(t)$ are subject to the following set of equations,
\begin{equation}
h(t)=-\frac{1}{d-2}\phi(t)+\frac{f}{d-2}X(t)\;,
\end{equation}
\begin{equation}
Y(t)=a+\frac{1}{2f}(4t(t-1)\phi'(t)+(2d-6)\phi(t)+2\Delta t^\Delta)\;,
\end{equation}
\begin{equation}
\begin{split}
X(t)={}&\frac{1}{2(d-1)f(d+f-2)}\bigg(2a(d-2)(2d-3)f+ \big[2(d-3)(d-2)^2+df\big]\phi(t)\\
{}&+(d-2)\bigg[ t^\Delta(-f+2(-\Delta^2+\Delta(d-2)+2\Delta^2 t))+2t(2t+d-3)(4t-3)\phi'(t)\\
{}&+4t^2(t-1)(2t-1)\phi''(t)\bigg]\bigg)\;,
\end{split}
\end{equation}
\begin{equation}
4t^2(t-1)\phi''(t)+(12t^2+(2d-14)t)\phi'(t)+(f+2d-6)\phi(t)+2fa+2\Delta(\Delta+1)t^\Delta=0\; ,
\end{equation}
where $a$ is an integration constant that will cancel out when we substitute the solution into the ansatz for $I_{\mu\nu}$. These equations come from the action of the modified Ricci operator $W_{\mu\nu}{}^{\rho\lambda}$\footnote{There is an error in (E.4) of \cite{Arutyunov:2002fh} that must be fixed in order to generalize to arbitrary $d$. The correct equation is \cite{Naqvi:1999va} \begin{equation}
{W_{\mu\nu}}^{\rho\lambda}\phi_{\rho \lambda}=-\triangledown_\rho\triangledown^\rho\phi_{\mu\nu}+\triangledown_\nu\triangledown^\rho\phi_{\rho\mu}+\triangledown_\nu\triangledown^\rho\phi_{\rho\nu}-\triangledown_\mu\triangledown^\nu\phi_{\rho}^\rho-((2-f)\phi_{\mu\nu}+\frac{2d-4+f}{2-d}g_{\mu\nu}\phi_{\rho}^\rho)
\end{equation}} on $A_{\mu\nu}$ and equating terms of the same structure. We omitted the tedious algebra here.

We start from the last equation and look for a polynomial solution for $\phi(t)$. As we will see shortly, a polynomial solution will lead to a truncation of the exchange diagram to finitely many $D$-functions. We find
\begin{equation}
\begin{split}
{}&\phi(t)=-\frac{2 a \delta  (\delta -d+1)}{(\delta -2) (\delta-d +3)}+\sum_{k=k_{\rm min}}^{ k_{\rm max}}a_k t^k\;,\\
{}&k_{\rm min}=\frac{\delta-2}{2}\;,\\
{}&k_{\rm max}=\Delta-1\;,\\
{}&a_{k-1}=\frac{(k+\frac{3-d+\delta}{2})(1+k-\frac{\delta}{2})}{(k-1)(k+1)}a_k\;,\\
{}&a_{\Delta-1}=-\frac{\Delta}{2\Delta-1}\;.
\end{split}
\end{equation}
For the polynomial solution to exist, $k_{\rm max}-k_{\rm min}=\Delta-\delta/2$ must be an non-negative integer. When $2\Delta=\delta-2$, which is the extremal case, we see the polynomial solution will stop from existing.

After obtaining the polynomial solution for $\phi(t)$, we can easily solve out $h(t)$, $X(t)$, $Y(t)$ from the rest three equations. And it is easy to see $I_{\mu\nu}(t)$ contains only finitely many terms of the following four types
\begin{equation}
g_{\mu\nu}t^n,\;\;\;\; \frac{P_\mu w_\nu}{w^2} t^n,\;\;\;\;  \frac{w_\mu w_\nu}{w^4} t^n,\;\;\;\;  P_\mu P_\nu t^n\;.
\end{equation}
We can get $A_{\mu\nu}$ from $I_{\mu\nu}(w'-x')$ with the following substitutions:
\begin{equation}
\begin{split}
{}& P'_\nu \frac{J_{\mu\nu}}{w^2}\rightarrow R_\mu\equiv P_\mu-2\frac{(w-x_1)_\mu}{(w-x_1)^2},\\
{}&\frac{J_{\mu\nu}}{w^2}\frac{(w'-x')_\mu}{(w'-x')^2}\rightarrow Q_\mu\equiv -\frac{(w-x_1)_\mu}{(w-x_1)^2}+\frac{(w-x_2)_\mu}{(w-x_2)^2}\;,\\
{}& \frac{J_{\mu\rho}}{w^2} g'_{\rho\lambda}\frac{J_{\lambda\nu}}{w^2}\rightarrow g_{\mu\nu}, \;\;\;\;\;\;\; t'^n\rightarrow x_{12}^{2n}K_n(x_1)K_n(x_2)\;.
\end{split}
\end{equation}
The last step is to contract $A_{\mu\nu}$ with $T_{\mu\nu}$. We list below the following handy contraction formulae,
\begin{equation}
\begin{split}
{}&Q_\mu Q_\mu=x_{12}^2 K_1(x_1) K_1(x_2)\;,\\
{}&Q_\mu \partial_\mu K_\Delta(x_i)=\Delta K_{\Delta+1}(x_i)(-x_{1i}^2K_1(x_1)+x_{2i}^2K_1(x_2))\;,\\
{}&R_\mu R_\mu=1\;,\\
{}& R_\mu \partial_\mu K_\Delta(x_i)=\Delta(K_\Delta(x_i)-2x_{1i}^2K_1(x_1)K_{\Delta+1}(x_i))\;,\\
{}&R_\mu Q_\mu = x_{12}^2K_1(x_1)K_1(x_2)\;.
\end{split}
\end{equation}
The above derivation amounts to an algorithm to write the requisite exchange diagrams as a sum of $D$-function. The explicit final result is too long to be reproduced here.

\section{Simplification of contact vertices} \label{contactvertices}
In this Appendix we show that the zero-derivative contact vertex can be absorbed into the two-derivative ones when the dimension of external scalar particle does not equal the spacetime dimension of the boundary theory. 

A zero-derivative contact vertex takes the form of
\begin{equation}
V_{0-\partial}=C_{\alpha_1\alpha_2\alpha_3\alpha_4}\int_{AdS_{d+1}}dX s^{\alpha_1}(X)s^{\alpha_2}(X)s^{\alpha_3}(X)s^{\alpha_4}(X)\, ,
\end{equation}
while a two-derivative contact vertex is
\begin{equation}
V_{2-\partial}=S_{\alpha_1\alpha_2\alpha_3\alpha_4}\int_{AdS_{d+1}}dX \triangledown s^{\alpha_1}(X)\triangledown s^{\alpha_2}(X)s^{\alpha_3}(X)s^{\alpha_4}(X)\;.
\end{equation}
Here $\alpha_i$ collectively denotes the R-symmetry index of ith field $s$.

Following the standard procedure in AdS supergravity calculation, we substitute in the on-shell value of scalar field
\begin{equation}
s^{\alpha}(X)=\int_{\mathbb{R}^d}dPK_{\Delta}(X,P)s^{\alpha}(P)
\end{equation}
so that it is determined by its boundary value $s^\alpha(P)$. Then the two types of contact vertices become
\begin{equation}
\begin{split}
V_{0-\partial}={}&C_{\alpha_1\alpha_2\alpha_3\alpha_4}\int_{AdS_{d+1}}dX \int_{\mathbb{R}^d}\prod dP_i \\
\times {}&K_{\Delta}(X,P_1)K_{\Delta}(X,P_2)K_{\Delta}(X,P_3)K_{\Delta}(X,P_4)s^{\alpha_1}(P_1)s^{\alpha_2}(P_2)s^{\alpha_3}(P_3)s^{\alpha_4}(P_4)\;,
\end{split}
\end{equation}
\begin{equation}
\begin{split}
V_{2-\partial}={}&S_{\alpha_1\alpha_2\alpha_3\alpha_4}\int_{AdS_{d+1}}dX \int_{\mathbb{R}^d}\prod dP_i \\
\times {}& \bigtriangledown K_{\Delta}(X,P_1) \bigtriangledown K_{\Delta}(X,P_2)K_{\Delta}(X,P_3)K_{\Delta}(X,P_4)s^{\alpha_1}(P_1)s^{\alpha_2}(P_2)s^{\alpha_3}(P_3)s^{\alpha_4}(P_4)\;.
\end{split}
\end{equation}
Because the external fields are identical, $C_{\alpha_1\alpha_2\alpha_3\alpha_4}$ is totally symmetric while $S_{\alpha_1\alpha_2\alpha_3\alpha_4}$ is only required to be symmetric under $\alpha_1\leftrightarrow \alpha_2$, $\alpha_3\leftrightarrow \alpha_4$ and $(\alpha_1\alpha_2)\leftrightarrow(\alpha_3\alpha_4)$. This in particular means that the totally symmetric $C_{\alpha_1\alpha_2\alpha_3\alpha_4}$ can be a $S_{\alpha_1\alpha_2\alpha_3\alpha_4}$. Let us see what the consequence is if we take $S_{\alpha_1\alpha_2\alpha_3\alpha_4}=C_{\alpha_1\alpha_2\alpha_3\alpha_4}$,
\begin{equation}
\begin{split}
V_{2-\partial}={}&  C_{\alpha_1\alpha_2\alpha_3\alpha_4}\int dX \int \prod dP_i\triangledown K_1\triangledown K_2 K_3 K_4 s^{\alpha_1}s^{\alpha_2}s^{\alpha_3}s^{\alpha_4}\\
={}&  \frac{1}{6}C_{\alpha_1\alpha_2\alpha_3\alpha_4}\int dX \int \prod dP_i\\
\times{}& (\triangledown K_1\triangledown K_2 K_3 K_4 +\triangledown K_1K_2\triangledown K_3 K_4 +\triangledown K_1K_2 K_3\triangledown K_4\\
 +{}&K_1 \triangledown K_2\triangledown K_3 K_4 +K_1\triangledown K_2K_3\triangledown K_4  +K_1 K_2 \triangledown K_3\triangledown K_4 )\\
\times{}&s^{\alpha_1}s^{\alpha_2}s^{\alpha_3}s^{\alpha_4}
\end{split}
\end{equation}
Here  $K_i\equiv K_\Delta(P_i)$ and we have used the total symmetry of $C_{\alpha_1\alpha_2\alpha_3\alpha_4}$ to symmetrize the expression. If  we  now perform the AdS integral first, each term can be written as a sum of $D$-functions. For example
\begin{equation}
\int_{AdS_{d+1}}dX\triangledown K_1\triangledown K_2K_3K_4=\Delta^2(D_{\Delta,\Delta,\Delta,\Delta}-2{x_{12}}^2D_{\Delta+1,\Delta+1,\Delta,\Delta})\;.
\end{equation}
The two-derivative vertex then becomes
\begin{eqnarray}
V_{2-\partial}& = &{} \frac{\Delta^2}{6}C_{\alpha_1\alpha_2\alpha_3\alpha_4}\int \prod dP_i s^{\alpha_1}s^{\alpha_2}s^{\alpha_3}s^{\alpha_4}\\
&&\times{} (6D_{\Delta,\Delta,\Delta,\Delta}-2{x_{12}}^2D_{\Delta+1,\Delta+1,\Delta,\Delta}-2{x_{13}}^2D_{\Delta+1,\Delta,\Delta+1,\Delta}-2{x_{14}}^2D_{\Delta+1,\Delta,\Delta,\Delta+1} \nonumber \\
{}&&-2{x_{23}}^2D_{\Delta,\Delta+1,\Delta+1,\Delta}-2{x_{24}}^2D_{\Delta,\Delta+1,\Delta,\Delta+1}-2{x_{34}}^2D_{\Delta,\Delta,\Delta+1,\Delta+1})\;. \nonumber
\end{eqnarray}
Using the identity
\begin{equation}
\frac{(2\Delta-d/2)}{\Delta} D_{\Delta,\Delta,\Delta,\Delta}={x^2_{14}}D_{\Delta+1,\Delta,\Delta,\Delta+1}+{x^2_{24}}D_{\Delta,\Delta+1,\Delta,\Delta+1}+{x^2_{34}}D_{\Delta,\Delta,\Delta+1,\Delta+1}\;,
\end{equation}
we simplify the expression to
 \begin{eqnarray}
V_{2-\partial}& =&  \frac{\Delta^2}{6}C_{\alpha_1\alpha_2\alpha_3\alpha_4}\int \prod dP_i s^{\alpha_1}s^{\alpha_2}s^{\alpha_3}s^{\alpha_4} (6D_{\Delta,\Delta,\Delta,\Delta}-\frac{(2\Delta-d/2)}{\Delta}\times4\times D_{\Delta,\Delta,\Delta,\Delta}) \nonumber \\
&=&\frac{\Delta (d-\Delta)}{3}C_{\alpha_1\alpha_2\alpha_3\alpha_4}\int \prod dP_i s^{\alpha_1}s^{\alpha_2}s^{\alpha_3}s^{\alpha_4} D_{\Delta,\Delta,\Delta,\Delta} \nonumber \\
&=&\frac{\Delta (d-\Delta)}{3} V_{0-\partial}\;.
\end{eqnarray}
We have therefore proved that when $\Delta\neq d$, we can absorb the contribution from zero-derivative contact vertices into the two-derivative ones.

\section{The $p=2$ case: a check of the domain-pinching mechanism}\label{peq2data}
We computed the $p=2$ correlator from supergravity, using the position space method of Section \ref{positionspace}. We found 
\begin{equation}
\footnotesize{
\begin{split}\label{peq2raw}
\mathcal{G}_{\rm sugra, conn}{}&(U,V;\sigma,\tau)=-\frac{2 U}{N^2 V}\times \bigg(-\sigma  \tau  \bar{D}_{3,2,1,2} U^2+\tau  \bar{D}_{3,2,1,2} U^2+V \sigma  \bar{D}_{3,2,2,1} U^2\\-{}&V \sigma  \tau  \bar{D}_{3,2,2,1} U^2
+V \sigma ^2 \bar{D}_{3,3,2,2} U^2+V \tau ^2 \bar{D}_{3,3,2,2} U^2+V \bar{D}_{3,3,2,2} U^2-4 V \sigma  \bar{D}_{3,3,2,2} U^2\\-{}&4 V \tau  \bar{D}_{3,3,2,2} U^2
-2 V \sigma  \tau  \bar{D}_{3,3,2,2} U^2+2 \tau ^2 \bar{D}_{2,1,1,2} U-2 \sigma  \tau  \bar{D}_{2,1,1,2} U-2 \tau  \bar{D}_{2,1,1,2} U\\
+{}&2 V \sigma ^2 \bar{D}_{2,1,2,1} U-2 V \sigma  \bar{D}_{2,1,2,1} U
-2 V \sigma  \tau  \bar{D}_{2,1,2,1} U-2 \tau ^2 \bar{D}_{2,1,2,3} U-\sigma  \tau  \bar{D}_{2,1,2,3} U\\
+{}&\tau  \bar{D}_{2,1,2,3} U-2 V \sigma ^2 \bar{D}_{2,1,3,2} U+V \sigma  \bar{D}_{2,1,3,2} U
-V \sigma  \tau  \bar{D}_{2,1,3,2} U+\sigma  \tau  \bar{D}_{2,2,1,3} U\\
-{}&\tau  \bar{D}_{2,2,1,3} U-6 V \sigma ^2 \bar{D}_{2,2,2,2} U-6 V \tau ^2 \bar{D}_{2,2,2,2} U-6 V \bar{D}_{2,2,2,2} U
+20 V \sigma  \bar{D}_{2,2,2,2} U\\+{}&20 V \tau  \bar{D}_{2,2,2,2} U+20 V \sigma  \tau  \bar{D}_{2,2,2,2} U-V^2 \sigma  \bar{D}_{2,2,3,1} U+V^2 \sigma  \tau  \bar{D}_{2,2,3,1} U\\
+{}&V \sigma ^2 \bar{D}_{2,2,3,3} U+V \tau ^2 \bar{D}_{2,2,3,3} U+V \bar{D}_{2,2,3,3} U-4 V \sigma  \bar{D}_{2,2,3,3} U-4 V \tau  \bar{D}_{2,2,3,3} U\\-{}&2 V \sigma  \tau  \bar{D}_{2,2,3,3} U
+V \sigma ^2 \bar{D}_{2,3,2,3} U+V \tau ^2 \bar{D}_{2,3,2,3} U+V \bar{D}_{2,3,2,3} U-4 V \sigma  \bar{D}_{2,3,2,3} U\\
-{}&2 V \tau  \bar{D}_{2,3,2,3} U-4 V \sigma  \tau  \bar{D}_{2,3,2,3} U
+V^2 \bar{D}_{2,3,3,2} U+V^2 \sigma ^2 \bar{D}_{2,3,3,2} U+V^2 \tau ^2 \bar{D}_{2,3,3,2} U\\-{}&2 V^2 \sigma  \bar{D}_{2,3,3,2} U-4 V^2 \tau  \bar{D}_{2,3,3,2} U
-4 V^2 \sigma  \tau  \bar{D}_{2,3,3,2} U-2 V \sigma ^2 \bar{D}_{3,1,2,2} U\\
-{}&2 \tau ^2 \bar{D}_{3,1,2,2} U
-V \sigma  \bar{D}_{3,1,2,2} U
+V \sigma  \tau  \bar{D}_{3,1,2,2} U
+\sigma  \tau  \bar{D}_{3,1,2,2} U\\-{}&\tau  \bar{D}_{3,1,2,2} U+2 V \sigma ^2 \bar{D}_{3,1,3,3} U+2 \tau ^2 \bar{D}_{3,1,3,3} U+V \sigma ^2 \bar{D}_{3,2,2,3} U+V \tau ^2 \bar{D}_{3,2,2,3} U\\
+{}&V \bar{D}_{3,2,2,3} U-2 V \sigma  \bar{D}_{3,2,2,3} U-4 V \tau  \bar{D}_{3,2,2,3} U-4 V \sigma  \tau  \bar{D}_{3,2,2,3} U+V \sigma ^2 \bar{D}_{3,2,3,2} U\\
+{}&V \tau ^2 \bar{D}_{3,2,3,2} U
+V \bar{D}_{3,2,3,2} U-4 V \sigma  \bar{D}_{3,2,3,2} U-2 V \tau  \bar{D}_{3,2,3,2} U-4 V \sigma  \tau  \bar{D}_{3,2,3,2} U\\
+{}&2 V \bar{D}_{1,1,2,2}-2 V \sigma  \bar{D}_{1,1,2,2}
-2 V \tau  \bar{D}_{1,1,2,2}
+V \sigma  \bar{D}_{1,2,2,3}-V \tau  \bar{D}_{1,2,2,3}\\
-{}&V^2 \sigma  \bar{D}_{1,2,3,2}+V^2 \tau  \bar{D}_{1,2,3,2}-2 V \bar{D}_{2,1,2,3}
-V \sigma  \bar{D}_{2,1,2,3}+V \tau  \bar{D}_{2,1,2,3}\\-{}&2 V \bar{D}_{2,1,3,2}+V \sigma  \bar{D}_{2,1,3,2}-V \tau  \bar{D}_{2,1,3,2}+2 V \bar{D}_{3,1,3,3}\bigg)\;.
\end{split} }
\end{equation}
We can get the Mellin transform of $\mathcal{G}_{\rm sugra, conn}(U,V;\sigma,\tau)$ by Mellin-transforming each $\bar{D}$-function in $\mathcal{G}_{\rm sugra, conn}$. Formally, the transformation reads
\begin{equation}
M(s,t;\sigma,\tau)=\int_0^\infty dU dV U^{-s/2-1}V^{-t/2+2-1}\mathcal{G}_{\rm sugra, conn}(U,V;\sigma,\tau)\;,
\end{equation}
 but notice each $\bar{D}$-function may come with a different fundamental domain of $s$ and $t$ in which the integrals converge. These fundamental domains are defined by the positivity condition of the Gamma function arguments.  Although no ambiguity arises when analytically continue the Mellin transformation outside this domain due to the absence of branch cuts, it is imperative to have the knowledge of the fundamental domain as the contour needs to be placed inside the fundamental domain in order to reproduce precisely the $\bar{D}$-function via the inverse Mellin-transformation.  To keep track of this information, in the following expression we simply keep the Gamma functions from each $\bar{D}$-function, and the domain information can be extracted by requiring that the arguments of the Gamma functions have positive real part. With this proviso, the reduced Mellin amplitude reads
 \begin{equation}
 \footnotesize{
\begin{split}\label{peq2mellin}
M(s,t;\sigma,\tau)={}&\frac{4}{N^2}\Gamma[2-\frac{s}{2}]\Gamma[2-\frac{t}{2}] \Gamma[\frac{1}{2} (s+t-4)]\\
\times{}&
\bigg\{\Gamma \left(2-\frac{s}{2}\right)\times\big[\Gamma \left(3-\frac{t}{2}\right)\times(\sigma  (\sigma -\tau +1) \Gamma \left(\frac{1}{2} (s+t-6)\right)\\
-{}&\left(\sigma ^2-2 \sigma  (2 \tau +1)+\tau ^2-4 \tau +1\right) \Gamma \left(\frac{1}{2} (s+t-4)\right))\\
+{}&\Gamma \left(2-\frac{t}{2}\right)\times (\left(-3 \sigma ^2+10 \sigma  (\tau +1)-3 \tau ^2+10 \tau -3\right) \Gamma \left(\frac{1}{2} (s+t-4)\right)\\
+{}& \left(\sigma ^2-4 \sigma  (\tau +1)+(\tau -1)^2\right) \Gamma \left(\frac{1}{2} (s+t-2)\right)\left.+\sigma  (\sigma -\tau -1) \Gamma \left(\frac{1}{2} (s+t-6)\right)\right)\\
+{}&\tau  \Gamma \left(1-\frac{t}{2}\right)\times((\sigma -\tau +1) \Gamma \left(\frac{1}{2} (s+t-4)\right)+(-\sigma +\tau +1) \Gamma \left(\frac{1}{2} (s+t-2)\right))\big]\\
+{}&\Gamma \left(3-\frac{s}{2}\right)\times \big[\Gamma \left(2-\frac{t}{2}\right)\times (\sigma  (\sigma +\tau -1) \Gamma \left(\frac{1}{2} (s+t-6)\right)\\
-{}&\left(\sigma ^2-2 \sigma  (\tau +2)+\tau ^2-4 \tau +1\right) \Gamma \left(\frac{1}{2} (s+t-4)\right))-\sigma ^2 \Gamma \left(3-\frac{t}{2}\right) \Gamma \left(\frac{1}{2} (s+t-6)\right)\\
+{}&\tau  \Gamma \left(1-\frac{t}{2}\right)\times((\sigma +\tau -1) \Gamma \left(\frac{1}{2} (s+t-4)\right)-\tau  \Gamma \left(\frac{1}{2} (s+t-2)\right))\big]\\
+{}&\Gamma \left(1-\frac{s}{2}\right)\times \big[\Gamma \left(3-\frac{t}{2}\right)\times((\sigma -\tau +1) \Gamma \left(\frac{1}{2} (s+t-4)\right)-\Gamma \left(\frac{1}{2} (s+t-2)\right))\\
+{}&\Gamma \left(2-\frac{t}{2}\right)\times((\sigma +\tau -1) \Gamma \left(\frac{1}{2} (s+t-4)\right)+(-\sigma +\tau +1) \Gamma \left(\frac{1}{2} (s+t-2)\right))\big]\bigg\}. 
\end{split}. }
\end{equation}
We can use this example to
illustrate
how the free field correlator $\mathcal{G}_{\rm free, conn}$ arises when one takes the inverse Mellin transform by the ``contour pinching mechanism'' described in Section \ref{subtlefree}. We will compare the expression that arises directly from the explicit supergravity calculation and the expression in the split form (\ref{splitform}), both written as inverse Mellin transformations. 
Each summand  in (\ref{peq2mellin}) contains a common Gamma function factor
$\Gamma[2-\frac{s}{2}]\Gamma[2-\frac{t}{2}]\Gamma[\frac{s+t-4}{2}]$
which sets common bounds for the boundaries of all the fundamental domains -- the real parts of $s$ and $t$ must be inside the big black-framed triangle in Figure. \ref{sugra}. A closer look shows that some the summands in (\ref{peq2mellin}) have smaller domains. Imposing positivity of the rest of the Gamma functions in each term shows that there are four types of domains: the red $\{(2,4),(4,2),(4,4)\}$, green $\{(2,2),(4,0),(4,2)\}$ and orange $\{(0,4),(2,2),(2,4)\}$ triangles of size two (where by size we mean the length of its projection onto the $\Re(s)$ axis or $\Re(t)$ axis) and the bigger grey triangle $\{(0,4),(4,0),(4,4)\}$ of size four. 
\begin{figure}[htbp]
\begin{center}
\includegraphics[scale=0.42]{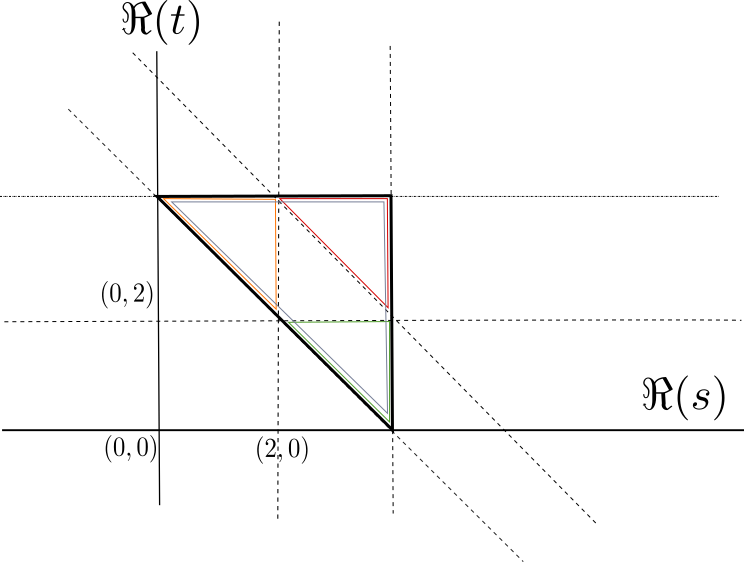}
\caption{The fundamental domains for the ``unmassaged'' supergravity result.}
\label{sugra}
\end{center}
\end{figure}

Now we take a look at the other form of the result where it has been split into two parts,
\begin{equation}\label{ward2}
\mathcal{G}_{\rm conn}=\mathcal{G}_{\rm free, conn}+R\mathcal{H}\;.
\end{equation}
The factor $R$ was introduced before and we repeat here for reader's convenience,
 \begin{equation}
 R=\tau{1}+(1-\sigma-\tau){V}+(-\tau-\sigma\tau+\tau^2){U}+(\sigma^2-\sigma-\sigma\tau){UV}+ \sigma{V^2}+\sigma\tau{U^2}\;.
 \end{equation}
 The first term $\mathcal{G}_{\rm free, conn}$ is the connected free field four-point function, which can be computed by Wick contractions,
 \begin{equation}
\mathcal{G}_{\rm free, conn}=\frac{4}{N^2}\frac{U}{V}\,(\tau+V\sigma+U\sigma\tau)\,.
\end{equation} 
 The function $\mathcal{H}$ was obtained in \cite{Dolan:2004iy},
 \begin{equation}
 \mathcal{H}=-\frac{4}{N^2}U^2\bar{D}_{2422}\;.
 \end{equation}
 We write $\mathcal{H}$ as an inverse Mellin transform,
  \begin{equation} \label{H2}
 \mathcal{H}=-\frac{4}{N^2}\times \frac{1}{4} \int_{\mathcal{C}}dsdt U^{s/2}V^{t/2-2}\Gamma[2-\frac{s}{2}]\Gamma[1-\frac{s}{2}]\Gamma[2-\frac{t}{2}]\Gamma[1-\frac{t}{2}]\Gamma[\frac{s+t}{2}-1]\Gamma[\frac{s+t}{2}]\, ,
 \end{equation}
 where $\mathcal{C}$ is associated with a point inside the fundamental domain
 \begin{equation}
 (s_0,t_0)\in \mathcal{D}=\{(s_0,t_0)|\Re(s)<2,\Re(t)<2,\Re(s)+\Re(t)>2\}\, ,
 \end{equation}
represented by the yellow size-two triangle in Figure \ref{split}. When multiplied by $R$, this domain will lead to six different domains generated by the six different shifts in $R$, namely, $1$, $U$, $V$, $UV$, $U^2$, $V^2$. They are the six colored triangles\footnote{In addition to the previously defined red, green, orange triangles, there are also size-two pink $\{(0,6),(2,4),(2,6)\}$, yellow $\{(0,2),(2,0),(2,2)\}$ and blue $\{(4,2),(6,0),(6,2)\}$ triangles.} in Figure \ref{split}. 
\begin{figure}[htbp]
\begin{center}
\includegraphics[scale=0.42]{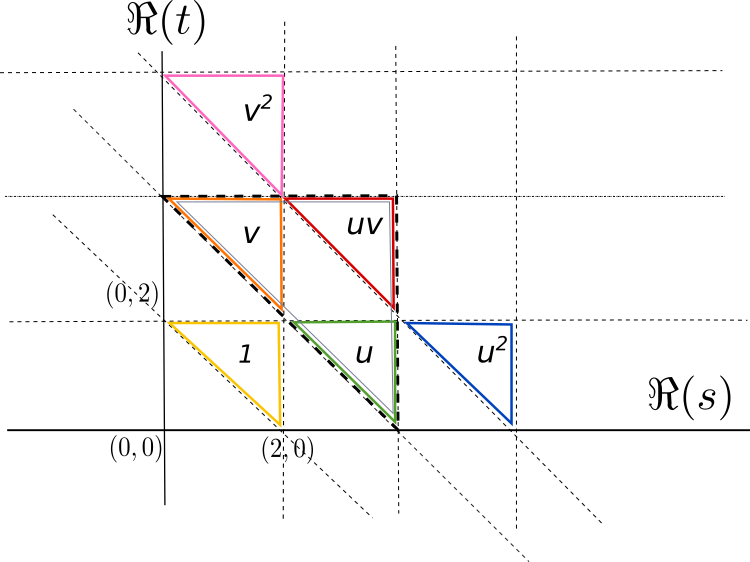}
\caption{The fundamental domains for the Mellin transform of $R {\cal H}$.}
\label{split}
\end{center}
\end{figure}

Having stated the results for the two sides of (\ref{ward2}) (the ``unmassaged'' lhs, whose Mellin transform is given by (\ref{peq2mellin}), and the ``massaged'' rhs, where the Mellin transform of ${\cal H}$ is given by (\ref{H2})), we will now try to match them. Compared to the supergravity answer, there are three more size-two triangles on the right side. They are in the colors of yellow, pink and blue, and are respectively due to the shifts caused by the terms $\tau$, $V^2\sigma$ and $U^2\sigma\tau$. Using the regularization procedure we introduced in Section \ref{subtlefree}, they can be eliminated by combining with terms from the other triangles that we want to keep.  Let us now describe in detail how this can be done.

We first pay attention to the terms multiplied by $\tau$ in $R$. We will combine it with terms multiplied by $-\tau V$ and $-\tau U$ from $R$. Naively the three shifted domains will not overlap. Under the regularization, these three domains grow a small overlap and allows us to add the integrands once the contours have all been moved there
  \begin{equation}
  \begin{split}
 \tau(1-V-U)\mathcal{H}={}&-\tau\frac{4}{N^2}\times \frac{1}{4} \int_{\mathcal{C}_{(2,2),\epsilon}}dsdt U^{s/2}V^{t/2-2}\times\left[\frac{st-4}{2}+\frac{s+t-3}{2}\epsilon+\frac{\epsilon^2}{4}\right]\\
 \times{}&\Gamma[2-\frac{s}{2}]\Gamma[1-\frac{s}{2}]\Gamma[2-\frac{t}{2}]\Gamma[1-\frac{t}{2}]\Gamma[\frac{s+t+\epsilon}{2}-2]\Gamma[\frac{s+t+\epsilon}{2}-1]\;.
 \end{split}
 \end{equation}
 Here $\mathcal{C}_{(2,2),\epsilon}$ denotes that we put the contour inside the size-$\epsilon$ triangle (not shown in the picture) at $(2,2)$ shared by these three triangles. We now analyze the terms in this integral.
 
 The $\epsilon^1$ term is the same integral as the one that we have encountered in the proof of the identity. It is evaluated to give
\begin{equation}
-\tau \frac{4}{N^2} UV^{-1}\;.
\end{equation} 
The $\epsilon^2$ term is easily seen to be zero. For the $\epsilon^0$ term, we rewrite it as
\begin{equation}
\frac{st-4}{2}=\frac{1}{2}(s-2)(t-2)+(s-2)+(t-2)\;.
\end{equation}
The point of this rewriting is that these zeros of $(s-2)$ and $(t-2)$ will cancel the same poles in the Gamma functions, such that one is allowed to ``open up the boundaries'' to enter a bigger domain. For example, consider the above term $(s-2)$. Its contour was originally placed at the size-$\epsilon$ domain at $(2,2)$ but now it can be moved into size-two green triangle because $(s-2)$ cancels the simple pole at $s=2$ from $\Gamma[1-\frac{s}{2}]$. Similarly the domain of the $\frac{1}{2}(s-2)(t-2)$ term can be extended to the size-four grey triangle and the $(t-2)$ term extended to the size-two orange triangle with the same reason. 

On the other hand, for the $\sigma V^2$ triangle, we will combine it into $\sigma(-V+V^2-UV)\mathcal{H}$. The goal of splitting the $\mathcal{O}(\epsilon^0)$ term here is to open up the boundaries into the orange, red and grey triangle and from the $\epsilon^1$ term one will get a monomial $-\sigma \frac{4}{N^2} U$. For the $\sigma\tau U^2$ triangle, one combines into $\sigma\tau(-U+U^2-UV)\mathcal{H}$. The $\epsilon$ term from the rewriting generates a monomial $-\sigma\tau \frac{4}{N^2} U^2V^{-1}$. Already, collecting these monomials, one get $-\mathcal{G}_{\rm free, conn}$, canceling precisely the free field part in the split formula.

To carry out the rest of the check, it is simplest to check by gathering terms with the same R-symmetry monomial. In the $p=2$ case one has six R-symmetry monomials and one can divide them into two groups: first check $1$, $\sigma^2$ and $\tau^2$, then $\tau$, $\sigma$, $\sigma\tau$. In fact, checking just one term in each class is enough, because both the supergravity result and the result written in a split form have crossing symmetry. These two classes of monomials form two orbits under the $S_3$ crossing symmetry group. One will need also to use the above trick of using zeros to open up boundaries (or the opposite, use poles to close). But the here one will find it is only necessary to shrink or expand between the size-four grey triangle and a size-two orange, red, green triangles. Because the manipulation is from a finite-size domain to another finite-size domain, the contour will always have room to escape and one will never get additional terms from the ``domain-pinching'' mechanism. We performed this explicit check and found a perfect match.

\section{The $p=3,4,5$ results from the position space method}\label{p345}
 \subsection*{p=3}
 The $p=3$ computation is very similar to the $p=2$ case. In total there are 6 exchange diagrams in the s-channel. They include the full $k=2$ multiplet $s_2$, $A_2$, $\varphi_2$ and 3 fields $s_4$, $A_4$, $\varphi_4$ from the $k=4$ multiplet. We have the following ansatz for the s-channel exchange amplitude
 \begin{equation}
 \mathcal{A}_{\rm s-channel}=\lambda_{s_2}\mathcal{A}_{s_2}+\lambda_{A_2}\mathcal{A}_{A_2}+\lambda_{\varphi_2}\mathcal{A}_{\varphi_2}+\lambda_{s_4}\mathcal{A}_{s_4}+\lambda_{A_4}\mathcal{A}_{A_4}+\lambda_{\varphi_4}\mathcal{A}_{\varphi_4}+\mathcal{A}_{\rm contact}
 \end{equation}
where 
\begin{equation}
\mathcal{A}_{\rm contact}=\bigg(\sum_{0\leq a+b\leq 3} c_{ab} \sigma^a \tau^b \bigg) \frac{3\pi^2 U^2}{8} (9\bar{D}_{3333}-8U\bar{D}_{4433})\;.
\end{equation}
Note here $\mathcal{A}_{\rm field}$ contains in it R-symmetry polynomial $Y_{nm}$ and the exchange formulae as sum of $\bar{D}$-functions can be found in Appendix \ref{exch}.

Imposing the superconformal Ward identity, we get the following solution  
\begin{equation}
\begin{split}
{}&\lambda_{s_2}=\frac{384}{\pi ^2 N^2},\;\;\;\;\;\;\lambda_{A_2}=-\frac{72}{\pi ^2 N^2},\;\;\;\;\;\;\lambda_{\varphi_2}=\frac{8}{\pi ^2 N^2},\\
{}&\lambda_{s_4}=\frac{1152}{5\pi ^2 N^2},\;\;\;\;\;\;\lambda_{A_4}=-\frac{144}{5\pi ^2 N^2},\;\;\;\;\;\;\lambda_{\varphi_4}=\frac{2}{\pi ^2 N^2},\\
{}&c_{00}=\frac{8}{\pi ^2 N^2},\;\;\;\;\;\;c_{01}=-\frac{26}{\pi ^2 N^2},\;\;\;\;\;\;c_{02}=-\frac{26}{\pi ^2 N^2},\\{}&c_{03}=\frac{8}{\pi ^2 N^2}\;\;\;\;\;\;c_{11}=-\frac{192}{\pi ^2 N^2},\;\;\;\;\;c_{12}=-\frac{20}{\pi ^2 N^2}\;.
\end{split}
\end{equation}

\subsection*{p=4}
$p=4$ is special in that we cannot use two-derivative contact vertices to absorb the contribution of zero-derivative ones by redefinition the parameters. So in this case we must include both types of contributions in the ansatz. The s-channel ansatz is given by

\begin{equation}
\begin{split}
 \mathcal{A}_{\rm s-channel}={}&\lambda_{s_2}\mathcal{A}_{s_2}+\lambda_{A_2}\mathcal{A}_{A_2}+\lambda_{\varphi_2}\mathcal{A}_{\varphi_2}\\
 +{}&\lambda_{s_4}\mathcal{A}_{s_4}+\lambda_{A_4}\mathcal{A}_{A_4}+\lambda_{\varphi_4}\mathcal{A}_{\varphi_4}+\lambda_{C_4}\mathcal{A}_{C_4}+\lambda_{\phi_4}\mathcal{A}_{\phi_4}\\
+{}&\lambda_{s_6}\mathcal{A}_{s_6}+\lambda_{A_6}\mathcal{A}_{A_6}+\lambda_{\varphi_6}\mathcal{A}_{\varphi_6}\\
+{}&\mathcal{A}_{\rm contact}
 \end{split}
 \end{equation}
where 
\begin{equation}
\mathcal{A}_{\rm contact}=\bigg(\sum_{0\leq a+b\leq 4} c_{ab} \sigma^a \tau^b \bigg) \frac{5\pi^2 U^2}{216} (4\bar{D}_{4444}-3U\bar{D}_{5544})+\bigg(\sum_{0\leq a+b\leq 4} c'_{ab} \sigma^a \tau^b \bigg) \frac{5\pi^2 U^2}{108} \bar{D}_{4444}.
\end{equation}
The superconformal Ward identity is expected  not to fix all the coefficients because we know certain crossing symmetric choice of the two-derivative contact coupling will give a zero contribution. As it turned out, all these unsolved coefficients are multiplied by a common factor 
\begin{equation}
-8\bar{D}_{4444}+\bar{D}_{4455}+\bar{D}_{4545}+V\bar{D}_{4554}+\bar{D}_{5445}+\bar{D}_{5454}+U\bar{D}_{5544}
\end{equation}
which is identically zero by $\bar{D}$-identities. These coefficients can be set to zero at our convenience.

The solution is 
\begin{equation}
\begin{split}
{}&\lambda_{s_2}= \frac{3456}{\pi ^2 N^2},\;\;\;\;\;\lambda_{A_2}= -\frac{384}{\pi ^2 N^2},\;\;\;\;\;\;\lambda _{\varphi_2}= \frac{18}{\pi ^2 N^2},\\
{}&\lambda_{s_4}= \frac{18432}{5 \pi ^2 N^2},\;\;\;\lambda_{A_4}= -\frac{1728}{5 \pi ^2 N^2},\;\;\;\lambda_{\varphi_4}= \frac{288}{25 \pi ^2 N^2},\;\;\;\lambda_{C_4}= -\frac{192}{25 \pi ^2 N^2},\;\;\;\lambda_{\phi_4}= \frac{576}{5 \pi ^2 N^2},\;\;\;\\
{}&\lambda_{s_6}= \frac{15552}{35 \pi ^2 N^2},\;\;\lambda_{A_6}= -\frac{5184}{175 \pi ^2 N^2},\;\;\lambda_{\varphi_6}= \frac{18}{25 \pi ^2 N^2},\\
{}&c_{12}= \frac{1728}{5 \pi ^2 N^2},c_{13}= \frac{576}{5 \pi ^2 N^2},c_{22}= \frac{2304}{5 \pi ^2 N^2},\\
{}&c'_{04}= \frac{216}{5 \pi ^2 N^2},c'_{12}= -\frac{16848}{5 \pi ^2 N^2},{c'}_{13}\to \frac{576}{5 \pi ^2 N^2},{c'}_{22}= -\frac{8928}{5 \pi ^2 N^2}
\end{split}
\end{equation}
with all the other unlisted coefficients being zero.

\subsection*{p=5}
The computation of $p=5$ is similar to that of $p=2$ and $p=3$. The ansatz is given by
\begin{equation}
\begin{split}
 \mathcal{A}_{\rm s-channel}={}&\lambda_{s_2}\mathcal{A}_{s_2}+\lambda_{A_2}\mathcal{A}_{A_2}+\lambda_{\varphi_2}\mathcal{A}_{\varphi_2}\\
 +{}&\lambda_{s_4}\mathcal{A}_{s_4}+\lambda_{A_4}\mathcal{A}_{A_4}+\lambda_{\varphi_4}\mathcal{A}_{\varphi_4}+\lambda_{C_4}\mathcal{A}_{C_4}+\lambda_{\phi_4}\mathcal{A}_{\phi_4}+\lambda_{t_4}\mathcal{A}_{t_4}\\
+{}&\lambda_{s_6}\mathcal{A}_{s_6}+\lambda_{A_6}\mathcal{A}_{A_6}+\lambda_{\varphi_6}\mathcal{A}_{\varphi_6}+\lambda_{C_6}\mathcal{A}_{C_6}+\lambda_{\phi_6}\mathcal{A}_{\phi_6}\\
+{}&\lambda_{s_8}\mathcal{A}_{s_8}+\lambda_{A_8}\mathcal{A}_{A_8}+\lambda_{\varphi_8}\mathcal{A}_{\varphi_8}\\
+{}&\mathcal{A}_{\rm contact}\;,
 \end{split}
 \end{equation}
where
\begin{equation}
\mathcal{A}_{\rm contact}=\bigg(\sum_{0\leq a+b\leq 5} c_{ab} \sigma^a \tau^b \bigg) \frac{7\pi^2 U^2}{11520} (25\bar{D}_{5555}-16U\bar{D}_{6655})\;.\end{equation}
The solution to this case is
\begin{equation}
\begin{split}
{}&\lambda_{s_2}= \frac{51200}{3\pi ^2 N^2},\;\;\;\;\;\lambda_{A_2}= -\frac{4000}{3 \pi ^2 N^2},\;\;\;\;\;\;\lambda _{\varphi_2}= \frac{32}{\pi ^2 N^2},\\
{}&\lambda_{s_4}= \frac{23040}{\pi ^2 N^2},\;\;\;\lambda_{A_4}= -\frac{1728}{\pi ^2 N^2},\;\;\;\lambda_{\varphi_4}= \frac{32}{ \pi ^2 N^2},\;\;\;\lambda_{C_4}= -\frac{320}{3 \pi ^2 N^2},\;\;\;\lambda_{\phi_4}= \frac{9216}{5 \pi ^2 N^2},\;\;\;\\
 {}&\lambda_{t_4}= \frac{512}{15 \pi ^2 N^2},\\
{}&\lambda_{s_6}= \frac{248832}{35 \pi ^2 N^2},\;\;\lambda_{A_6}= -\frac{2880}{7 \pi ^2 N^2},\;\;\lambda_{\varphi_6}= \frac{288}{49 \pi ^2 N^2},\;\;\;\lambda_{C_6}= -\frac{288}{49 \pi ^2 N^2},\;\;\;\lambda_{\phi_6}= \frac{4608}{35 \pi ^2 N^2},\\
{}&\lambda_{s_8}= \frac{20480}{63 \pi ^2 N^2},\;\;\lambda_{A_8}= -\frac{6400}{441 \pi ^2 N^2},\;\;\lambda_{\varphi_8}= \frac{8}{49 \pi ^2 N^2},\\
{}& c_{00}= -\frac{1600}{7 \pi ^2 N^2},\;\;\;c_{01}= -\frac{15800}{7 \pi ^2 N^2},\;\;\;c_{02}= \frac{67400}{7 \pi ^2 N^2},\;\;\;c_{03}= \frac{67400}{7 \pi ^2 N^2},\;\;\;\\
{}& c_{04}= -\frac{15800}{7 \pi ^2 N^2},\;\;\;c_{05}= -\frac{1600}{7 \pi ^2 N^2},\;\;\;c_{11}= \frac{176000}{7 \pi ^2 N^2},\;\;\; c_{12}= \frac{941400}{7 \pi ^2 N^2},\;\;\;\\
{}&c_{13}= \frac{184000}{7 \pi ^2 N^2},\;\;\;c_{14}= -\frac{14800}{7 \pi ^2 N^2},\;\;\;c_{22}= \frac{968400}{7 \pi ^2 N^2},\;\;\;c_{00}= \frac{76400}{7 \pi ^2 N^2}.\;\;\;
\end{split}
\end{equation}

\newpage

\bibliography{reflong} 

\providecommand{\href}[2]{#2}\begingroup\raggedright\begin{thebibliography}{10}

\bibitem{Rastelli:2016nze}
L.~Rastelli and X.~Zhou, ``{Mellin amplitudes for $AdS_5\times S^5$},''
  \href{http://dx.doi.org/10.1103/PhysRevLett.118.091602}{{\em Phys. Rev.
  Lett.} {\bfseries 118} no.~9, (2017) 091602},
\href{http://arxiv.org/abs/1608.06624}{{\ttfamily arXiv:1608.06624 [hep-th]}}.

\bibitem{Freedman:1998tz}
D.~Z. Freedman, S.~D. Mathur, A.~Matusis, and L.~Rastelli, ``{Correlation
  functions in the CFT(d) / AdS(d+1) correspondence},''
  \href{http://dx.doi.org/10.1016/S0550-3213(99)00053-X}{{\em Nucl. Phys.}
  {\bfseries B546} (1999) 96--118},
\href{http://arxiv.org/abs/hep-th/9804058}{{\ttfamily arXiv:hep-th/9804058
  [hep-th]}}.

\bibitem{Lee:1998bxa}
S.~Lee, S.~Minwalla, M.~Rangamani, and N.~Seiberg, ``{Three point functions of
  chiral operators in D = 4, N=4 SYM at large N},'' {\em Adv. Theor. Math.
  Phys.} {\bfseries 2} (1998) 697--718,
\href{http://arxiv.org/abs/hep-th/9806074}{{\ttfamily arXiv:hep-th/9806074
  [hep-th]}}.

\bibitem{Intriligator:1998ig}
K.~A. Intriligator, ``{Bonus symmetries of N=4 superYang-Mills correlation
  functions via AdS duality},''
  \href{http://dx.doi.org/10.1016/S0550-3213(99)00242-4}{{\em Nucl. Phys.}
  {\bfseries B551} (1999) 575--600},
\href{http://arxiv.org/abs/hep-th/9811047}{{\ttfamily arXiv:hep-th/9811047
  [hep-th]}}.

\bibitem{Intriligator:1999ff}
K.~A. Intriligator and W.~Skiba, ``{Bonus symmetry and the operator product
  expansion of N=4 SuperYang-Mills},''
  \href{http://dx.doi.org/10.1016/S0550-3213(99)00430-7}{{\em Nucl. Phys.}
  {\bfseries B559} (1999) 165--183},
\href{http://arxiv.org/abs/hep-th/9905020}{{\ttfamily arXiv:hep-th/9905020
  [hep-th]}}.

\bibitem{Eden:1999gh}
B.~Eden, P.~S. Howe, and P.~C. West, ``{Nilpotent invariants in N=4 SYM},''
  \href{http://dx.doi.org/10.1016/S0370-2693(99)00705-4}{{\em Phys. Lett.}
  {\bfseries B463} (1999) 19--26},
\href{http://arxiv.org/abs/hep-th/9905085}{{\ttfamily arXiv:hep-th/9905085
  [hep-th]}}.

\bibitem{Petkou:1999fv}
A.~Petkou and K.~Skenderis, ``{A Nonrenormalization theorem for conformal
  anomalies},'' \href{http://dx.doi.org/10.1016/S0550-3213(99)00514-3}{{\em
  Nucl. Phys.} {\bfseries B561} (1999) 100--116},
\href{http://arxiv.org/abs/hep-th/9906030}{{\ttfamily arXiv:hep-th/9906030
  [hep-th]}}.

\bibitem{Howe:1999hz}
P.~S. Howe, C.~Schubert, E.~Sokatchev, and P.~C. West, ``{Explicit construction
  of nilpotent covariants in N=4 SYM},''
  \href{http://dx.doi.org/10.1016/S0550-3213(99)00768-3}{{\em Nucl. Phys.}
  {\bfseries B571} (2000) 71--90},
\href{http://arxiv.org/abs/hep-th/9910011}{{\ttfamily arXiv:hep-th/9910011
  [hep-th]}}.

\bibitem{Heslop:2001gp}
P.~J. Heslop and P.~S. Howe, ``{OPEs and three-point correlators of protected
  operators in N=4 SYM},''
  \href{http://dx.doi.org/10.1016/S0550-3213(02)00023-8}{{\em Nucl. Phys.}
  {\bfseries B626} (2002) 265--286},
\href{http://arxiv.org/abs/hep-th/0107212}{{\ttfamily arXiv:hep-th/0107212
  [hep-th]}}.

\bibitem{Baggio:2012rr}
M.~Baggio, J.~de~Boer, and K.~Papadodimas, ``{A non-renormalization theorem for
  chiral primary 3-point functions},''
  \href{http://dx.doi.org/10.1007/JHEP07(2012)137}{{\em JHEP} {\bfseries 07}
  (2012) 137},
\href{http://arxiv.org/abs/1203.1036}{{\ttfamily arXiv:1203.1036 [hep-th]}}.

\bibitem{DHoker:1999jke}
E.~D'Hoker, D.~Z. Freedman, S.~D. Mathur, A.~Matusis, and L.~Rastelli,
  ``{Extremal correlators in the AdS / CFT correspondence},''
\href{http://arxiv.org/abs/hep-th/9908160}{{\ttfamily arXiv:hep-th/9908160
  [hep-th]}}.

\bibitem{Bianchi:1999ie}
M.~Bianchi and S.~Kovacs, ``{Nonrenormalization of extremal correlators in N=4
  SYM theory},'' \href{http://dx.doi.org/10.1016/S0370-2693(99)01211-3}{{\em
  Phys. Lett.} {\bfseries B468} (1999) 102--110},
\href{http://arxiv.org/abs/hep-th/9910016}{{\ttfamily arXiv:hep-th/9910016
  [hep-th]}}.

\bibitem{Eden:1999kw}
B.~Eden, P.~S. Howe, C.~Schubert, E.~Sokatchev, and P.~C. West, ``{Extremal
  correlators in four-dimensional SCFT},''
  \href{http://dx.doi.org/10.1016/S0370-2693(99)01442-2}{{\em Phys. Lett.}
  {\bfseries B472} (2000) 323--331},
\href{http://arxiv.org/abs/hep-th/9910150}{{\ttfamily arXiv:hep-th/9910150
  [hep-th]}}.

\bibitem{Erdmenger:1999pz}
J.~Erdmenger and M.~Perez-Victoria, ``{Nonrenormalization of next-to-extremal
  correlators in N=4 SYM and the AdS / CFT correspondence},''
  \href{http://dx.doi.org/10.1103/PhysRevD.62.045008}{{\em Phys. Rev.}
  {\bfseries D62} (2000) 045008},
\href{http://arxiv.org/abs/hep-th/9912250}{{\ttfamily arXiv:hep-th/9912250
  [hep-th]}}.

\bibitem{Eden:2000gg}
B.~U. Eden, P.~S. Howe, E.~Sokatchev, and P.~C. West, ``{Extremal and
  next-to-extremal n point correlators in four-dimensional SCFT},''
  \href{http://dx.doi.org/10.1016/S0370-2693(00)01181-3}{{\em Phys. Lett.}
  {\bfseries B494} (2000) 141--147},
\href{http://arxiv.org/abs/hep-th/0004102}{{\ttfamily arXiv:hep-th/0004102
  [hep-th]}}.

\bibitem{Fleury:2016ykk}
T.~Fleury and S.~Komatsu, ``{Hexagonalization of Correlation Functions},''
  \href{http://dx.doi.org/10.1007/JHEP01(2017)130}{{\em JHEP} {\bfseries 01}
  (2017) 130},
\href{http://arxiv.org/abs/1611.05577}{{\ttfamily arXiv:1611.05577 [hep-th]}}.

\bibitem{Eden:2016xvg}
B.~Eden and A.~Sfondrini, ``{Tessellating cushions: four-point functions in
  $\mathcal{N} $ = 4 SYM},''
  \href{http://dx.doi.org/10.1007/JHEP10(2017)098}{{\em JHEP} {\bfseries 10}
  (2017) 098},
\href{http://arxiv.org/abs/1611.05436}{{\ttfamily arXiv:1611.05436 [hep-th]}}.

\bibitem{Basso:2017khq}
B.~Basso, F.~Coronado, S.~Komatsu, H.~T. Lam, P.~Vieira, and D.-l. Zhong,
  ``{Asymptotic Four Point Functions},''
\href{http://arxiv.org/abs/1701.04462}{{\ttfamily arXiv:1701.04462 [hep-th]}}.

\bibitem{Maldacena:1997re}
J.~M. Maldacena, ``{The Large N limit of superconformal field theories and
  supergravity},'' \href{http://dx.doi.org/10.1023/A:1026654312961}{{\em Int.
  J. Theor. Phys.} {\bfseries 38} (1999) 1113--1133},
  \href{http://arxiv.org/abs/hep-th/9711200}{{\ttfamily arXiv:hep-th/9711200
  [hep-th]}}.
[Adv. Theor. Math. Phys.2,231(1998)].

\bibitem{Witten:1998qj}
E.~Witten, ``{Anti-de Sitter space and holography},'' {\em Adv. Theor. Math.
  Phys.} {\bfseries 2} (1998) 253--291,
\href{http://arxiv.org/abs/hep-th/9802150}{{\ttfamily arXiv:hep-th/9802150
  [hep-th]}}.

\bibitem{Gubser:1998bc}
S.~S. Gubser, I.~R. Klebanov, and A.~M. Polyakov, ``{Gauge theory correlators
  from noncritical string theory},''
  \href{http://dx.doi.org/10.1016/S0370-2693(98)00377-3}{{\em Phys. Lett.}
  {\bfseries B428} (1998) 105--114},
\href{http://arxiv.org/abs/hep-th/9802109}{{\ttfamily arXiv:hep-th/9802109
  [hep-th]}}.

\bibitem{Arutyunov:2000py}
G.~Arutyunov and S.~Frolov, ``{Four point functions of lowest weight CPOs in
  N=4 SYM(4) in supergravity approximation},''
  \href{http://dx.doi.org/10.1103/PhysRevD.62.064016}{{\em Phys. Rev.}
  {\bfseries D62} (2000) 064016},
\href{http://arxiv.org/abs/hep-th/0002170}{{\ttfamily arXiv:hep-th/0002170
  [hep-th]}}.

\bibitem{Arutyunov:2002fh}
G.~Arutyunov, F.~A. Dolan, H.~Osborn, and E.~Sokatchev, ``{Correlation
  functions and massive Kaluza-Klein modes in the AdS / CFT correspondence},''
  \href{http://dx.doi.org/10.1016/S0550-3213(03)00448-6}{{\em Nucl. Phys.}
  {\bfseries B665} (2003) 273--324},
\href{http://arxiv.org/abs/hep-th/0212116}{{\ttfamily arXiv:hep-th/0212116
  [hep-th]}}.

\bibitem{Arutyunov:2003ae}
G.~Arutyunov and E.~Sokatchev, ``{On a large N degeneracy in N=4 SYM and the
  AdS / CFT correspondence},''
  \href{http://dx.doi.org/10.1016/S0550-3213(03)00353-5}{{\em Nucl. Phys.}
  {\bfseries B663} (2003) 163--196},
\href{http://arxiv.org/abs/hep-th/0301058}{{\ttfamily arXiv:hep-th/0301058
  [hep-th]}}.

\bibitem{Berdichevsky:2007xd}
L.~Berdichevsky and P.~Naaijkens, ``{Four-point functions of different-weight
  operators in the AdS/CFT correspondence},''
  \href{http://dx.doi.org/10.1088/1126-6708/2008/01/071}{{\em JHEP} {\bfseries
  01} (2008) 071},
\href{http://arxiv.org/abs/0709.1365}{{\ttfamily arXiv:0709.1365 [hep-th]}}.

\bibitem{Uruchurtu:2008kp}
L.~I. Uruchurtu, ``{Four-point correlators with higher weight superconformal
  primaries in the AdS/CFT Correspondence},''
  \href{http://dx.doi.org/10.1088/1126-6708/2009/03/133}{{\em JHEP} {\bfseries
  03} (2009) 133},
\href{http://arxiv.org/abs/0811.2320}{{\ttfamily arXiv:0811.2320 [hep-th]}}.

\bibitem{Uruchurtu:2011wh}
L.~I. Uruchurtu, ``{Next-next-to-extremal Four Point Functions of N=4 1/2 BPS
  Operators in the AdS/CFT Correspondence},''
  \href{http://dx.doi.org/10.1007/JHEP08(2011)133}{{\em JHEP} {\bfseries 08}
  (2011) 133},
\href{http://arxiv.org/abs/1106.0630}{{\ttfamily arXiv:1106.0630 [hep-th]}}.

\bibitem{DHoker:2000xhf}
E.~D'Hoker, J.~Erdmenger, D.~Z. Freedman, and M.~Perez-Victoria, ``{Near
  extremal correlators and vanishing supergravity couplings in AdS / CFT},''
  \href{http://dx.doi.org/10.1016/S0550-3213(00)00534-4}{{\em Nucl. Phys.}
  {\bfseries B589} (2000) 3--37},
\href{http://arxiv.org/abs/hep-th/0003218}{{\ttfamily arXiv:hep-th/0003218
  [hep-th]}}.

\bibitem{Arutyunov:1999fb}
G.~Arutyunov and S.~Frolov, ``{Scalar quartic couplings in type IIB
  supergravity on AdS(5) x S**5},''
  \href{http://dx.doi.org/10.1016/S0550-3213(00)00210-8}{{\em Nucl. Phys.}
  {\bfseries B579} (2000) 117--176},
\href{http://arxiv.org/abs/hep-th/9912210}{{\ttfamily arXiv:hep-th/9912210
  [hep-th]}}.

\bibitem{Arutyunov:2000ima}
G.~Arutyunov and S.~Frolov, ``{On the correspondence between gravity fields and
  CFT operators},'' \href{http://dx.doi.org/10.1088/1126-6708/2000/04/017}{{\em
  JHEP} {\bfseries 04} (2000) 017},
\href{http://arxiv.org/abs/hep-th/0003038}{{\ttfamily arXiv:hep-th/0003038
  [hep-th]}}.

\bibitem{DHoker:1998bqu}
E.~D'Hoker and D.~Z. Freedman, ``{Gauge boson exchange in AdS(d+1)},''
  \href{http://dx.doi.org/10.1016/S0550-3213(98)00852-9}{{\em Nucl. Phys.}
  {\bfseries B544} (1999) 612--632},
\href{http://arxiv.org/abs/hep-th/9809179}{{\ttfamily arXiv:hep-th/9809179
  [hep-th]}}.

\bibitem{DHoker:1998ecp}
E.~D'Hoker and D.~Z. Freedman, ``{General scalar exchange in AdS(d+1)},''
  \href{http://dx.doi.org/10.1016/S0550-3213(99)00169-8}{{\em Nucl. Phys.}
  {\bfseries B550} (1999) 261--288},
\href{http://arxiv.org/abs/hep-th/9811257}{{\ttfamily arXiv:hep-th/9811257
  [hep-th]}}.

\bibitem{Liu:1998th}
H.~Liu, ``{Scattering in anti-de Sitter space and operator product
  expansion},'' \href{http://dx.doi.org/10.1103/PhysRevD.60.106005}{{\em Phys.
  Rev.} {\bfseries D60} (1999) 106005},
\href{http://arxiv.org/abs/hep-th/9811152}{{\ttfamily arXiv:hep-th/9811152
  [hep-th]}}.

\bibitem{DHoker:1999bve}
E.~D'Hoker, D.~Z. Freedman, S.~D. Mathur, A.~Matusis, and L.~Rastelli,
  ``{Graviton and gauge boson propagators in AdS(d+1)},''
  \href{http://dx.doi.org/10.1016/S0550-3213(99)00524-6}{{\em Nucl. Phys.}
  {\bfseries B562} (1999) 330--352},
\href{http://arxiv.org/abs/hep-th/9902042}{{\ttfamily arXiv:hep-th/9902042
  [hep-th]}}.

\bibitem{DHoker:1999pj}
E.~D'Hoker, D.~Z. Freedman, S.~D. Mathur, A.~Matusis, and L.~Rastelli,
  ``{Graviton exchange and complete four point functions in the AdS / CFT
  correspondence},''
  \href{http://dx.doi.org/10.1016/S0550-3213(99)00525-8}{{\em Nucl. Phys.}
  {\bfseries B562} (1999) 353--394},
\href{http://arxiv.org/abs/hep-th/9903196}{{\ttfamily arXiv:hep-th/9903196
  [hep-th]}}.

\bibitem{DHoker:1999aa}
E.~D'Hoker, D.~Z. Freedman, and L.~Rastelli, ``Ads / cft four point functions:
  How to succeed at z integrals without really trying,'' {\em Nucl. Phys.}
  {\bfseries B562} (1999) 395--411.

\bibitem{Arutyunov:1999en}
G.~Arutyunov and S.~Frolov, ``{Some cubic couplings in type IIB supergravity on
  AdS(5) x S**5 and three point functions in SYM(4) at large N},''
  \href{http://dx.doi.org/10.1103/PhysRevD.61.064009}{{\em Phys. Rev.}
  {\bfseries D61} (2000) 064009},
\href{http://arxiv.org/abs/hep-th/9907085}{{\ttfamily arXiv:hep-th/9907085
  [hep-th]}}.

\bibitem{Elvang:2015rqa}
H.~Elvang and Y.-t. Huang, {\em {Scattering Amplitudes in Gauge Theory and
  Gravity}}.
\newblock Cambridge University Press, 2015.
\newblock
\url{http://www.cambridge.org/mw/academic/subjects/physics/theoretical-physics-and-mathematical-physics/scattering-amplitudes-gauge-theory-and-gravity?format=AR}.
\newblock

\bibitem{nima}
N.~Arkani-Hamed, J.~Bourjaily, F.~Cachazo, A.~Goncharov, A.~Postnikov, and
  J.~Trnka, {\em "{Grassmannian Geometry of Scattering Amplitudes}"}.
\newblock Cambridge University Press, 2016.

\bibitem{Mack:2009mi}
G.~Mack, ``{D-independent representation of Conformal Field Theories in D
  dimensions via transformation to auxiliary Dual Resonance Models. Scalar
  amplitudes},''
\href{http://arxiv.org/abs/0907.2407}{{\ttfamily arXiv:0907.2407 [hep-th]}}.

\bibitem{Penedones:2010ue}
J.~Penedones, ``{Writing CFT correlation functions as AdS scattering
  amplitudes},'' \href{http://dx.doi.org/10.1007/JHEP03(2011)025}{{\em JHEP}
  {\bfseries 03} (2011) 025},
\href{http://arxiv.org/abs/1011.1485}{{\ttfamily arXiv:1011.1485 [hep-th]}}.

\bibitem{Paulos:2011ie}
M.~F. Paulos, ``{Towards Feynman rules for Mellin amplitudes},''
  \href{http://dx.doi.org/10.1007/JHEP10(2011)074}{{\em JHEP} {\bfseries 10}
  (2011) 074},
\href{http://arxiv.org/abs/1107.1504}{{\ttfamily arXiv:1107.1504 [hep-th]}}.

\bibitem{Fitzpatrick:2011ia}
A.~L. Fitzpatrick, J.~Kaplan, J.~Penedones, S.~Raju, and B.~C. van Rees, ``{A
  Natural Language for AdS/CFT Correlators},''
  \href{http://dx.doi.org/10.1007/JHEP11(2011)095}{{\em JHEP} {\bfseries 11}
  (2011) 095},
\href{http://arxiv.org/abs/1107.1499}{{\ttfamily arXiv:1107.1499 [hep-th]}}.

\bibitem{Costa:2012cb}
M.~S. Costa, V.~Goncalves, and J.~Penedones, ``{Conformal Regge theory},''
  \href{http://dx.doi.org/10.1007/JHEP12(2012)091}{{\em JHEP} {\bfseries 12}
  (2012) 091},
\href{http://arxiv.org/abs/1209.4355}{{\ttfamily arXiv:1209.4355 [hep-th]}}.

\bibitem{Aharony:2016dwx}
O.~Aharony, L.~F. Alday, A.~Bissi, and E.~Perlmutter, ``{Loops in AdS from
  Conformal Field Theory},''
\href{http://arxiv.org/abs/1612.03891}{{\ttfamily arXiv:1612.03891 [hep-th]}}.

\bibitem{Dolan:2006ec}
F.~A. Dolan, M.~Nirschl, and H.~Osborn, ``{Conjectures for large N
  superconformal N=4 chiral primary four point functions},''
  \href{http://dx.doi.org/10.1016/j.nuclphysb.2006.05.009}{{\em Nucl. Phys.}
  {\bfseries B749} (2006) 109--152},
\href{http://arxiv.org/abs/hep-th/0601148}{{\ttfamily arXiv:hep-th/0601148
  [hep-th]}}.

\bibitem{Dolan:2002zh}
F.~A. Dolan and H.~Osborn, ``{On short and semi-short representations for
  four-dimensional superconformal symmetry},''
  \href{http://dx.doi.org/10.1016/S0003-4916(03)00074-5}{{\em Annals Phys.}
  {\bfseries 307} (2003) 41--89},
\href{http://arxiv.org/abs/hep-th/0209056}{{\ttfamily arXiv:hep-th/0209056
  [hep-th]}}.

\bibitem{Kim:1985ez}
H.~J. Kim, L.~J. Romans, and P.~van Nieuwenhuizen, ``{The Mass Spectrum of
  Chiral N=2 D=10 Supergravity on S**5},''
\href{http://dx.doi.org/10.1103/PhysRevD.32.389}{{\em Phys. Rev.} {\bfseries
  D32} (1985) 389}.

\bibitem{Lee:1999pj}
S.~Lee, ``{AdS(5) / CFT(4) four point functions of chiral primary operators:
  Cubic vertices},''
  \href{http://dx.doi.org/10.1016/S0550-3213(99)00614-8}{{\em Nucl. Phys.}
  {\bfseries B563} (1999) 349--360},
\href{http://arxiv.org/abs/hep-th/9907108}{{\ttfamily arXiv:hep-th/9907108
  [hep-th]}}.

\bibitem{Arutyunov:2017dti}
G.~Arutyunov, S.~Frolov, R.~Klabbers, and S.~Savin, ``{Towards 4-point
  correlation functions of any 1/2-BPS operators from supergravity},''
\href{http://arxiv.org/abs/1701.00998}{{\ttfamily arXiv:1701.00998 [hep-th]}}.

\bibitem{Fitzpatrick:2012cg}
A.~L. Fitzpatrick and J.~Kaplan, ``{AdS Field Theory from Conformal Field
  Theory},'' \href{http://dx.doi.org/10.1007/JHEP02(2013)054}{{\em JHEP}
  {\bfseries 02} (2013) 054},
\href{http://arxiv.org/abs/1208.0337}{{\ttfamily arXiv:1208.0337 [hep-th]}}.

\bibitem{Costa:2014kfa}
M.~S. Costa, V.~Gon{\c c}alves, and J.~Penedones, ``{Spinning AdS
  Propagators},'' \href{http://dx.doi.org/10.1007/JHEP09(2014)064}{{\em JHEP}
  {\bfseries 09} (2014) 064},
\href{http://arxiv.org/abs/1404.5625}{{\ttfamily arXiv:1404.5625 [hep-th]}}.

\bibitem{Goncalves:2014rfa}
V.~Gon{\c c}alves, J.~Penedones, and E.~Trevisani, ``{Factorization of Mellin
  amplitudes},'' \href{http://dx.doi.org/10.1007/JHEP10(2015)040}{{\em JHEP}
  {\bfseries 10} (2015) 040},
\href{http://arxiv.org/abs/1410.4185}{{\ttfamily arXiv:1410.4185 [hep-th]}}.

\bibitem{Paulos:2012nu}
M.~F. Paulos, M.~Spradlin, and A.~Volovich, ``{Mellin Amplitudes for Dual
  Conformal Integrals},'' \href{http://dx.doi.org/10.1007/JHEP08(2012)072}{{\em
  JHEP} {\bfseries 08} (2012) 072},
\href{http://arxiv.org/abs/1203.6362}{{\ttfamily arXiv:1203.6362 [hep-th]}}.

\bibitem{Nandan:2013ip}
D.~Nandan, M.~F. Paulos, M.~Spradlin, and A.~Volovich, ``{Star Integrals,
  Convolutions and Simplices},''
  \href{http://dx.doi.org/10.1007/JHEP05(2013)105}{{\em JHEP} {\bfseries 05}
  (2013) 105},
\href{http://arxiv.org/abs/1301.2500}{{\ttfamily arXiv:1301.2500 [hep-th]}}.

\bibitem{Lowe:2016ucg}
D.~A. Lowe, ``{Mellin transforming the minimal model CFTs: AdS/CFT at strong
  curvature},'' \href{http://dx.doi.org/10.1016/j.physletb.2016.07.029}{{\em
  Phys. Lett.} {\bfseries B760} (2016) 494--497},
\href{http://arxiv.org/abs/1602.05613}{{\ttfamily arXiv:1602.05613 [hep-th]}}.

\bibitem{Paulos:2016fap}
M.~F. Paulos, J.~Penedones, J.~Toledo, B.~C. van Rees, and P.~Vieira, ``{The
  S-matrix Bootstrap I: QFT in AdS},''
\href{http://arxiv.org/abs/1607.06109}{{\ttfamily arXiv:1607.06109 [hep-th]}}.

\bibitem{Nizami:2016jgt}
A.~A. Nizami, A.~Rudra, S.~Sarkar, and M.~Verma, ``{Exploring Perturbative
  Conformal Field Theory in Mellin space},''
\href{http://arxiv.org/abs/1607.07334}{{\ttfamily arXiv:1607.07334 [hep-th]}}.

\bibitem{Gopakumar:2016cpb}
R.~Gopakumar, A.~Kaviraj, K.~Sen, and A.~Sinha, ``{A Mellin space approach to
  the conformal bootstrap},''
\href{http://arxiv.org/abs/1611.08407}{{\ttfamily arXiv:1611.08407 [hep-th]}}.

\bibitem{Gopakumar:2016wkt}
R.~Gopakumar, A.~Kaviraj, K.~Sen, and A.~Sinha, ``{Conformal Bootstrap in
  Mellin Space},''
\href{http://arxiv.org/abs/1609.00572}{{\ttfamily arXiv:1609.00572 [hep-th]}}.

\bibitem{Dey:2016mcs}
P.~Dey, A.~Kaviraj, and A.~Sinha, ``{Mellin space bootstrap for global
  symmetry},''
\href{http://arxiv.org/abs/1612.05032}{{\ttfamily arXiv:1612.05032 [hep-th]}}.

\bibitem{Rastelli:2017ecj}
L.~Rastelli and X.~Zhou, ``{The Mellin Formalism for Boundary CFT$_d$},''
  \href{http://dx.doi.org/10.1007/JHEP10(2017)146}{{\em JHEP} {\bfseries 10}
  (2017) 146},
\href{http://arxiv.org/abs/1705.05362}{{\ttfamily arXiv:1705.05362 [hep-th]}}.

\bibitem{Dey:2017fab}
P.~Dey, K.~Ghosh, and A.~Sinha, ``{Simplifying large spin bootstrap in Mellin
  space},''
\href{http://arxiv.org/abs/1709.06110}{{\ttfamily arXiv:1709.06110 [hep-th]}}.

\bibitem{Yuan:2017vgp}
E.~Y. Yuan, ``{Loops in the Bulk},''
\href{http://arxiv.org/abs/1710.01361}{{\ttfamily arXiv:1710.01361 [hep-th]}}.

\bibitem{Fitzpatrick:2012yx}
A.~L. Fitzpatrick, J.~Kaplan, D.~Poland, and D.~Simmons-Duffin, ``{The Analytic
  Bootstrap and AdS Superhorizon Locality},''
  \href{http://dx.doi.org/10.1007/JHEP12(2013)004}{{\em JHEP} {\bfseries 12}
  (2013) 004},
\href{http://arxiv.org/abs/1212.3616}{{\ttfamily arXiv:1212.3616 [hep-th]}}.

\bibitem{Komargodski:2012ek}
Z.~Komargodski and A.~Zhiboedov, ``{Convexity and Liberation at Large Spin},''
  \href{http://dx.doi.org/10.1007/JHEP11(2013)140}{{\em JHEP} {\bfseries 11}
  (2013) 140},
\href{http://arxiv.org/abs/1212.4103}{{\ttfamily arXiv:1212.4103 [hep-th]}}.

\bibitem{Caron-Huot:2017vep}
S.~Caron-Huot, ``{Analyticity in Spin in Conformal Theories},''
\href{http://arxiv.org/abs/1703.00278}{{\ttfamily arXiv:1703.00278 [hep-th]}}.

\bibitem{Li:2017lmh}
D.~Li, D.~Meltzer, and D.~Poland, ``{Conformal Bootstrap in the Regge Limit},''
\href{http://arxiv.org/abs/1705.03453}{{\ttfamily arXiv:1705.03453 [hep-th]}}.

\bibitem{Alday:2017gde}
L.~F. Alday, A.~Bissi, and E.~Perlmutter, ``{Holographic Reconstruction of AdS
  Exchanges from Crossing Symmetry},''
\href{http://arxiv.org/abs/1705.02318}{{\ttfamily arXiv:1705.02318 [hep-th]}}.

\bibitem{Kulaxizi:2017ixa}
M.~Kulaxizi, A.~Parnachev, and A.~Zhiboedov, ``{Bulk Phase Shift, CFT Regge
  Limit and Einstein Gravity},''
\href{http://arxiv.org/abs/1705.02934}{{\ttfamily arXiv:1705.02934 [hep-th]}}.

\bibitem{Fitzpatrick:2011hu}
A.~L. Fitzpatrick and J.~Kaplan, ``{Analyticity and the Holographic
  S-Matrix},'' \href{http://dx.doi.org/10.1007/JHEP10(2012)127}{{\em JHEP}
  {\bfseries 10} (2012) 127},
\href{http://arxiv.org/abs/1111.6972}{{\ttfamily arXiv:1111.6972 [hep-th]}}.

\bibitem{Eden:2000bk}
B.~Eden, A.~C. Petkou, C.~Schubert, and E.~Sokatchev, ``{Partial
  nonrenormalization of the stress tensor four point function in N=4 SYM and
  AdS / CFT},'' \href{http://dx.doi.org/10.1016/S0550-3213(01)00151-1}{{\em
  Nucl. Phys.} {\bfseries B607} (2001) 191--212},
\href{http://arxiv.org/abs/hep-th/0009106}{{\ttfamily arXiv:hep-th/0009106
  [hep-th]}}.

\bibitem{Nirschl:2004pa}
M.~Nirschl and H.~Osborn, ``{Superconformal Ward identities and their
  solution},'' \href{http://dx.doi.org/10.1016/j.nuclphysb.2005.01.013}{{\em
  Nucl. Phys.} {\bfseries B711} (2005) 409--479},
\href{http://arxiv.org/abs/hep-th/0407060}{{\ttfamily arXiv:hep-th/0407060
  [hep-th]}}.

\bibitem{Ponomarev:2017qab}
D.~Ponomarev, ``{A Note on (Non)-Locality in Holographic Higher Spin
  Theories},'' \href{http://dx.doi.org/10.3390/universe4010002}{{\em Universe}
  {\bfseries 4} no.~1, (2018) 2},
\href{http://arxiv.org/abs/1710.00403}{{\ttfamily arXiv:1710.00403 [hep-th]}}.

\bibitem{Aprile:2017xsp}
F.~Aprile, J.~M. Drummond, P.~Heslop, and H.~Paul, ``{Unmixing Supergravity},''
\href{http://arxiv.org/abs/1706.08456}{{\ttfamily arXiv:1706.08456 [hep-th]}}.

\bibitem{Aprile:2018efk}
F.~Aprile, J.~Drummond, P.~Heslop, and H.~Paul, ``{The double-trace spectrum of
  $N=4$ SYM at strong coupling},''
\href{http://arxiv.org/abs/1802.06889}{{\ttfamily arXiv:1802.06889 [hep-th]}}.

\bibitem{Dolan:2001tt}
F.~A. Dolan and H.~Osborn, ``{Superconformal symmetry, correlation functions
  and the operator product expansion},''
  \href{http://dx.doi.org/10.1016/S0550-3213(02)00096-2}{{\em Nucl. Phys.}
  {\bfseries B629} (2002) 3--73},
\href{http://arxiv.org/abs/hep-th/0112251}{{\ttfamily arXiv:hep-th/0112251
  [hep-th]}}.

\bibitem{Symanzik:1972wj}
K.~Symanzik, ``{On Calculations in conformal invariant field theories},''
\href{http://dx.doi.org/10.1007/BF02824349}{{\em Lett. Nuovo Cim.} {\bfseries
  3} (1972) 734--738}.

\bibitem{Beem:2013sza}
C.~Beem, M.~Lemos, P.~Liendo, W.~Peelaers, L.~Rastelli, and B.~C. van Rees,
  ``{Infinite Chiral Symmetry in Four Dimensions},''
  \href{http://dx.doi.org/10.1007/s00220-014-2272-x}{{\em Commun. Math. Phys.}
  {\bfseries 336} no.~3, (2015) 1359--1433},
\href{http://arxiv.org/abs/1312.5344}{{\ttfamily arXiv:1312.5344 [hep-th]}}.

\bibitem{Alday:2017xua}
L.~F. Alday and A.~Bissi, ``{Loop Corrections to Supergravity on $AdS_5 \times
  S^5$},''
\href{http://arxiv.org/abs/1706.02388}{{\ttfamily arXiv:1706.02388 [hep-th]}}.

\bibitem{Aprile:2017bgs}
F.~Aprile, J.~M. Drummond, P.~Heslop, and H.~Paul, ``{Quantum Gravity from
  Conformal Field Theory},''
\href{http://arxiv.org/abs/1706.02822}{{\ttfamily arXiv:1706.02822 [hep-th]}}.

\bibitem{Raju:2010by}
S.~Raju, ``{BCFW for Witten Diagrams},''
  \href{http://dx.doi.org/10.1103/PhysRevLett.106.091601}{{\em Phys. Rev.
  Lett.} {\bfseries 106} (2011) 091601},
\href{http://arxiv.org/abs/1011.0780}{{\ttfamily arXiv:1011.0780 [hep-th]}}.

\bibitem{Raju:2012zr}
S.~Raju, ``{New Recursion Relations and a Flat Space Limit for AdS/CFT
  Correlators},'' \href{http://dx.doi.org/10.1103/PhysRevD.85.126009}{{\em
  Phys. Rev.} {\bfseries D85} (2012) 126009},
\href{http://arxiv.org/abs/1201.6449}{{\ttfamily arXiv:1201.6449 [hep-th]}}.

\bibitem{longads7}
L.~Rastelli and X.~Zhou, ``{Holographic Four-Point Functions in the (2, 0)
  Theory},''
\href{http://arxiv.org/abs/1712.02788}{{\ttfamily arXiv:1712.02788 [hep-th]}}.

\bibitem{ads3}
L.~Rastelli, K.~Roumpedakis, and X.~Zhou {\em in progress} .

\bibitem{Galliani:2017jlg}
A.~Galliani, S.~Giusto, and R.~Russo, ``{Holographic 4-point correlators with
  heavy states},''
\href{http://arxiv.org/abs/1705.09250}{{\ttfamily arXiv:1705.09250 [hep-th]}}.

\bibitem{Zhou:2017zaw}
X.~Zhou, ``{On Superconformal Four-Point Mellin Amplitudes in Dimension
  $d>2$},''
\href{http://arxiv.org/abs/1712.02800}{{\ttfamily arXiv:1712.02800 [hep-th]}}.

\bibitem{Naqvi:1999va}
A.~Naqvi, ``{Propagators for massive symmetric tensor and p forms in
  AdS(d+1)},'' \href{http://dx.doi.org/10.1088/1126-6708/1999/12/025}{{\em
  JHEP} {\bfseries 12} (1999) 025},
\href{http://arxiv.org/abs/hep-th/9911182}{{\ttfamily arXiv:hep-th/9911182
  [hep-th]}}.

\bibitem{Dolan:2004iy}
F.~A. Dolan and H.~Osborn, ``{Conformal partial wave expansions for N=4 chiral
  four point functions},''
  \href{http://dx.doi.org/10.1016/j.aop.2005.07.005}{{\em Annals Phys.}
  {\bfseries 321} (2006) 581--626},
\href{http://arxiv.org/abs/hep-th/0412335}{{\ttfamily arXiv:hep-th/0412335
  [hep-th]}}.

\end{thebibliography}\endgroup
\bibliographystyle{utphys}

\end{document}